\documentclass[useAMS,usenatbib]{mnras}
\usepackage{graphicx,subfig}
\usepackage{float}
\usepackage{longtable}
\usepackage{bm}
\usepackage{amsmath}
\usepackage{amsfonts,amssymb}
\usepackage{color}
\definecolor{v}{rgb}{0.6, 0.2, 0.8}
\definecolor{oceanboatblue}{rgb}{0.0, 0.47, 0.75} 
\definecolor{tomas}{rgb}{0.2, 0.8, 0.2}
\usepackage{soul} 
\usepackage{ulem}
\usepackage{enumerate}
\usepackage{float}
\usepackage{subfig}
\usepackage{tabularx}
\usepackage{multicol}        
\usepackage{bm}	
\hypersetup{draft}

\begin{document}

\title[Dark energy with new SL systems]{Testing dark energy models with a new sample of strong-lensing systems}
\author[H. Amante, Maga\~na, Motta, Garc\'{\i}a-Aspeitia, Verdugo]{Mario H. Amante$^1$\thanks{E-mail:mario.herrera@fisica.uaz.edu.mx}, Juan Maga\~na$^{2,3,4}$ \thanks{E-mail:jmagana@astro.puc.cl}, V. Motta$^{4}$\thanks{E-mail: veronica.motta@uv.cl}, Miguel A. Garc\'ia-Aspeitia$^{1,5}$\thanks{E-mail:aspeitia@fisica.uaz.edu.mx} and
\newauthor
Tom\'as Verdugo$^{6}\thanks{E-mail:tomasv@astro.unam.mx}$\\
$^{1}$ Unidad Acad\'emica de F\'isica, Universidad Aut\'onoma de Zacatecas, \\Calzada Solidaridad esquina con Paseo a la Bufa S/N C.P. 98060,Zacatecas, M\'exico.\\
$^2$Instituto de Astrof\'isica, Pontificia Universidad Cat\'olica de Chile, Av. Vicu\~na Mackenna, 4860, Santiago, Chile.\\
$^3$Centro de Astro-Ingenier\'ia, Pontificia Universidad Cat\'olica de Chile, Av. Vicu\~na Mackenna, 4860, Santiago, Chile.\\
$^4$Instituto de F\'isica y Astronom\'ia, Facultad de Ciencias, \\Universidad de Valpara\'iso, Avda. Gran Breta\~na 1111, Valpara\'iso, Chile. \\
$^{5}$Consejo Nacional de Ciencia y Tecnolog\'ia, Av. Insurgentes Sur 1582. \\ Colonia Cr\'edito Constructor, Del. Benito Ju\'arez C.P. 03940, Ciudad de M\'exico, M\'exico.\\
$^{6}$Instituto de Astronom\'ia, Universidad Nacional Aut\'onoma de M\'exico, Apartado postal 106, C.P. 22800,  Ensenada, B.C., M\'exico}

\date{Accepted YYYYMMDD. Received YYYYMMDD; in original form YYYYMMDD}
\pubyear{2017}


\maketitle

\begin{abstract}
Inspired by a new compilation of strong lensing systems, which consist of 204 points in the redshift range  $0.0625< z_{l} < 0.958$ for the lens and $0.196< z_{s} < 3.595$ for the source, we constrain three models that generate a late cosmic acceleration: the $\omega$-cold dark matter model, the Chevallier-Polarski-Linder and the Jassal-Bagla-Padmanabhan parametrizations. Our compilation contains only those systems with early type galaxies acting as lenses, with spectroscopically measured stellar velocity dispersions, estimated Einstein radius, and both the lens and source redshifts. We assume an axially symmetric mass distribution in the lens equation, using a correction to alleviate differences between the measured velocity dispersion ($\sigma$) and the dark matter halo velocity dispersion ($\sigma_{DM}$) as well as other systematic errors that may affect the measurements. 
We have considered different sub-samples to constrain the cosmological parameters of each model. Additionally, we generate a mock data of SLS to asses the impact of the chosen mass profile on the accuracy of Einstein radius estimation. Our results show that cosmological constraints are very sensitive to the selected data: some cases show convergence problems in the estimation of cosmological parameters (e.g. systems with observed distance ratio $D^{obs}<0.5$), others show high values for the chi-square function (e.g. systems with a lens equation $D^{obs} >1$ or high velocity dispersion $\sigma > 276$ km s$^{-1}$). However, we obtained a fiduciary sample with 143 systems which improves the  constraints on each tested cosmological model.
\end{abstract}

\begin{keywords}
gravitational lensing: strong, dark energy, cosmology: observations, cosmology: theory, cosmological parameters.
\end{keywords}

\section{Introduction}

 Contrasting cosmological models with modern observations is fundamental to understand the nature of the $\sim96\%$ of our Universe \citep{Planck:2016,Planck_CP:2018} known as the dark sector, which refers to $\sim26\%$ of the total content in dark matter (DM) the one responsible for the formation of large-scale structure, and $\sim69\%$ in dark energy (DE), the possible cause for the current accelerated expansion \citep{Schmidt,Perlmutter,Riess}. 
In the most accepted paradigm, DM is a non-relativistic matter in the decoupling epoch (i.e. cold), and the traditional way to treat the DE nature is through the addition of an effective cosmological constant (CC) in the energy-momentum tensor of the Einstein field equations. The CC origin is related to the quantum vacuum fluctuations, but this hypothesis is plagued by severe pathologies due to its inability to renormalize the energy density of quantum vacuum, obtaining a discrepancy of $\sim120$ orders of magnitude between the theoretical estimations and the cosmological observations \citep{Zeldovich,Weinberg}. The CC also has the coincidence problem, i.e.  why the Universe transition, from a decelerated to an accelerated phase, is produced at late times.

The CC theoretical problems have led the community to propose a variety of ideas to reproduce the late cosmic acceleration. Some of them postulate the existence of DE, for example, quintessence \citep{Ratra:1988,WETTERICH:1988}, phantom \citep{Chiba:2000,Caldwell:2002} fields, Chaplygin gas \citep{Chaplygin,Kamenshchik:2001,Bilic:2001,Hernandez-Almada:2018osh}, $w(z)$ parameterizations for dynamical DE \citep{Chevallier:2000qy,Linder:2002dt,Jassal}, interacting dark energy \citep{Calrdera:2009}, etc \citep[for a thorough review of all these alternatives see][]{Copeland:2006wr,Li:2011}. 
Other models modify the Einsteinian gravity to resemble the DE like brane models \citep{Aspeitia:2009bj,Garcia:2018,Garcia-Aspeitia:2018fvw}, $f(R)$ models \citep{Buchdahl,STAROBINSKY198099,Sotiriou:2008rp}, scalar-tensor theories \citep{PhysRev.124.925,Galiautdinov:2016qqy,Langlois:2017dyl}, Unimodular gravity \citep{Perez:2017krv,Garcia-Aspeitia:2019yni}, among others.

On the other hand, observational data are used to test these models.  Among the most frequently used are the cosmic microwave background radiation  \citep[CMB,][]{Planck:2016,Planck_CP:2018}, 
baryonic acoustic oscillations \citep[BAO,][]{Eisenstein:2005,Blake:2012,Alam:2017,Bautista:2017},
type Ia Supernovae \citep[SNe Ia,][]{Scolnic:2017caz}
and observational Hubble data \cite[OHD,][]{Jimenez:2001gg,Moresco:2016JCAP,Magana:2017nfs}. Consistency in the cosmological parameters among different techniques, rather than more accurate measurements, is desirable to better understand the nature of DE. In the last years, several efforts have been made by the community to include gravitational lens systems in the study of the Universe's evolution. Some of the pioneers are \citet{Futamase:2000qx,Biesiada:2006zf}, who used only one strong-lens system to study some of the most popular cosmological models.   \citet{2008A&A...477..397G} introduced a methodology to estimate cosmological parameters using Strong Lensing Systems (SLS) \citep[see also][]{Jullo:2010Sci,Magana:2015ApJ,Magana:2018ApJ}. They apply the relation between the Einstein radius and the central stellar velocity dispersion assuming an isothermal profile for the total density distribution of the lens (elliptical) galaxy. Their simulations found that the method is accurate enough to obtain information about the underlying cosmology. They concluded that the stellar velocity dispersion and velocity dispersion of the isothermal lens model are very similar in the $w$ cold dark matter ($w$CDM) model.
\citet{Biesiada:2010} used the same procedure  comparing a distance ratio, $D^{obs}$, constructed from SLS observations such as the Einstein radius and spectroscopic velocity dispersion of the lens galaxy, with a theoretical counterpart, $D^{th}$. By using a sample containing 20 SLS, they demonstrated that this technique is useful to provide insights into DE. \citet{Cao:2012} updated the sample to $80$ systems and proposed a modification that takes into account deviations  from sphericity, i.e. from the singular isothermal sphere (SIS). Later on, \citet{Cao:2015qja} considered lens profile deviations due to the redshift evolution of elliptical galaxies by using
spherically symmetric power-law mass distributions for the lenses and also increased the compilation up to $118$ points. They also explore the consequences of using aperture-corrected velocity dispersions on the parameter estimations. Some authors have pointed out the need for a sufficiently large sample to test DE models with higher precision \citep{Yennapureddy:2018rdz}. For instance, \cite{Melia:2014oqa} have emphasized that a sample of $\sim$ 200 SLS can discern the $R_{h} = ct$ model from the standard one. \citet{Qi:2018arXiv} simulated strong lensing data to constrain the curvature of the Universe and found that, by increasing the sample (16000 lenses) and combining with compact radio quasars, it could be constrained with an accuracy of $\sim 10^{-3}$. 
 Recently, \citet{Leaf:2018lfu} have revisited this cosmological tool with the largest sample of SLS (158) until now, including 40 new systems presented by \citet{2017ApJ...851...48S}. The authors proposed a new approach to improve this technique by introducing in the observational distance ratio error ($\delta D^{obs}$), a parameter $\sigma_{x}$ to take into account the SIE scatter and any other source of errors in the measurements. In their analysis, they excluded $29$ SLS  that are outside the region $0<D_{obs}<1$, and the system SL2SJ085019-034710 \citep{Sonnenfeld:2013xga}  which seems to be an extreme outlier for their models. Their results show that a $\sigma_x = 12.2\%$  provides more statistically significative cosmological constraints. Finally, \citet{Chen:2018jcf} used 157 SLS to analyze the $\Lambda$CDM model. They considered a lens mass distribution  $\rho(r)=\rho_0 r^{-\gamma}$ and three possibilities for the $\gamma$ parameter: a constant value, a dependence with the lens redshift ($z_l$), and a dependence with both the surface mass density and the lens redshift. They concluded that although $\Omega_{0m}$,  used as the only free parameter in $\Lambda$CDM scenario, is very sensitive to the lens mass model, it provides weak constraints which are also in tension with Planck measurements.

In this work we compile a new sample of 204 strong gravitational lens systems, which have been measured by different surveys:
Sloan Lens ACS survey \citep[SLACS][]{bolton2006}; BOSS Emission-Line Lens Survey  \citep[BELLS][]{ brownstein2012};  
CfA-Arizona Space Telescope LEns Survey \citep[CASTLES][]{munoz1998};  
Lenses Structure and Dynamics survey \citep[LSD][]{Treu2004}; CFHT Strong Lensing Legacy Survey \citep[SL2S][]  {cabanac2007};  
Strong-lensing Insights into Dark Energy Survey \citep[STRIDES][]{treu2018}. We have added 47 systems to the last compilation \citet{Chen:2018jcf} with the aim to constrain the parameters for the $w$ cold dark matter ($\omega$CDM) model, the  Chevallier-Polarski-Linder (CPL) and Jassal-Bagla-Padmanabhan (JBP) parameterizations of the DE equation of state. 

The paper is organized as follows. In Sec. \ref{data} we show the data for the strong lensing cosmological observations. In Sec. \ref{MG} we present the Friedmann equations for the models and parametrizations mentioned previously. In section \ref{MS} we introduce the criteria to assess the goodness-of-fit for each case. In section \ref{Res} our results are shown and finally in section \ref{Con} we present the conclusions and perspectives.

\section{Strong lensing as a cosmological test} \label{data}
\subsection{Methodology}
Strong lensing systems have been used over the years to constrain cosmological parameters and supply an alternative way to understand the nature of dark energy \citep{Biesiada:2006zf,Biesiada:2010,Cao:2012,Cao:2015qja,Jullo:2010Sci,Magana:2015ApJ,Magana:2018ApJ}. In this paper we gather new  data from SLS, making a catalog with 204 systems. This compilation allow us to analyze cosmological models with more precision and compare with other astrophysical tools as standard candles or standard rulers \citep{Scolnic:2017caz, Blake:2012,Alam:2017,Bautista:2017}. When a galaxy acts as a lens, the  separation among the multiple-images depends on the deflector mass and the angular diameter distances to the lens and to the source. When a lens is described by a Singular Isothermal Sphere (SIS), the Einstein radius is defined as \citep{Schneider_book:1992} 
\begin{equation}
\theta_{E} = 4 \pi \frac{\sigma^2_{SIS} D_{ls}}{c^2 D_{s}}, \label{thetaESIS}
\end{equation}
where $\sigma_{\mathrm{SIS}}$ is the velocity dispersion of the lensing galaxy, $c$ is the speed of light, $D_{s}$ is the angular diameter distance to the source, and $D_{ls}$ the angular diameter distance between the lens and the source.
The enclosed projected mass inside the Einstein radius ($\theta_E$) is independent of the mass profile \citep{Schneider_book:1992}, generally estimated using an isothermal-ellipsoid mass distribution (SIE). 
Furthermore, it has been demostrated that the lensing mass distribution of early-type galaxies is very close to isothermal \citep{kochanek1995,munoz2001,rusin2002,Treu:2002ee,koopmans2003,rusin2003b,2008A&A...477..397G}.

Since the angular diameter distance $D$, in terms of redshift $z$ is defined as
\begin{equation}
D(z)=\frac{c}{H_0(1+z)}\int_0^z\frac{dz^{\prime}}{E(z^{\prime})},
\end{equation}
\noindent 
being $H_{0}$ the Hubble constant, then the Einstein radius depends on the cosmological model through the dimensionless Friedmann equation $E(z)$. By defining a theoretical distance ratio $D^{th}\equiv D_{ls}/D_{s}$, we obtain

\begin{equation}
D^{th}\left(z_{l}, z_{s}; \bf{\Theta} \right) = \frac{\int^{z_{s}}_{z_{l}}\frac{\mathrm{dz}^{\prime}}{E(z^{\prime},\mathbf{\Theta})}}{\int^{z_{s}}_{0}\frac{\mathrm{dz}^{\prime}}{E(z^{\prime},\mathbf{\Theta})}}, \label{Dth}
\end{equation}
where $\mathbf{\Theta}$ is the free parameter vector for any underlying cosmology, $z_{l}$ and $z_{s}$ are the redshifts to the lens and source respectively.
On the other hand, its observable counterpart can be computed as
\begin{equation}
D^{obs} = \frac{c^2 \theta_{E}}{4 \pi \sigma^2}, \label{Dlens}
\end{equation} 
where $\sigma$ is the measured velocity dispersion of the lens dark matter halo. Therefore, the compilation of SLS with  their measurements for $\sigma$ and $\theta_{E}$ can be used to estimate cosmological parameters \citep{2008A&A...477..397G} by minimizing the following chi-square function,
\begin{equation}
\chi_{\mbox{SL}}^2(\mathbf{\Theta}) = \sum_{i=1}^{N_{SL}} \frac{ \left[ D^{th}\left(z_{l}, z_{s}; \bf{\Theta} \right)  -D^{obs}(\theta_{E},\sigma^2)\right]^2 }{ (\delta D^{\rm{obs}})^2},
\label{eq:chisquareSL}
\end{equation}
where the sum is over all the ($N_{SL}$) lens systems and $\delta D^{\rm{obs}}$ is the uncertainty of  each $D^{obs}$ measurement which can be computed employing the standard way of error propagation as 
\begin{equation}
\delta D^{\rm{obs}}= D^{\rm{obs}} \sqrt{\left( \frac{\delta \theta_E}{\theta_E} \right)^2 + 4 \left( \frac{\delta \sigma}{\sigma} \right)^2},
\label{eq:deltadobs}
\end{equation}
being $\delta \theta_E$ and $\delta \sigma$ the error reported for the Einstein radius and velocity dispersion respectively.\\

One of the advantages of this method is its independence of the Hubble constant $H_{0}$, as it is eliminated in the ratio of two angular diameter distances  (see Eq. \ref{thetaESIS}). Hence, the tension of $H_{0}$ between some of the most reliable measurements \citep{Riess:2016jrr,Aghanim:2018eyx,Riess:2019} is not a problem for this method since it is not neccesary to assume any $H_{0}$ initial value. 
Some disadvantages are: its dependency on the lens model fitted to the data to obtain the Einstein radius \cite[e.g.][]{Cao:2015qja}, the spectroscopically measured stellar velocity dispersion ($\sigma$) that might not be the same as the dark matter halo velocity dispersion $\sigma_{DM}$, and any other systematic error that could change the separation between images or the observed $\sigma$. Consequently, we take into account these uncertainties by introducing  the parameter $f$ into the relation $\sigma_{DM}=f \sigma$, thus Eq. \eqref{thetaESIS} is
\begin{equation}
D^{obs} = \frac{c^2 \theta_{E}}{4 \pi \sigma^2 f^2}. \label{Df}
\end{equation}
\cite{Ofek:2003sp} estimate that those systematics might affect the image separation $\Delta\theta$ up to $\sim 20\%$ (since $\Delta\theta \propto \sigma^{2}$), and thus assume the constraints $(0.8)^{1/2}<f<(1.2)^{1/2}$. Moreover, \citet{Treu:2006ApJ} claim that, for systems with velocity dispersion between $200-300$km s$^{-1}$, there is a relation between the measurement of $\sigma_{spec}$ from spectroscopy and those estimated from the lens model 
\begin{equation}
<f_{SIE}>=\frac{\sigma_{spec}}{\sigma_{SIE}}\approx 1.010\pm 0.017, \label{Treu2006}
\end{equation}
where $\sigma_{SIE}$ is the velocity dispersion obtained from a singular isothermal ellipsoid, and it is also consistent with \citet{Ofek:2003sp} results. 
Notice that \citet{Treu:2006ApJ} relation cannot be used in our case because: a) $\sigma$ for several objects fall outside the interval of validity, and b) it was obtained assuming a $\Lambda$CDM model, and thus it could introduce another source of bias in our estimations. 
Therefore, hereafter we use \citet{Ofek:2003sp} estimation for $f$. We also investigate the repercussions of assuming $f$ as an independent parameter for each SLS to estimate cosmological parameters (see Appendix \ref{Fseparada}). In addition, we analyze a mock catalog of 788 SLS to asess the impact on the Einstein radius estimation when using an isothermal profile instead of a more complex model \eqref{Df} (see Appendix \ref{APII}). 

On the other hand, for some SLS we obtain incorrect $D_{obs}>1$ values (i.e. $D_{ls}>D_{s}$, see Table \ref{tab:SLDgtone}). \citet{Leaf:2018lfu} point out that these values are theoretically unphysical, thus they should be either disregarded or corrected by introducing an extra source of error (e.g. $\delta D^{\rm{obs}}$). However, as the source of such behavior is unknown, we choose to keep these observed systems throughout our analysis (without introducing the suggested $\delta D^{\rm{obs}}$ error) and, instead, offer the parameter estimations with and without those systems for comparison. In the following section we present the data that will be used in the chi-square function (Eq. \ref{eq:chisquareSL}) to test DE models. \\

\subsection{Data}\label{sec:data}
In this section we describe our new compilation of SLS. To construct $D^{obs}$ we have choosen only systems with spectroscopically data well measured from different surveys. We have considered 19 SLS from the CASTLES, 107 from SLACS, 38 from BELLS, 4 from LSD, 35 from SL2S and one system from the DES survey. The final list has a total of 204 systems, being the largest sample of SLS to date. We use spectroscopy to select those lenses with lenticular (S0) or elliptical (E) morphologies which have been modeled assuming a SIS ($\sim 3\%$) or SIE ($\sim 97\%$) lens model. Many systems have not been taken into account due to several issues. For instance, the system PG1115+080 \citep{tonry1998}  from the CASTLES survey has been discarded because the lens mass model is steeper than isothermal. In addition, the system MGJ0751+2716 \citep{2018MNRAS.478.4816S} was also discarded because the main lens belongs to a group of galaxies. From the SLACS survey \citep{Bolton:2008xf,
Auger2009}, we remove the systems SDSSJ1251-0208, SDSSJ1432+6317, SDSSJ1032+5322 and SDSSJ0955+0101 since the lens galaxies are  late-type. The same reason is applied to the systems SDSSJ1611+1705 and SDSSJ1637+1439 from the BELLS survey \citep{0004-637X-833-2-264}. We have also discarded the systems SDSSJ2347$-$0005 and SDSSJ0935$-$0003 from the SLACS survey and the system SDSSJ111040.42$+$364924.4 from the BELLS survey because they have large measured velocity dispersions ($ \sim 400$km s$^{-1}$ or bigger values), suggesting the lens might be part of a group of galaxies or that there is substructure in the line-of sight. For those systems without reported velocity dispersion error, we assumed the average error of the measurements in the survey subsample as follows. For the 9 systems from CASTLES we consider the average error on  $\sigma$ for this survey, i.e. a 14 \%. In the case of the system DES J2146-0047 \citep{Agnello2015}, we have assumed a  10\% error on $\sigma$, which is the average error of the entire sample. The LSD survey \citep{Koopmans:2002qh, Treu2004} reports $\sigma$ corrected by circular aperture using the expression obtained by \cite{1995MNRAS.273.1097J, 1995MNRAS.276.1341J}. A close inspection of  the $\sigma_{spec}$ values, with and without aperture correction, presented by \cite{Cao:2015qja} show the difference is smaller than reported error.
Thus, we decided to use the observed values ($\sigma$) and the reported error for the sample without the apperture correction. 
On the other hand, in those systems in which the Einstein radius error was not reported, we followed \cite{Cao:2015qja} and assumed an error of $\delta\theta_{E}= 0.05$, which is the average value of the systems with reported errors in this sample. 

Our final sample (FS) is presented in Table \ref{tab:SL}, having a total of 204 data points whose lens and source redshifts are in the ranges $0.0625< z_{l} < 0.958$ and $0.196< z_{s} < 3.595$, respectively. All the systems for which we assumed a velocity dispersion error are marked in the sample.

In addition, we have constructed the following subsamples to test the impact on the parameter estimation for the different DE models. We divide the sample into different regions according to the observed value for the distance ratio $D^{obs}$, because there are systems that do not fall in a physical region. We also split the sample into different regions according to the redshift of the lens galaxy to check for any significant changes in the estimation of cosmological parameters  associated to the deflector position. Finally, following \cite{Chen:2018jcf}, we also separate the systems in distinct sub-samples according to the measured velocity dispersion. We name the sub-samples as follows:
\begin{itemize}
    \item SS1: 172 data points with  $D^{obs} \leq 1$ 
    \item SS2: 29 data points with $D^{obs} < 0.5$
    \item SS3: $143$ data points in $0.5 \leq D^{obs} \leq 1$
    \item SS4: $32$ data points with $D^{obs} > 1$
    \item SS5: $64$ data points with $\sigma < 210$ km s$^{-1}$
    \item SS6: $53$ data points in 210 km s$^{-1}$ $\leq \sigma < 243$ km s$^{-1}$
    \item SS7: $49$ data points in 243 km s$^{-1}$$ \leq \sigma \leq 276$ km s$^{-1}$
    \item SS8: 38 data points with $\sigma > 276$ km s$^{-1}$
    \item SS9: $52$ data points with $D^{obs} \leq 1$ and $z_l < 0.2$
    \item SS10: 48 data points with $D^{obs} \leq 1$ and $0.2 \leq z_l \leq 0.4$
    \item SS11: $72$ data points with $D^{obs} \leq 1$ and $z_l>0.4$
\end{itemize}

\section{Cosmological models} \label{MG}
Hereafter we consider the following models with flat geometry that contain dark and baryonic matter and dark energy, neglecting the density parameter of radiation because its contribution at low redshifts is of the order of $\sim10^{-5}$. 

\begin{itemize}

\item  The $\omega$CDM cosmology.- This model is the simplest extension to the CC.  The dark energy has a constant EoS but it deviates from $w_0=-1$, and should satisfy $\omega_0<-1/3$ to obtain an accelerated Universe. The equation $E(z)$ can be written as: 
\begin{eqnarray}
E(z)_{\omega}^2=\Omega_{m0}(1+z)^3+(1 -\Omega_{m0})(1+z)^{3(1+\omega_0)},
\end{eqnarray}
where $\Omega_{m0}$ is the matter density parameter at $z=0$ and the deceleration parameter reads
\begin{eqnarray}
&&q(z)_{\omega}=\frac{1}{2E(z)^2_{\omega}}\Big[3\Omega_{m0}(1+z)^3\nonumber\\&&+3(1+\omega_0)(1-\Omega_{m0})(1+z)^{3(1+\omega_0)}\Big]-1.
\end{eqnarray}
\item The CPL parametrization.- An approach to study dynamical DE models is through a parametrization of its equation of state. One of the most popular is proposed by \cite{Chevallier:2000qy,Linder:2002dt}, and reads as
\begin{equation}
\omega(z)=\omega_0+\omega_1\frac{z}{(1+z)},
\label{eq:eosCPL}
\end{equation}
where $\omega_{0}$ is the EoS at redshift $z=0$ and $\omega_{1}=\mathrm{d}w/\mathrm{d}z|_{z=0}$. The dimensionless $E(z)$ for the CPL parametrization is 
\begin{eqnarray}
&&E(z)_{CPL}^2=\Omega_{m0}(1+z)^3+\nonumber\\&&(1-\Omega_{m0})(1+z)^{3(1+\omega_0+\omega_{1})}\exp\left(\frac{-3\omega_1z}{1+z}\right), \label{CPLE}
\end{eqnarray}
the $q(z)$ is written in the form
\begin{eqnarray}
&&q(z)_{CPL}=\frac{1}{2E(z)^2_{CPL}}\Big[3\Omega_{m0}(1+z)^3\nonumber\\&&+3(1-\Omega_{m0})(1+w_0+w_1)(1+z)^{3(1+\omega_0+\omega_{1})}\times\nonumber\\&&\exp\left(\frac{-3\omega_1z}{1+z}\right)-3w_1(1-\Omega_{m0})(1+z)^{3(1+\omega_0+\omega_{1})-1}\times\nonumber\\&&\exp\left(\frac{-3\omega_1z}{1+z}\right)\Big]-1.
\end{eqnarray}

\item The JBP parametrization.- \cite{Jassal} proposed the following \textit{ansatz} to parametrize the dark energy EoS
\begin{equation}
\omega(z)=\omega_0+\omega_1\frac{z}{(1+z)^2},
\label{eq:eosJBP}
\end{equation}
where $\omega_{0}$ is the EoS at redshift $z=0$ and $\omega_{1}=\mathrm{d}w/\mathrm{d}z|_{z=0}$. The dimensionless $E(z)$ for the JBP parametrization is 
\begin{eqnarray}
&&E(z)_{JBP}^2=\Omega_{m0}(1+z)^3+\nonumber\\&&(1-\Omega_{m0})(1+z)^{3(1+\omega_0)}\exp\left(\frac{3\omega_1z^2}{2(1+z)^2}\right), \label{JBPE}
\end{eqnarray}
the deceleration parameter reads
\begin{eqnarray}
&&q(z)_{JBP}=\frac{1}{2E(z)^2_{JBP}}\Big[3\Omega_{m0}(1+z)^3\nonumber\\&&3(1-\Omega_{m0})(1+\omega_1)(1+z)^{3(1+\omega_1)}\times\nonumber\\&&\exp\left(\frac{3\omega_1z^2}{2(1+z)^2}\right)+3\omega_1(1-\Omega_{m0})z(1+z)^{3(1+\omega_1)-2}\times\nonumber\\&&\exp\left(\frac{3\omega_1z^2}{2(1+z)^2}\right)\Big]-1.
\end{eqnarray}
\end{itemize}
Therefore, it may be possible to reconstruct the $q(z)$ by constraining the parameters for each model and 
determine whether the Universe experiments an accelerated phase at late times.

\section{Model Selection} \label{MS}

To compare among the different DE models, we use the Akaike information criterion \cite[AIC,][]{Akaike:1974}
and Bayesian information criterion  \cite[BIC,][]{Schwarz:1978} defined as:
\begin{eqnarray}
\mbox{AIC}&=&\chi^{2}_{min} +2k, \label{AIC}\\ 
\mbox{BIC}&=&\chi^{2}_{min} + k \ln N, \label{BIC} 
\end{eqnarray}
\noindent 
where $\chi^{2}_{min}$ is the chi-square obtained from the best fit of the parameters, $k$ is the number of parameters and $N$ the number of data points used in the fit. 
A model with smaller AIC and BIC is more favored. Notice that the AIC (BIC) absolute value is irrelevant, the important quantity is the relative value of AIC (BIC) for the model $i$ with respect to the minimum AIC$_{min}$ (BIC$_{min}$) 
among all the models. Table \ref{MScriterion} shows the $\Delta\mathrm{AIC} = \mathrm{AIC}_{i} - \mathrm{AIC}_{min}$ and $\Delta\mathrm{BIC} = \mathrm{BIC}_{i} - \mathrm{BIC}_{min}$ criteria. \citep[see ][and references therein for further details] {Shi:2012ma}.

\begin{table}
\caption{ $\Delta$AIC and $\Delta$BIC criteria.}
\centering
\begin{tabular}{cc}
\hline
$\Delta$AIC	&	Empirical support for model $i$ 	\\
\hline 
0 - 2 &	Substantial \\
4 - 7 &	 Considerably less \\
$>$ 10 & Essentially none\\ \\
\hline
$\Delta$BIC & Evidence against model $i$	\\
\hline
0 - 2 & Not worth more than a bare mention \\
2 - 6 &	Positive \\
6 - 10 & Strong \\
$>$ 10 & Very strong \\
\hline
\end{tabular}
\label{MScriterion}
\end{table}

In addition, to measure the quality of our cosmological constraints we use the FOM estimator
\begin{equation}
\mbox{FoM}=\frac{1}{\sqrt{\det \mbox{Cov} (f1,f2,f3,...)}},
\label{FoM}
\end{equation}
\noindent
where $\mbox{Cov} (f1,f2,f3,...)$ is the covariance matrix of the
cosmological parameters {$f_{i}$} \citep{Wang:2008}. This indicator is a generalization of those proposed by \citet{Albrecht:2006}, and larger values imply stronger constraints on the cosmological parameters since they correspond to a smaller error ellipse.

\section{Results} \label{Res}
In the parameter estimation we have considered the Gaussian likelihood $\mathcal{L}(\mathbf{\Theta})\propto e^{-\chi_{\mathrm{SL}}^{2}(\mathbf{\Theta})}/2$, where the $\chi_{\mathrm{SL}}^{2}(\mathbf{\Theta})$ is given by Eq. (\ref{eq:chisquareSL}).
The free parameters $\mathbf{\Theta}$ for each model were estimated through a MCMC Bayesian statistical analysis. We used the Affine Invariant Markov chain Monte Carlo (MCMC) Ensemble sampler from the \texttt{emcee} \citep{ForemanMackey:2012ig} Python module. We considered 1000 (burn-in-phase) steps to approach the region of the mean value, 5000 MCMC steps and 1000 walkers initialized close to the region of maximum probability according to other astrophysical observations. We check the convergence of the chains using the Gelman-Rubin test proposed by \cite{gelman1992}, stopping the burn-in-phase if all the parameters are less than $1.07$.\\

In each model, we have considered thirteen tests in the Bayesian analysis. The first test was performed employing the FS using the SIS approach given by Eq. \ref{Dlens}; the second test was done using the SS1 sample; and the third test was carried out on the FS sample adding a new parameter ($f$) that takes into account unknown systematics as was previously mentioned in equation \eqref{Df}. We also performed tests with the SLS data binned in $D^{obs}$: the samples SS2, SS3 and SS4. In addition, we estimated the model parameters with the SLS data binned in $\sigma$ using: the SS5 sample; the SS6 sample; the SS7 sample; and the SS8 sample. Finally, we constrained the model parameters with the SLS data binned in the lens redshift ($z_l$) for: the SS9 sample, the SS10 sample, and the SS11 sample.
These tests were carried out assuming a gaussian prior for $\Omega_{0m}=0.3111 \pm 0.0056$, according to the most recent observations from Planck 2018 \citep{Planck_CP:2018}. We also assume the following priors: $-4 <\omega_0< 1$, $-5 < \omega_1 < 5$ and $ (0.8)^2 <f < (1.2)^2$. Table \ref{tab:wCDMCPLyJBPin} provides the mean values for the free parameters of each model using the different cases mentioned above. The error propagation was performed through a Monte Carlo approach at 1$\sigma$ confidence level.  We also present the values for $\chi^{2}_{min}$ and $\chi^{2}_{red} = \chi^{2}_{min}/(N - k) $, where $N$ and $k$ are the number of data points and parameters used in each scenario (see section \ref{MS}). Table \ref{tab:AICBICwCDMCPLyJBPin} gives the AIC, BIC, and FOM values for each cosmological model using
the results of the FS, FS+f, SS3 analysis.

\subsection{The $\omega$CDM constraints}
\begin{figure*}
\centering
\begin{tabular}{cc}
\subfloat[FS with and without systems $D^{obs}>1$.]{\label{figur:1}\includegraphics[width=67mm]{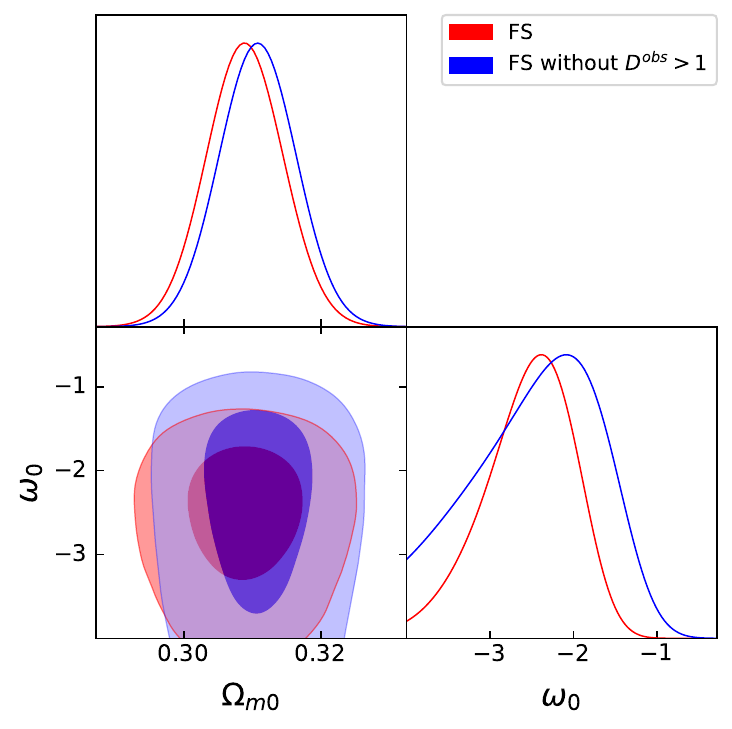}}
  \subfloat[SLS data binned in $D^{obs}$.]{\label{figur:2}\includegraphics[width=67mm]{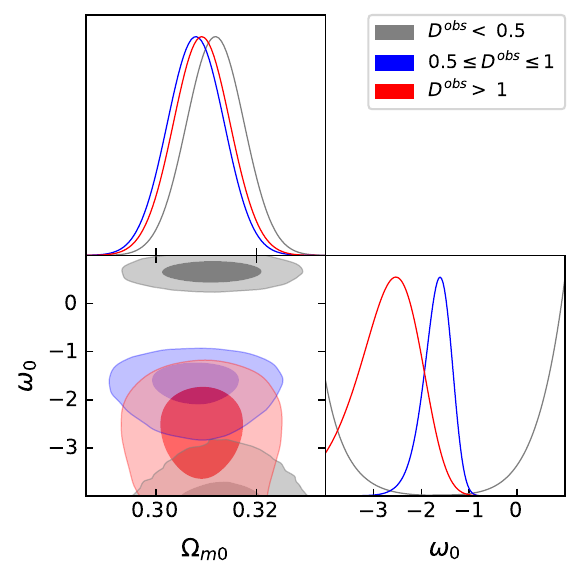}}
  \\
  \subfloat[SLS data binned in $\sigma$. ]{\label{figur:3}\includegraphics[width=67mm]{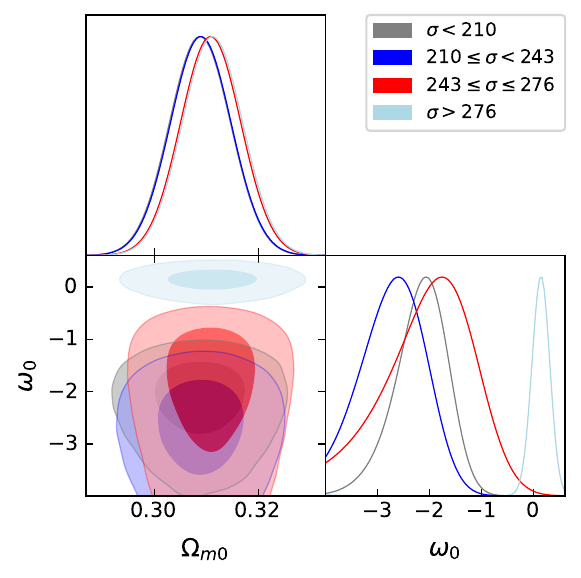}}
  \subfloat[SLS data binned in $z_l$.]{\label{figur:4}\includegraphics[width=67mm]{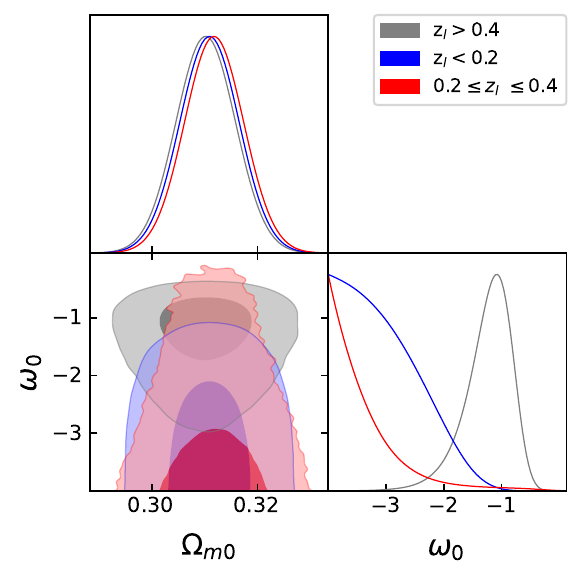}}
  \\
  \subfloat[FS with $f$.]{\label{figur:5}\includegraphics[width=72mm]{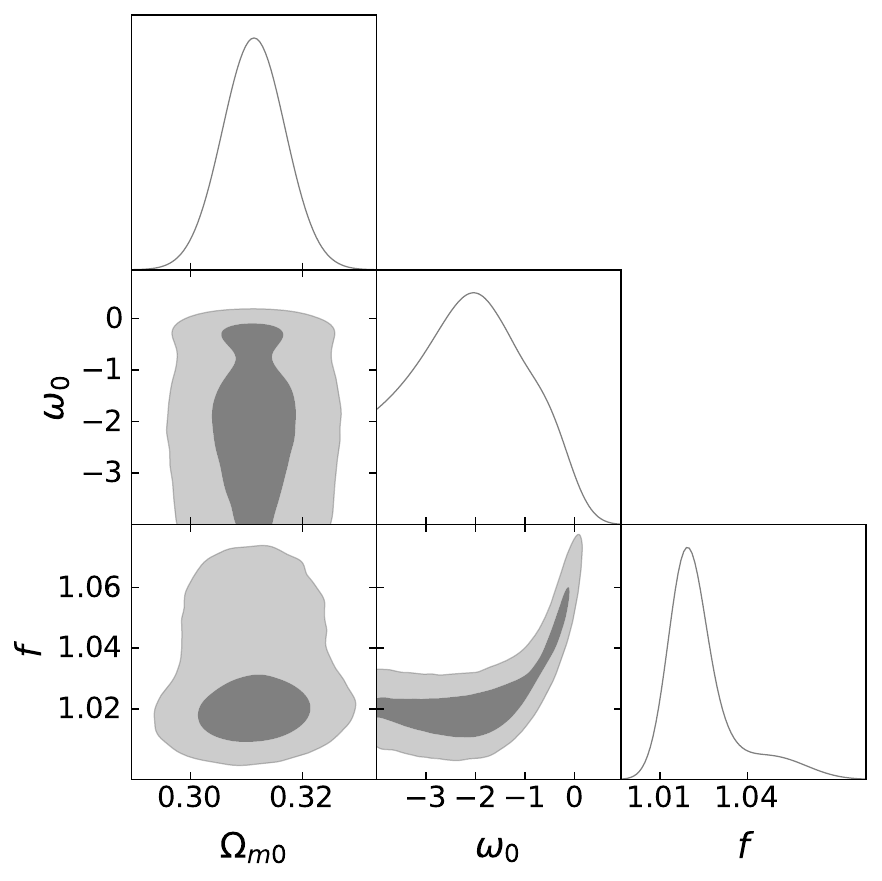}} & \\
  \end{tabular}
\caption{1D marginalized posterior distributions and the 2D $68\%$,  $99.7\%$ confidence levels for the $\Omega_{m0}$ and $\omega_{0}$ parameters of the $\omega$CDM model and $f$ parameter (lower panel) for the different samples presented in Table \ref{tab:wCDMCPLyJBPin}.}
\label{figwCDM}
 \end{figure*}

\begin{figure}
  \centering
    \includegraphics[width=8.5cm,scale=0.55]{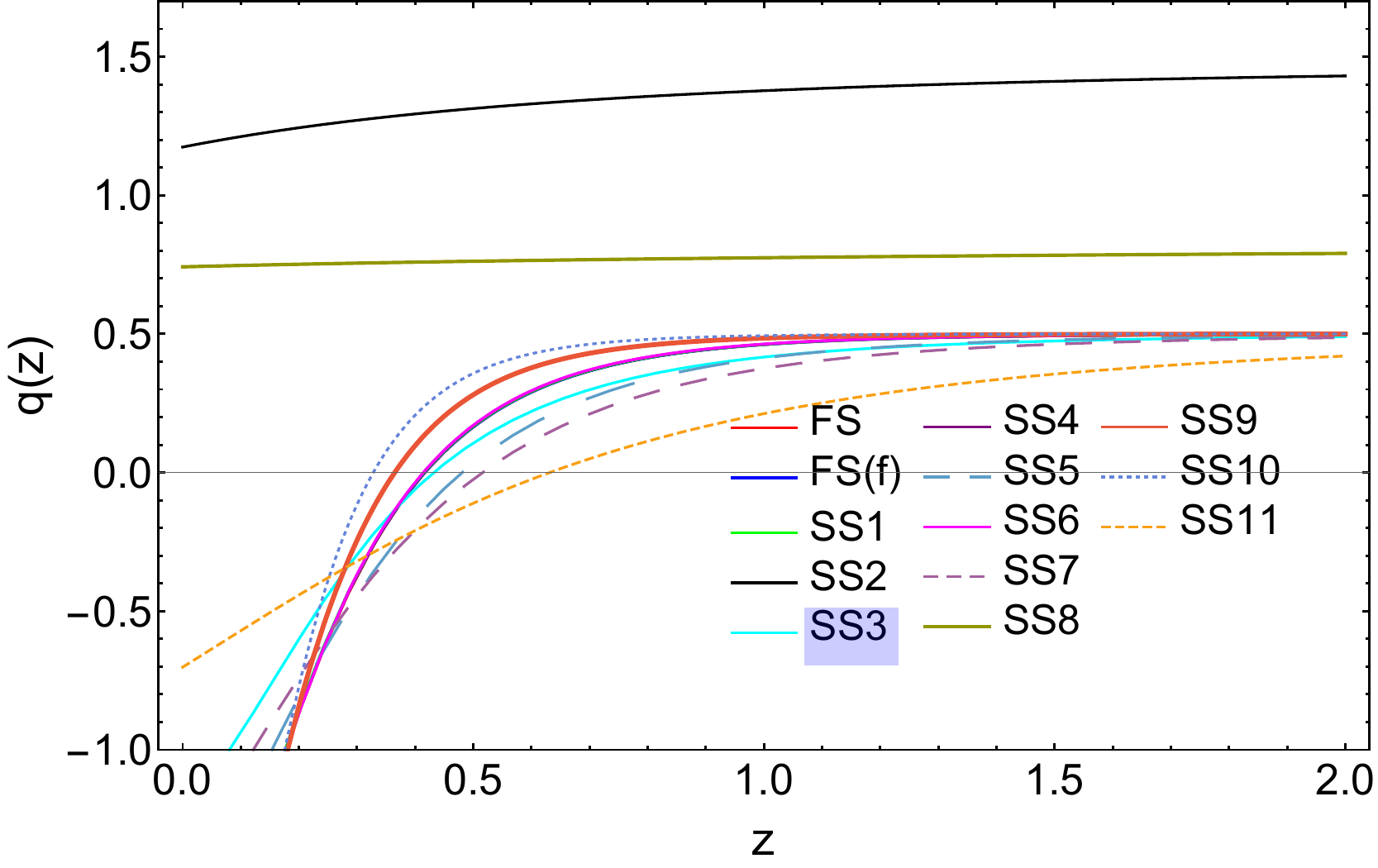} 
    \includegraphics[width=8.5cm,scale=0.55]{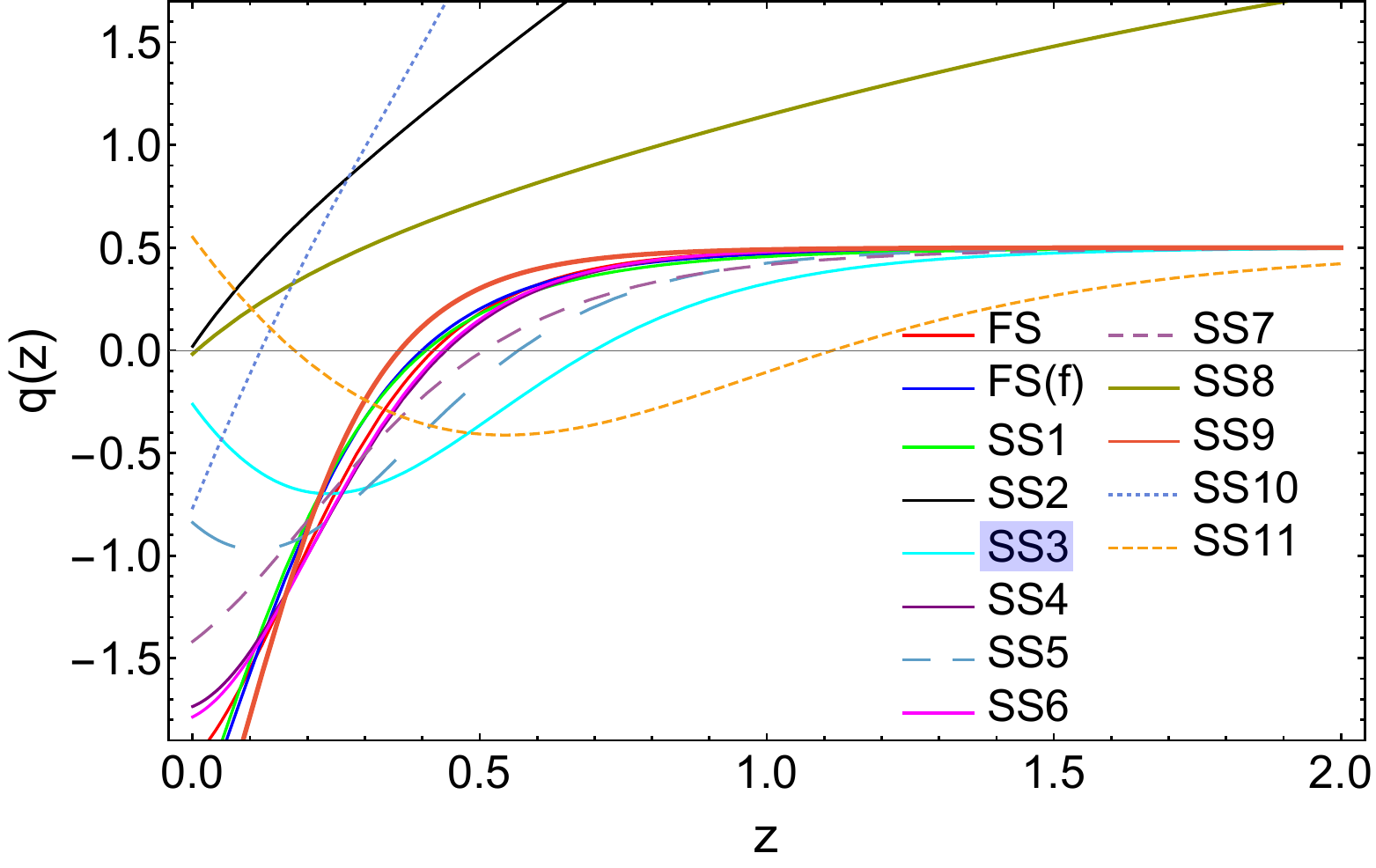} 
    \includegraphics[width=8.5cm,scale=0.55]{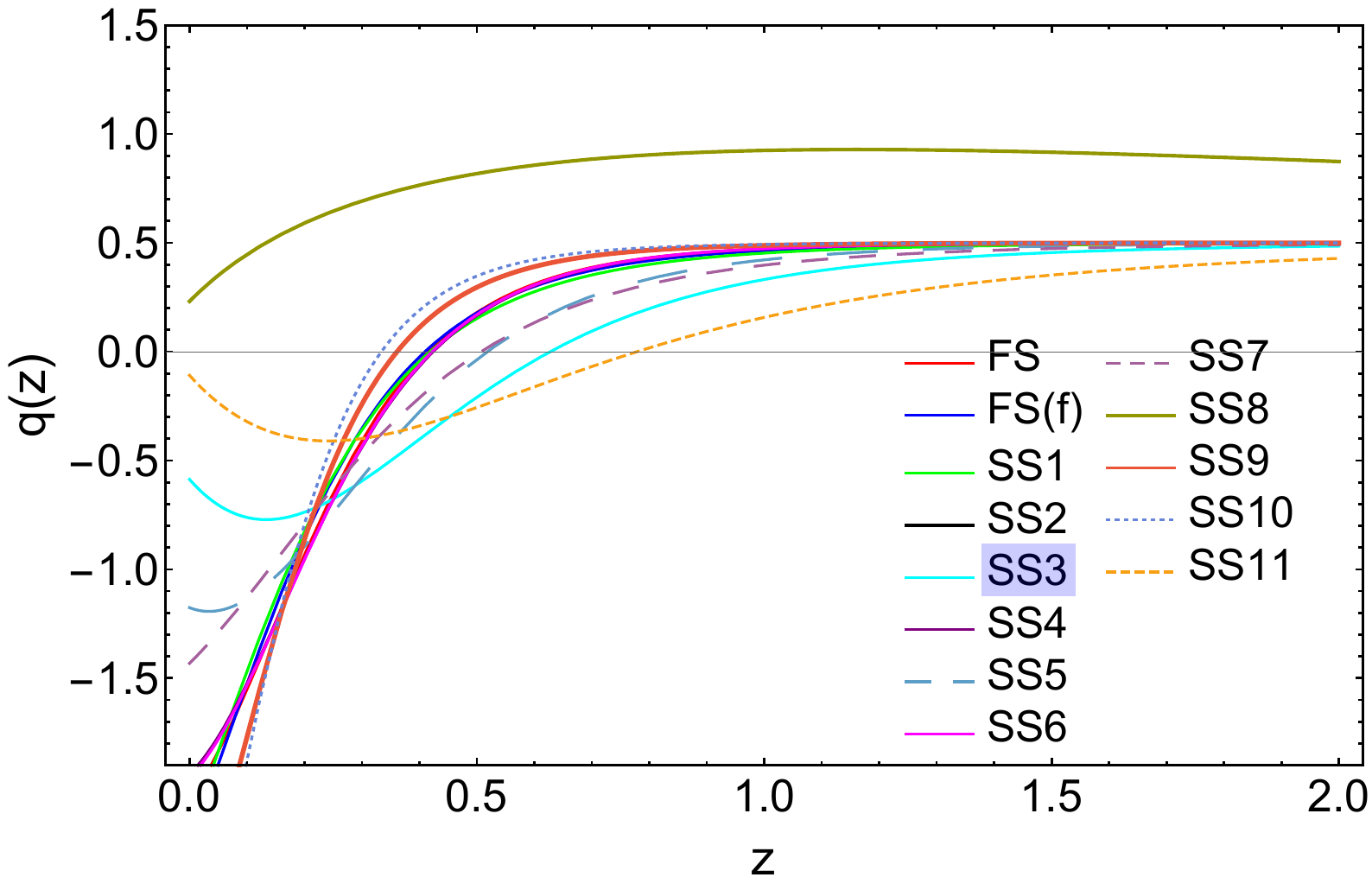} 
\caption{ Reconstruction of the $q(z)$ parameter, using data constraints for $\omega$CDM (upper panel), CPL (middle panel) and JBP (lower panel) models. In the JBP case (lower panel), notice that SS2 it is outside the range shown in the figure, in this case SS2 does not predicts an accelerated universe and therefore it is not inside of the labels. Remarking SS3 as the best sample for our analysis.}
\label{q_all}
\end{figure}

For the $\omega$CDM model, we considered two free parameters $\Omega_{0m}$ and $\omega_0$, aditionally we consider an extra parameter $f$ only in one case (FS). As we mentioned earlier, a Gaussian prior on $\Omega_{0m}$ is assumed, hence our attention is focused on the $\omega_0$ parameter. We found consistency for the constraints obtained in the first three tests (see figure \ref{figwCDM}, upper-left and lower panels), i.e. the $\omega_{0}$ value is not affected whether a new extra parameter $f$ is considered or the systems with $D^{obs} > 1$ are excluded. However, the $\chi^{2}_{min}$ and $\chi^{2}_{red}$ values  reflect the goodness of the constraints: improvement when we exclude the systems in the region of $D^{obs} > 1$, and without significant change when we consider the corrective parameter $f$, in agreement with previous studies \citep[$f \sim 1$; $f=1.010 \pm 0.017  $ for][respectively]{Cao:2012,Treu:2006ApJ}.  
However, when different sub-samples are considered (see figure \ref{figwCDM}, upper-right and middle panels) $\omega_0$ seems to have different values,  the majority pointing to an Universe with a phantom DE, and two cases (SS2 and SS8) where $\omega_0$ is positive, which does not satisfies the accelerated condition $\omega_{0}<-1/3$, yielding an unphysical $q(z)$ (see Figure \ref{q_all}, upper panel). In the following, we discuss in further detail some key results obtained from the sub-samples.

The SS2 was done using systems in the region of $D^{obs}<0.5$, this sub-sample consist of 29 points, eight of them have the peculiarity that their theoretical lens equation Eq. \eqref{Dth} provides $D^{th} <0.5$ only when $\omega_0$ becomes positive. However, three of these systems (MG0751$+$2716, SL2SJ$085019-$034710, and SL2SJ$02325-$040823)  never enter the aforementioned region. These are consequences of the functional form for $D^{th}$, which gives the same result for different values of $\omega_0$. As a  repercussion, cosmological parameters show convergence problems with the MCMC, e.g. notice that the $\Omega_{m0}\mbox{-}\omega_{0}$ posterior distribution presents double contours (see figure \ref{figwCDM}, upper-right panel). 

The SS8 considers systems with velocity dispersions for the lens galaxies in the region of $\sigma > 276$km s$^{-1}$. This case gives larger $\chi^{2}_{red}$ values than previous tests, indicating an Universe without an acceleration stage within $3\sigma$ error. The $D^{obs}>1$ (SS4) region presents the highest $\chi^{2}_{red}$ value for the $\omega$CDM model. These issues indicate that some systems within this regions could have some uncertainties affecting the measured quantities $\theta_E$ and $\sigma$. Appendix \ref{tab:SLDgtone} shows different kinds of systematics affecting the measurements for the $D^{obs}>1$ region.

Finally, we examined three different regions according to the lens galaxy redshift, excluding those systems with $D^{obs}>1$. We found that:  $0.2 \leq z_l \leq 0.4$ presents the worst constraints, although with similar values to those obtained using the complete sample; $z_l < 0.2$ and $z_l>0.4$ show the second and third best constraints (see $\chi^{2}_{red}$ values in table \ref{tab:wCDMCPLyJBPin}).  
In general, we obtain a good fit to the data and the worst results (higher $\chi^{2}_{red}$) are those corresponding to the sub-samples:
$D^{obs} >1$ (SS4), $\sigma < 210$km s$^{-1}$ (SS5) and $\sigma>276$km s$^{-1}$ (SS8). The best $\chi^{2}_{red}$ value is obtained using SS3, i.e when $0.5<D^{obs}<1$. 

For most of the tests, the estimated values for the $\omega_0$ constraints  deviate from the concordance $\Lambda$CDM model ($\omega_{0}=1$) towards the phantom region. When the complete sample is used, our $\omega_{0}$ bounds ($\sim -2.5$) are inconsistent with $\omega_0=-0.978 \pm 0.059$,  $\omega_0=-1.56 \pm 0.54$ and $\omega_0=-1.026 \pm 0.041$, obtained by \cite{Abbott:2018wog,Aghanim:2018eyx, Scolnic:2017caz} respectively. However, our best $\omega_0=-1.653^{+0.264}_{-0.322}$ constraint obtained for $0.5 \leq D^{obs} <1$ (SS3) is consistent with the aforementioned values and those obtained by \citet{Cao:2012} ($\omega_0 = -1.15^{+0.34}_{-0.35}$) using the same method (46 SLS) with a corrective parameter $f$, and \cite{Cao:2015qja} ($\omega_0 = -1.48^{+0.54}_{-0.94}$) assuming a generalization of the mass distribution (118 SLS). 
To judge the quality of our constraints we also calculate the FOM (Eq. \ref{FoM}) for each test. The strongest constraints (i.e highest FOM) are obtained when $ 0.5 \leq D^{obs} < 1$ (see table \ref{tab:AICBICwCDMCPLyJBPin}). On the other hand, the $D^{obs}<0.5$ region provides the least reliable constraints, as expected from the convergence problems and the double confidence contours (see Figure \ref{figwCDM} for details).

We present the reconstruction of the $\omega$CDM deceleration parameter in figure \ref{q_all} (upper panel) using the constraints obtained in each test. Notice that both SS2 and SS8 constraints exhibit unphysical values, i.e. they do not provide an acceleration stage and their $q(z)$ are in disagreement with the standard theoretical prediction at high redshifts ($q(z)\to 0.5$). 
The constraints obtained from other samples provide an accelerated phase at late times.  However, only SS11 yields a $q_{0}=q(z=0)$ value in agreement with the standard model ($q_{0\Lambda CDM}\sim-0.5$), while the remaining sample values are in tension with the standard one.

\subsection{The CPL constraints}
\begin{figure*}
\centering
\begin{tabular}{cc}
\subfloat[FS with and without systems $D^{obs}>1$.]{\label{figur:1}\includegraphics[width=67mm]{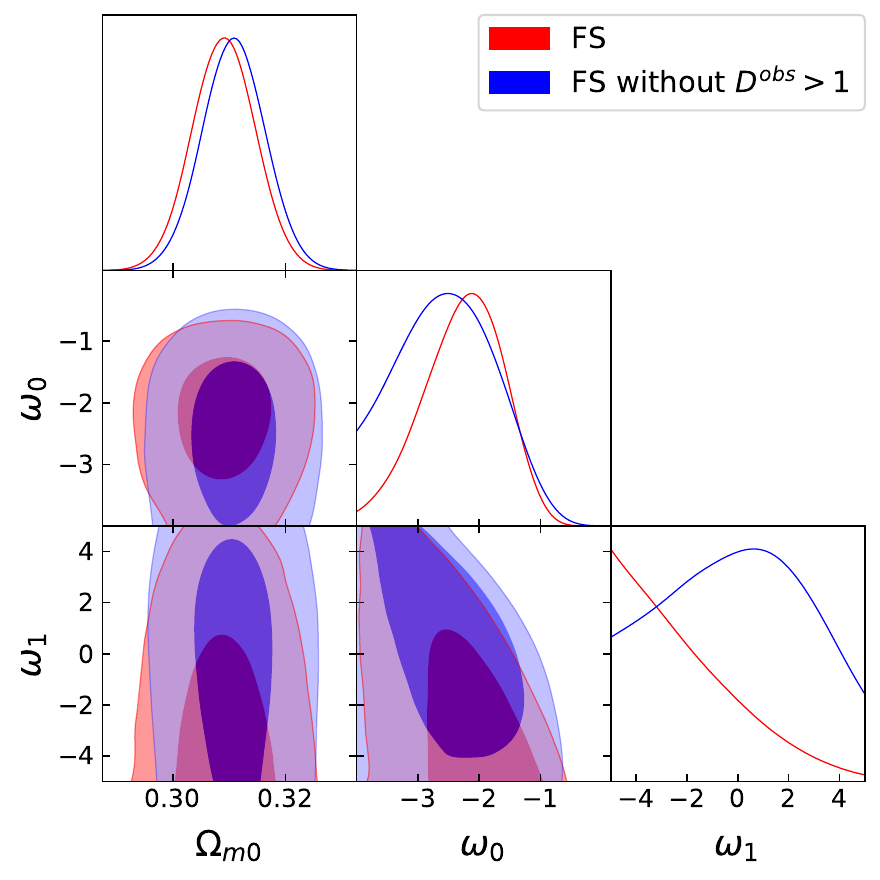}}
  \subfloat[SLS data binned in $D^{obs}$.]{\label{figur:2}\includegraphics[width=67mm]{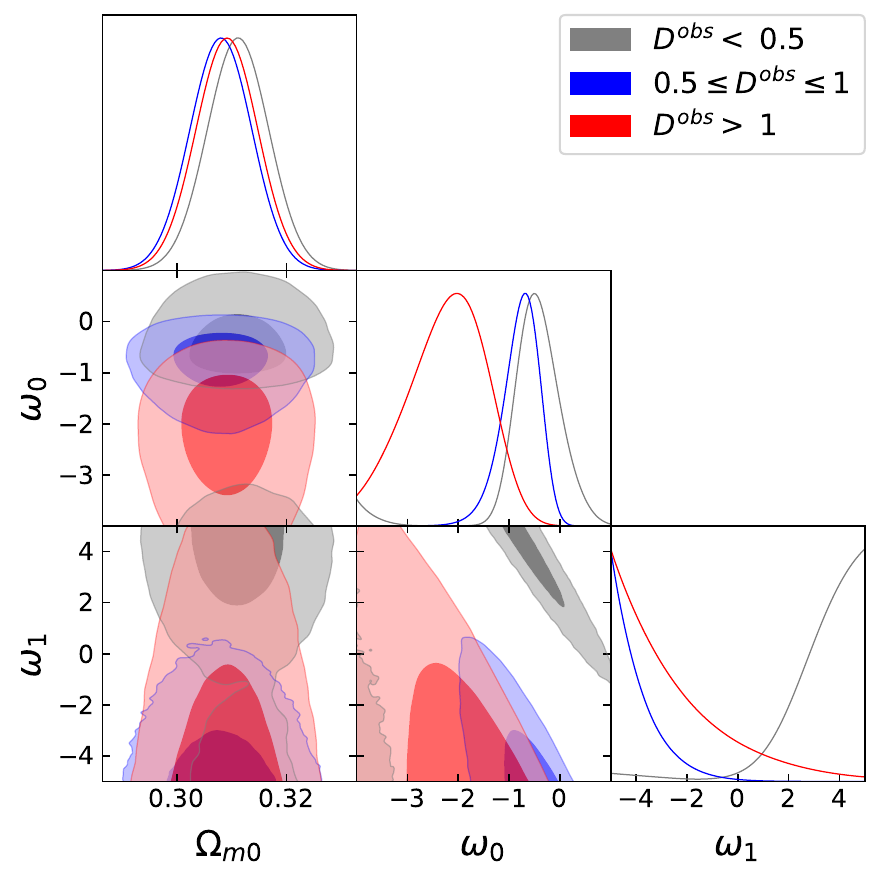}}
  \\
  \subfloat[SLS data binned in $\sigma$.]{\label{figur:3}\includegraphics[width=67mm]{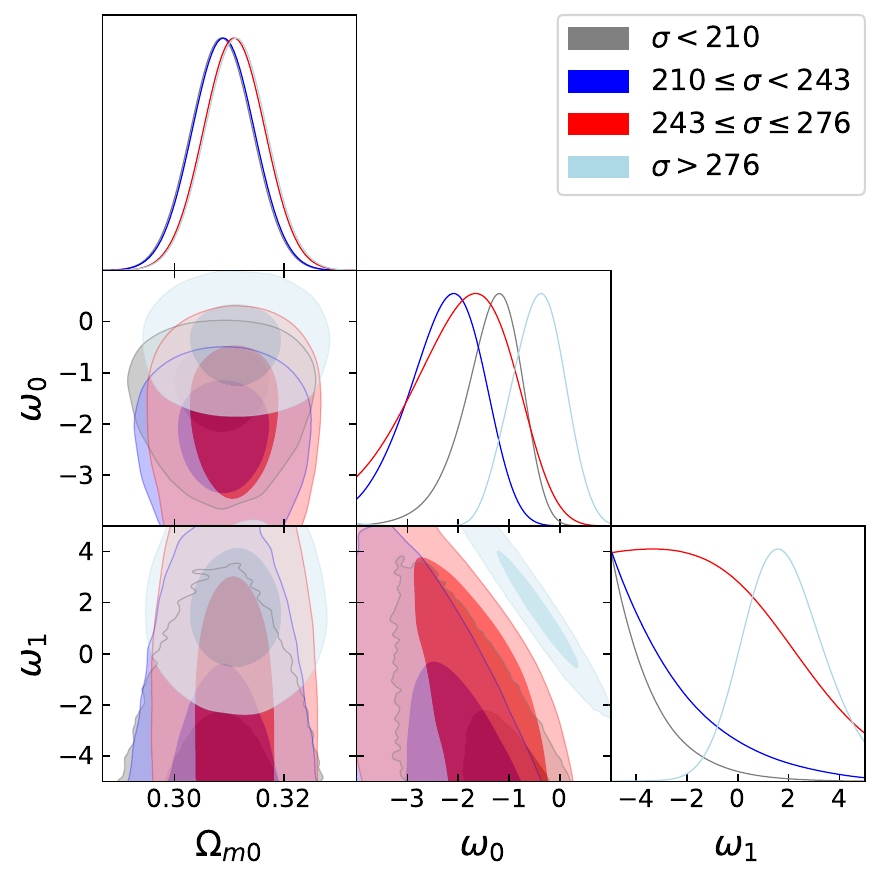}}
  \subfloat[SLS data binned in $z_l$.]{\label{figur:4}\includegraphics[width=67mm]{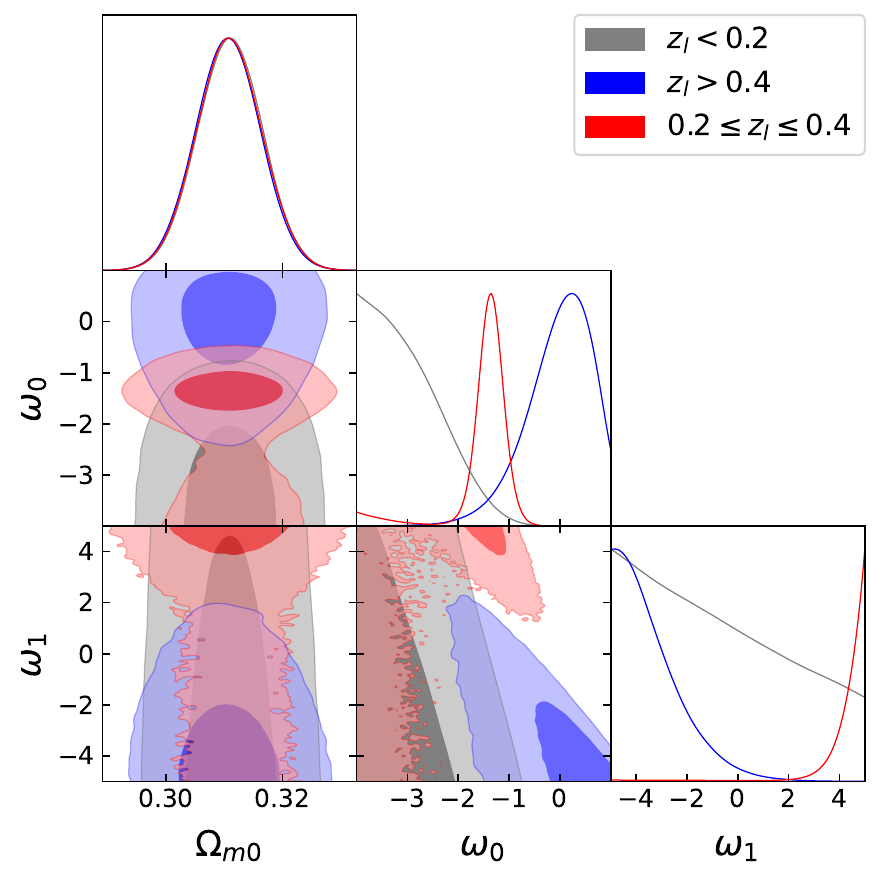}}
  \\
  \subfloat[FS with $f$.]{\label{figur:5}\includegraphics[width=72mm]{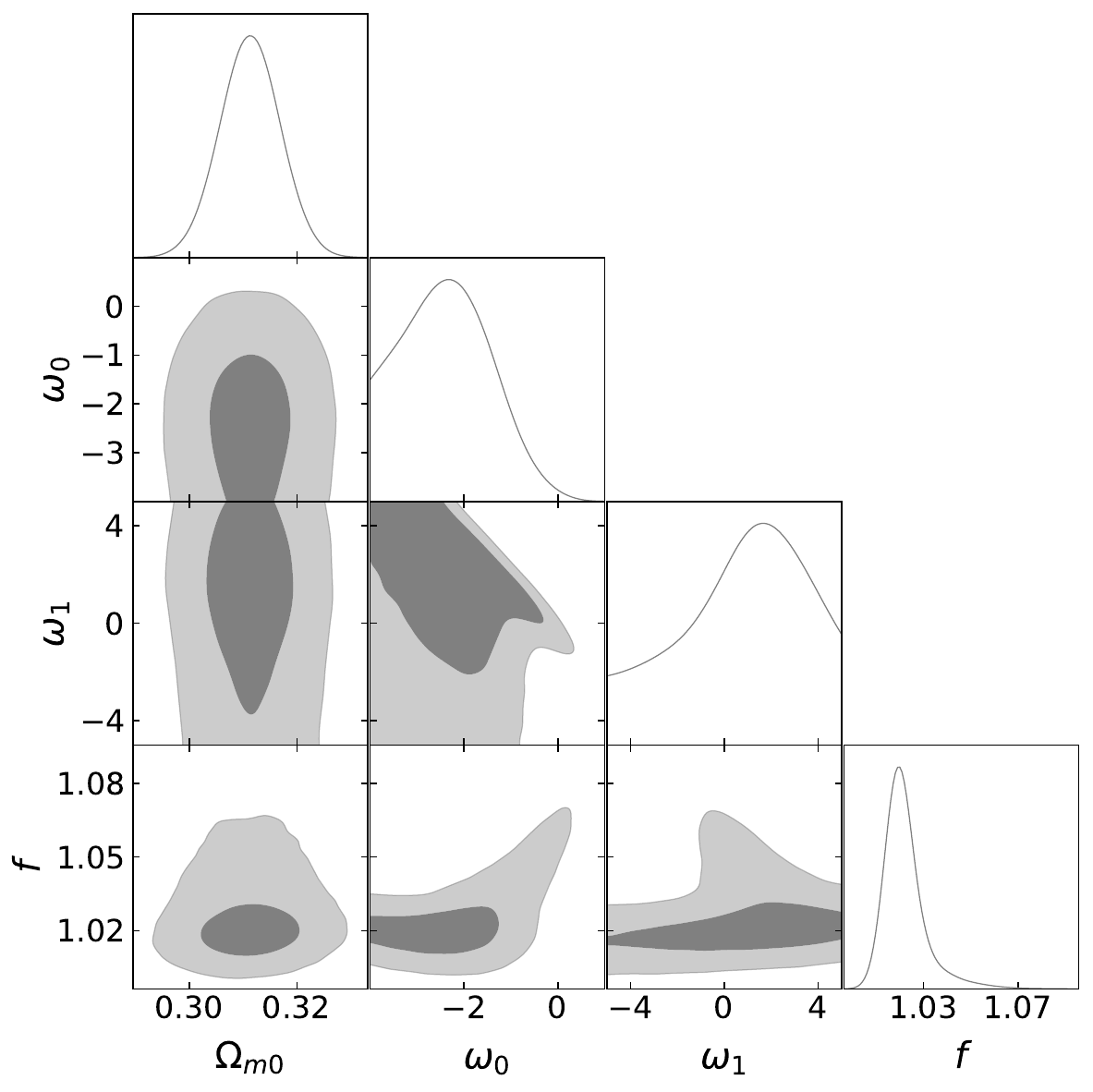}} & \\
  \end{tabular}
\caption{1D marginalized posterior distributions and the 2D $68\%$,  $99.7\%$ confidence levels for the $\Omega_{m0}$, $\omega_{0}$ and $\omega_{1}$ parameters of the CPL model and $f$ parameter (lower panel) for the different samples presented in Table \ref{tab:wCDMCPLyJBPin}.}
\label{figCPL}
 \end{figure*}

Notice that despite the CPL parametrization for the DE EoS (Eq. \ref{eq:eosCPL}) adds an extra free parameter compared to the $\omega$CDM one, the range of values for $\chi^{2}_{min}$ and  $\chi^{2}_{red}$ are roughly similar. The $\omega_0$ constraints estimated with the FS (with and without $f$) and the SS1 are very similar ($\sim -2.3,-2.7$), a straightforward  comparison among $\omega_1$ values is not possible because the bounds are different for each scenario (see figure \ref{figCPL}, upper-left and lower panels), but showing consistency at 1$\sigma$ of confidence level. When sub-samples are considered (SS2 to SS11, figure \ref{figCPL}, upper-right and middle panels), $\omega_0$ is positive only for the region $z_l >0.4$ (see middle-right panel), and adopts negative values for the other sub-samples. The $\omega_1$ parameter is very sensitive to all cases having different mean values in each test, however most being consistent at 1$\sigma$ of confidence level. 

When a new parameter $f$, (FS) is added, no substantial improvements on $\chi^{2}_{red}$ value is shown. The result for the corrective parameter $f$ is consistent with the ones obtained by \citet{Cao:2012,Treu:2006ApJ}. Once again, the $\omega_0$ and $\omega_1$ constraints seem to get worse in the region $D^{obs}<0.5$ (SS2), showing convergence problems reflected in double contours in the correlation of the cosmological parameters (see figure \ref{figCPL}, upper-right panel). 
Notice that $\chi^{2}_{red}$ values higher than the one obtained for the entire sample (table \ref{tab:wCDMCPLyJBPin}) are achieved only in the regions: $D^{obs}>1$ (SS4), $\sigma<210$~km s$^{-1}$ (SS5) and $\sigma>276$~km s$^{-1}$ (SS8). On the other hand, the smallest $\chi^{2}_{red}$ value is reached in the region of $0.5<D^{obs}<1$ (SS3), suggesting a better model fitting. 

When the complete sample is used, the $\omega_0$ and $\omega_1$ constraints are not consistent with the observations of \citet{Scolnic:2017caz, Aghanim:2018eyx} ($\omega_0= -1.009 \pm 0.159$, $ \omega_1=-0.129 \pm 0.755$ and $\omega_0= -0.961 \pm 0.077$,  $\omega_1=-0.28 \pm 0.29$, respectively) but show consistency at 1$\sigma$ with those obtained by \citet{Cao:2012} ($\omega_0= -0.024 \pm 2.42 \, \omega_1=-6.35 \pm 9.75$) using 46 SLS. In spite of this, some sub-samples are consistent with the works of \citet{Scolnic:2017caz, Aghanim:2018eyx}, showing different behaviors for the DE (phantom and quintessence regime). The $\omega_0$ parameter adopt a positive value in the region $z_l > 0.4$, a similar value is reported in the work of \citet{Cao:2012} using a sample with 80 SLS. For all the tests, the FOM estimator gives tight constraints in the $0.2 \leq z_l \leq 0.4$ region (SS10), and poor constraints  in the region 243km s$^{-1}$ $< \sigma \leq $ 276km s$^{-1}$ (SS7).

We reconstruct the $q(z)$ function for CPL model using the constraints obtained for each test (see figure \ref{q_all}, middle panel). Notice that SS2, SS8 and SS10 yield constraints that result in unphysical behaviours for $q(z)$. On the contrary, those provided by SS3, SS5 and SS11 source an acceleration-deceleration stage which is characteristic of models where the DE EoS is parametrized, i.e. a slowing down of cosmic acceleration \citep{Cardenas:2012,Cardenas:2013,Magana:2014,Wang:2016ApJ,Zhang:2018}. Although SS11 constraints produce an acceleration phase in the Universe, it does not ocurrs at $z=0$.

\subsection{The JBP constraints}
\begin{figure*}
\centering
\begin{tabular}{cc}
\subfloat[FS with and without systems $D^{obs}>1$.]{\label{figur:1}\includegraphics[width=67mm]{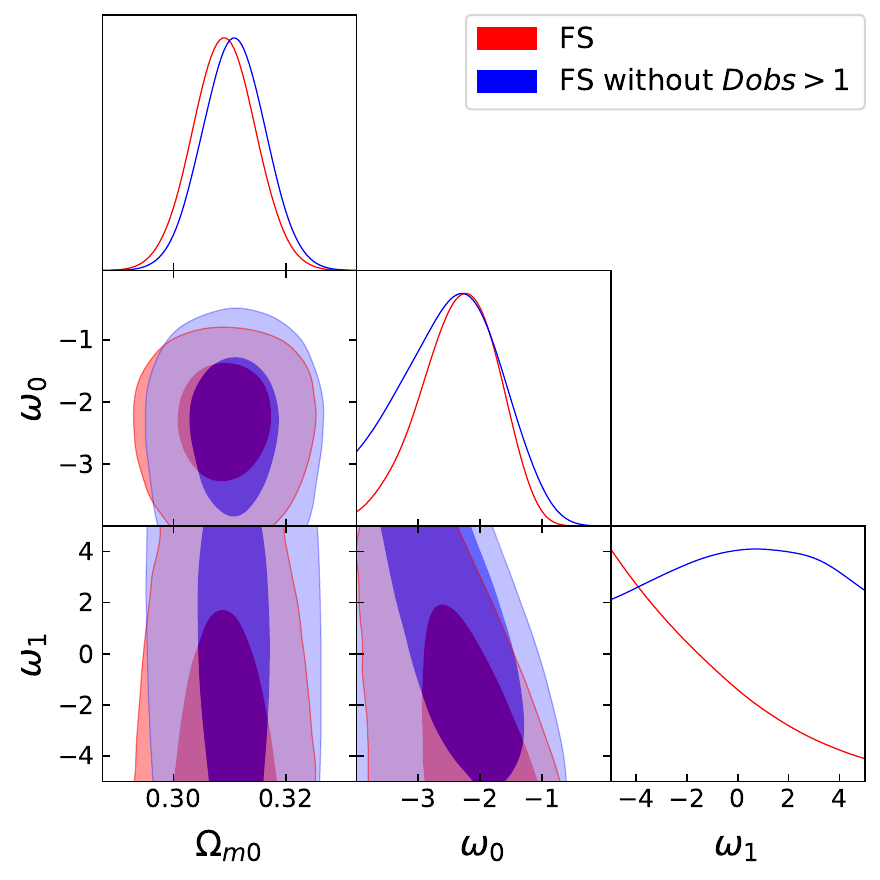}}
  \subfloat[SLS data binned in $D^{obs}$.]{\label{figur:2}\includegraphics[width=67mm]{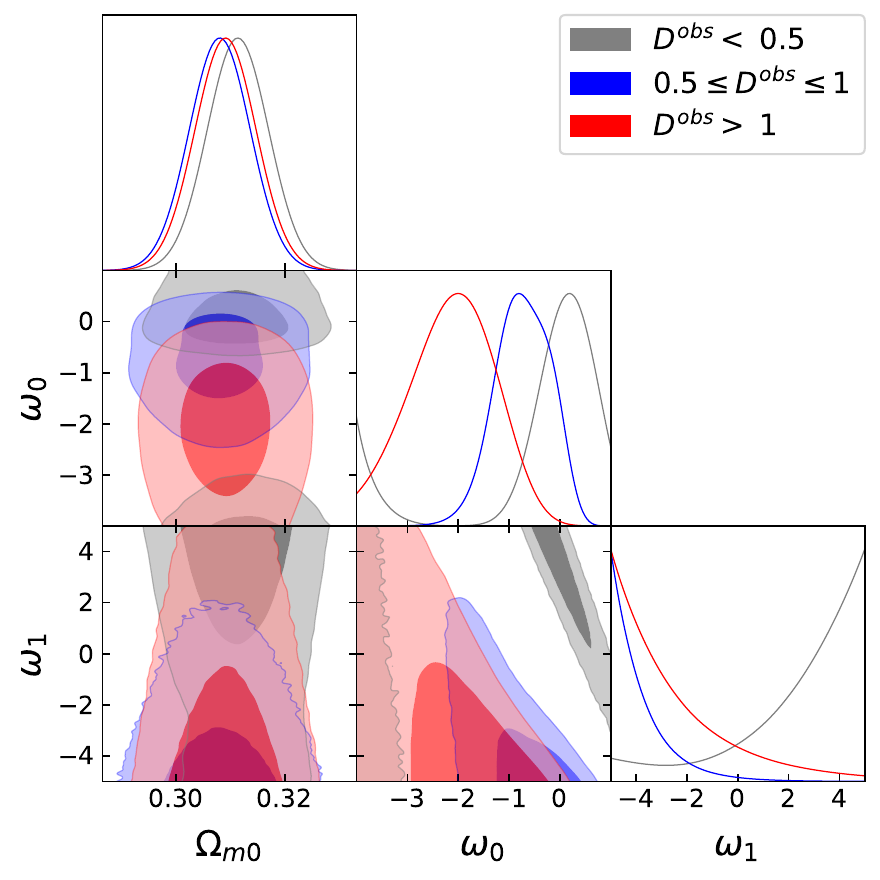}}
  \\
  \subfloat[SLS data binned in $\sigma$.]{\label{figur:3}\includegraphics[width=67mm]{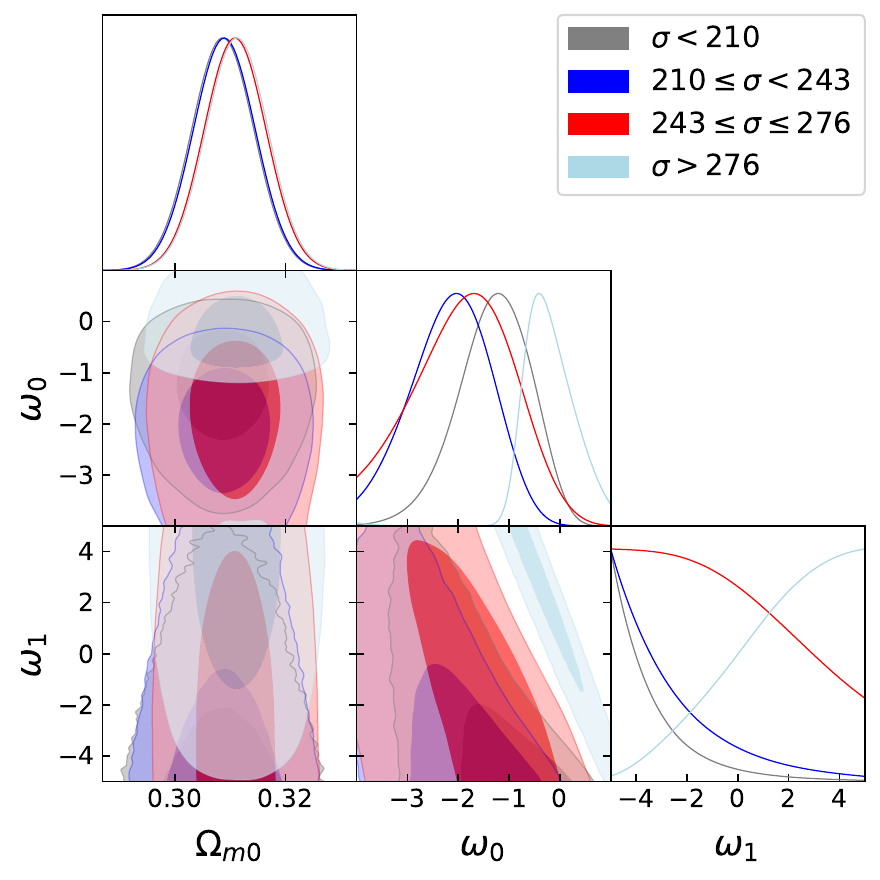}}
  \subfloat[SLS data binned in $z_l$.]{\label{figur:4}\includegraphics[width=67mm]{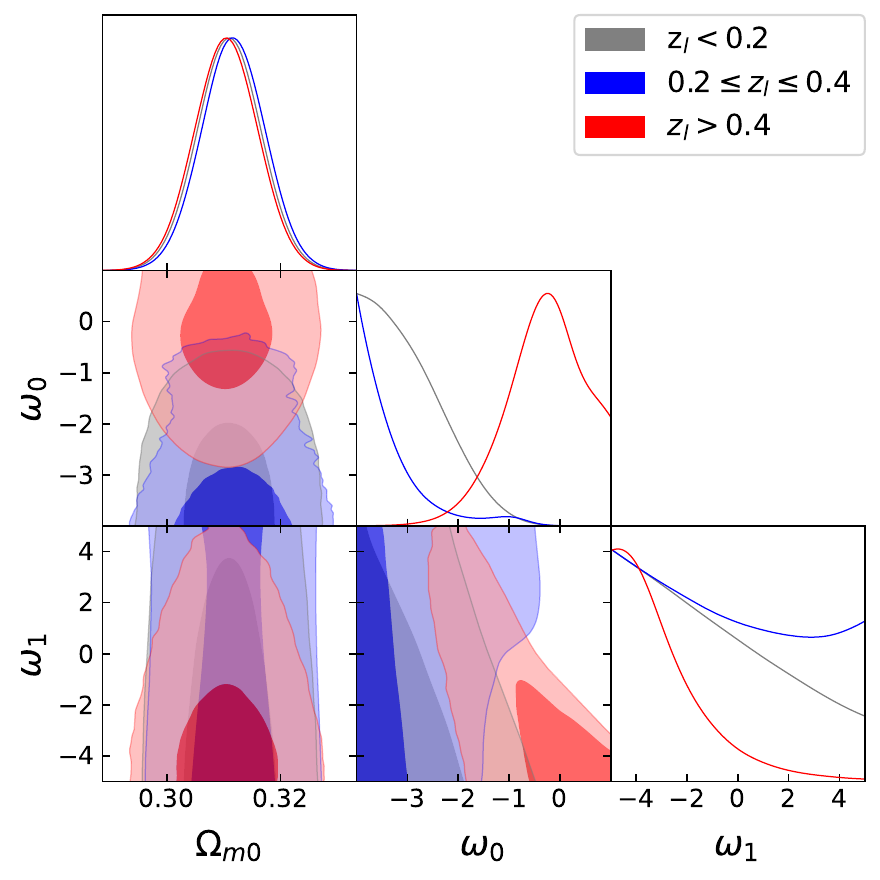}}
  \\
  \subfloat[FS with $f$.]{\label{figur:5}\includegraphics[width=72mm]{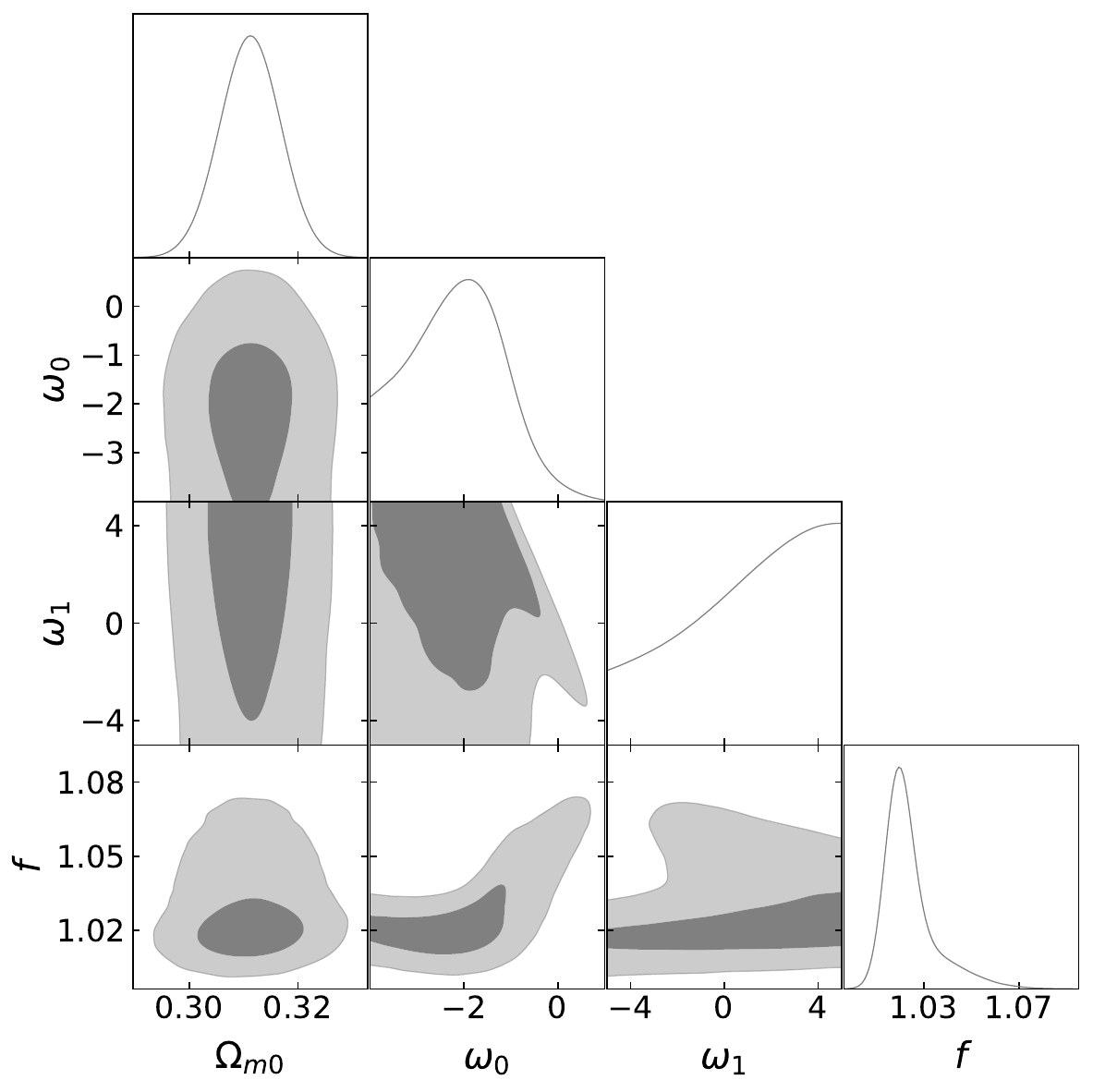}} & \\
  \end{tabular}
\caption{1D marginalized posterior distributions and the 2D $68\%$,  $99.7\%$ confidence levels for the $\Omega_{m0}$, $\omega_{0}$ and $\omega_{1}$ parameters of the JBP model and $f$ parameter (lower panel) for the different samples presented in Table \ref{tab:wCDMCPLyJBPin}.}
\label{figJBP}
 \end{figure*}

The free parameters of the JBP parametrization are $\Omega_{0m}$, $\omega_0$ and $\omega_1$. It is worth to notice that the range of values for $\chi^{2}_{min}$ and  $\chi^{2}_{red}$ are similar to those obtained for the $\omega$CDM and CPL models. For the FS (with and without $f$) and SS1, although the $\omega_0$ constraints differ slightly, they locate on the phantom regime (see Table \ref{tab:wCDMCPLyJBPin}, figure \ref{figJBP} upper-left and lower panels). The $\omega_1$ parameter has remarkable changes in its value for each case being consistent at 1$\sigma$. The corrective parameter $f$ does not introduce significant improvements in the $\chi^{2}_{red}$ value, and it is consistent with the values reported in \citet{Cao:2012,Treu:2006ApJ}. When different sub-samples are used (e.g. SS2 to SS11), the cosmological parameter constraints are sensitive to the selected data (see figure  \ref{figJBP}, upper-right and middle panels). The $\omega_0$ parameter prefers negative values leading to an accelerated expansion stage in all the scenarios excluding the $D^{obs}<0.5$ region (SS2). The $\omega_1$ parameter also adopt distinct values in all the cases. Like for $w$CDM and CPL models, we obtain large  $\chi^{2}_{red}$ values for the  regions: $D^{obs}>1$ (SS4), $\sigma<210$ km s$^{-1}$ (SS5) and $\sigma>276$ km s$^{-1}$ (SS8). Similarly, the best value for $\chi^{2}_{red}$ is also achieved in the region $0.5<D^{obs}<1$ (SS3). The $D^{obs} < 0.5$ (SS2) region presents convergence problems as well, showing double contours in all the free parameters and being  incompatible with an accelerated Universe. For the first three tests, the JBP constraints are inconsistent with those obtained by \citet{Wang:2016ApJ} ($\omega_0 = -0.648 \pm 0.252 , \ \omega_1= -3.419 \pm 2.290$) and \citet{Magana:2017usz}, ($\omega_0 = -0.80 \pm 0.45 , \ \omega_1= -3.78 \pm 3.73$). 
However, the constraint obtained for $\omega_0$  in the region $0.5 \leq D^{obs} \leq 1$ is consistent with those estimated by \citet{Magana:2017usz}  and \citet{Wang:2016ApJ}, although $\omega_1$ is only consistent with the value obtained by \citet{Magana:2017usz}. The FOM estimator gives tight constraints in the $0.5 \leq D^{obs} < 1$ (SS3) region and weak ones in $D^{obs}<0.5$ (SS2) and $0.2 \leq  z_l \leq 0.4$ (SS10) regions. 

The figure \ref{q_all} shows the  reconstruction of the deceleration parameter for the JBP model using the constraints derived from each test. It is noticeable the $q(z)$ behavior showing a slowing down of the cosmic acceleration at late times for the SS3 and SS11. This behavior is in agreement with those found by several authors for these parametrizations \citep[see for example][]{Magana:2014,Wang:2016ApJ}. Notice also that SS2 and SS8 present a non standard behavior, i.e. they never cross the acceleration region. The remaining cases are in good agreement with the standard knowledge.

\subsection{The impact of the $f$ parameter} 
\label{sec_f}

The mean values of the $f$ parameter obtained with different models ($ 1.021^{+0.012}_{-0.006}$, $ 1.021^{+0.008}_{-0.006}$, and $1.021^{+0.011}_{-0.006}$ for $\omega$CDM, CPL, and JBP respectively) are consistent with each other and also in agreement with those reported by \cite{Treu:2006ApJ}, and \cite{Cao:2012}. This indicates that the possible systematic errors affecting the image separation produced by the lens for our sample is estimated at most at 5 \% when the value of $f$ is considered equal for all SLS. On the other hand, if we assume $f$ as an independent free parameter in each SLS, the value of $f$ becomes larger for some measured systems. However, if we restrict the analysis to the region between $0.5 \leq D^{obs} \leq 1$ (i.e., sample SS3) the scatter decreases (see Appendix \ref{Fseparada} for details).

The deviation on the value of $f$ could be related to the fact that some systems are not properly described by an isothermal model as found by \cite{10.1093/mnras/sty2833} for the BELLS gallery in almost all cases at the 2$\sigma$ level. These deviations are also found by \cite{Barnabe:2011gb}, and by \cite{ Vegetti:2014lqa} for the SLACS survey employing more sophisticated modelling. In this sense, in Appendix\,\ref{APII} we studied the effect of assuming a SIE (or a SIS) acting as a lens when the mass distribution follows a power-law. Although the scatter in $f$ is even greater, with some systems outside the range proposed by Ofek et al. (2003), none of the mock systems shows a nonphysical value for $D^{obs}$.
It is important to note that in most  previous studies the lens models are close to isothermal. Thus, in this work we are using Eq.\eqref{Dlens} as a first approximation, and Eq. \eqref{Df} to take into account all the above mentioned irregularities \citep[see for example][]{Cao:2012}. Except those systematics that produces unphysically values on the observational lens equation.

\subsection{The restricted sample SS3 as a fiduciary sample}

Table \ref{tab:wCDMCPLyJBPin} shows that there are four regions showing non-physical behaviours for the EoS of the dark energy (i.e. it does not satisfy $\omega_0<-1/3$) or higher values for the $\chi^{2}_{min}$ function with respect to the FS, we label these sub-samples as unreliable (marked with a letter U in the table)  since they have similar aspects among the cosmological models presented in this work being the following regions: SS2 ($D^{obs}<0.5$) wich present convergence problems in the estimation of cosmological parameters and a non-accelerating Universe at $z=0$ for the $\omega$CDM and JBP models, SS4 ($D^{obs} > 1$) showing a non physical value for the lens equation and the highest $\chi^{2}_{min}$ value for all the models, SS5 ($\sigma < 210$ km s$^{-1}$) and  SS8 ($\sigma > 276$ km s$^{-1}$) showing higher $\chi^{2}_{min}$ values than those obtained for the FS and also showing a non-accelerating Universe in SS8 for the $\omega$CDM model. It is worth to notice that the region SS11 ($D^{obs} \leq 1$ and $z_l>0.4$) also presents a non-accelerating Universe at $z=0$ for the CPL model, however is consistent with an accelerated one at $1 \sigma$ of confidence level and do not show an increasing $\chi^{2}_{min}$ function in comparisson with the FS, hence we do not discard this region from the following. Even though cosmological parameters has different mean values for the remaining samples, most are consistent at 1 $\sigma$ of confidence level for all the models.

However, to constrain cosmological parameters, we recommend the use of the restricted sample (SS3), which shows the best constraints and a lower dispersion in the value of the corrective parameter $f$, when is considered as an independent parameter in each SLS (see appendix \ref{Fseparada}). The SS3 constraints for the different models, also shows consistency with CMB  \citep{Aghanim:2018eyx} and SNe Ia \citep{Scolnic:2017caz} measurements. Therefore, we favour the SS3 sample as the fiduciary sample.

\subsection{Comparing cosmological models}
\label{comparisson}

In this section, we discuss the comparison among cosmological models for only three samples: FS, FS+$f$ and SS3. We add to the present discussion the FS, FS+$f$ samples, because they are the complete samples (besides, as we appreciate in Table \ref{tab:wCDMCPLyJBPin}, $\chi^{2}_{red}$ values are very similar among the three models in them). Nevertheless, we highlight the results for the fiduciary SS3 sample.

In Table \ref{tab:AICBICwCDMCPLyJBPin} we present the model selection's criteria, for the constraints obtained from the FS sample (with and without the $f$ parameter), and the  SS3 sample. To discern among models (in any sample), it is necesary to compare the different criterias;  the favored model is obtained trhough the compromise between the AIC, BIC and FOM estimators. When we perform the model comparison using the FS constraints, the $\omega$CDM model produce the lowest AIC and BIC values. By measuring the relative differences $\Delta$AIC = AIC$_{i}$ - AIC$_{min}$, and $\Delta$BIC = BIC$_{i}$ - BIC$_{min}$ (where the subindex $i$ refers to the different models and AIC$_{min}$ (BIC$_{min}$) is the lowest AIC (BIC) value) there is substantial support for the three models ($\Delta$AIC$<2$ or $\sim 2$). However, exists positive evidence against the CPL and JBP models ($4.2<\Delta$BIC$<4.8$ respectively). By looking the FOM criteria, the higest values is obtained from the $\omega$CDM model. Therefore, all estimators suggest that  $\omega$CDM is the favored model from the FS constraints. Similarly, the AIC, BIC, and FOM values from the FS+$f$ constraints suggest the same, the  $\omega$CDM model is the prefered one. 

As mentioned before, we strongly suggest the use of SS3 sample to estimate cosmological parameters. In particular, for the constraints obtained for sample SS3, the lowest AIC and BIC values are obtained from the CPL model. Indeed, the relative difference $\Delta$AIC with respect to this model ($\sim 3.1$ and $6.2$ for JBP and $\omega$CDM  model respectively) point out to a considerably less support for the these models. Moreover, the $\Delta$BIC $\sim 3.1$ and $\sim 3.2$ for the JBP and $\omega$CDM  model respectively, suggest a positive evidence against  both models. 
By analyzing the FOM criteria, the highest value is obtained for the CPL model, i.e. it produce the strongest constraints. Thus, all the criteria suggest that the CPL is the favored model from the fiduciary SS3 sample. However, it is important to mention that does not exist enough evidence to rule out the $\omega$CDM model. Therefore, when the SLS data are used as cosmological test in the range $0.5 \leq D^{obs} \leq 1$, a dynamical dark energy with a CPL parameterization is favored to explain the late cosmic acceleration.

\begin{table*}
\caption{Mean values for the $\omega$CDM, CPL and JBP parameters ($\Omega_{m0}$, $w_0$, $w_1$) and the $f$ corrective parameter derived from different test employing the SLS data. We labeled some subsamples as U, wich present weak constraints for all the models and unreliable results for the bayesian analysis.}
\centering
\begin{tabular}{|lcccccc|}
\hline
Data set: data point number & $\chi^{2}_{min}$ & $\chi^{2}_{red}$& $\Omega_{m0}$ & $w_{0}$ & $w_{1}$& $f$ \\
\hline
\multicolumn{7}{|c|}{$\omega$CDM model}\\
\multicolumn{7}{|c|}{}\\
FS (all systems: 204) &  570.359 & 2.824 & $  0.309^{+0.006}_{-0.006}$ & $  -2.478^{+0.479}_{-0.587}$ &  --- & ---   \\
SS1 ($D^{obs}\leq 1$: 172) & 409.759 & 2.396 & $ 0.311^{+0.006}_{-0.006}$ & $ -2.593^{+0.794}_{-0.860}$ & --- & ---  \\
FS (all systems with $f$: 204) & 559.619 & 2.784 & $ 0.311^{+0.006}_{-0.006}$ & $ -2.013^{+1.229}_{-1.167}$ & ---& $1.021^{+0.012}_{-0.006}$ \\
SS2 ($D^{obs}<0.5: 29)^{U}$ & 52.943 & 1.961 & $ 0.312^{+0.006}_{-0.006}$ & $ 0.572^{+0.174}_{-4.403}$& ---& ---   \\
SS3 ($0.5 \leq D^{obs} \leq 1$: 143) & 263.795 & 1.871 & $0.308^{+0.006}_{-0.006}$ & $-1.653^{+0.264}_{-0.322}$& ---& ---   \\
SS4 ($D^{obs} > 1:  32)^{U}$& 383.875 & 2.258 & $0.311^{+0.006}_{-0.006}$ & $-2.336^{+0.684}_{-0.894}$& ---& ---   \\
SS5 ($\sigma < 210$ km s$^{-1}: 64)^{U}$ & 201.766 & 3.254 & $0.309^{+0.006}_{-0.006}$ & $-2.100^{+0.416}_{-0.525}$& ---& ---   \\
SS6 (210 km s$^{-1}$ $\leq \sigma < 243$ km s$^{-1}$: 53) & 118.092 & 2.316 & $0.309^{+0.006}_{-0.006}$ & $-2.633^{+0.550}_{-0.641}$& ---& ---  \\
SS7 (243 km s$^{-1}$$ \leq \sigma \leq 276$ km s$^{-1}$: 49)& 107.252  & 2.282 & $0.311^{+0.006}_{-0.006}$ & $-1.854^{+0.672}_{-0.958}$& ---& ---  \\
SS8 ($\sigma > 276$ km s$^{-1}: 38)^{U}$ & 109.457  &  3.040 &   $0.311^{+0.006}_{-0.006}$& $ 0.148^{+0.122}_{-0.134}$& ---& ---   \\ 
SS9 ($D^{obs} \leq 1$ and $z_l < 0.2$: 52) & 110.663 & 2.213 & $0.311^{+0.006}_{-0.006}$ & $-3.134^{+0.807}_{-0.603}$& ---& ---   \\
SS10 ($D^{obs} \leq 1$ and $0.2 \leq z_l \leq 0.4$: 48) &  112.860 & 2.453 & $ 0.312^{+0.006}_{-0.006}$ & $-3.558^{+0.701}_{-0.327}$& ---& ---   \\
SS11 ($D^{obs} \leq 1$ and $z_l>0.4$: 72) & 149.099 & 2.129 & $0.311^{+0.006}_{-0.006}$ & $-1.163^{+0.318}_{-0.446}$& ---& ---  \\
\multicolumn{7}{|c|}{}\\
\multicolumn{7}{|c|}{CPL model}\\
\multicolumn{7}{|c|}{}\\
FS (all systems: 204) &  569.317 & 2.832 & $ 0.309^{+0.006}_{-0.006}$ & $-2.224^{+0.612}_{-0.707}$ & $-2.548^{+2.882}_{-1.765}$  & -- \\
SS1 ($D^{obs}\leq 1$: 172) & 409.780 & 2.410 & $0.311^{+0.006}_{-0.006}$ & $-2.768^{+0.833}_{-0.764}$ & $0.163^{+2.914}_{-3.270}$  & --\\
FS (all systems with $f$: 204) & 559.260 & 2.796 & $ 0.311^{+0.006}_{-0.006}$ & $ -2.371^{+0.984}_{-0.984}$ & $1.045^{+2.338}_{-3.219}$& $ 1.021^{+0.008}_{-0.006}$\\
SS2 ($D^{obs}<0.5: 29)^{U}$ & 46.235 & 1.778 & $0.311^{+0.006}_{-0.006}$ & $-0.505^{+0.456}_{-0.342}$ &
$3.585^{+0.989}_{-1.535}$ & ---\\
SS3 ($0.5 \leq D^{obs} \leq 1$: 143) & 255.552 & 1.825 & $0.308^{+0.006}_{-0.006}$ & $-0.735^{+0.319}_{-0.395}$ &
$-4.202^{+1.208}_{-0.592}$ & ---\\
SS4 ($D^{obs} > 1: 32)^{U}$ & 384.067 & 2.273 & $0.311^{+0.006}_{-0.006}$ & $-2.530^{+0.823}_{-0.832}$ &
$-0.056^{+2.871}_{-3.077}$& --- \\
SS5 ($\sigma < 210$ km s$^{-1}: 64)^{U}$ & 196.854 & 3.227 & $0.309^{+0.006}_{-0.006}$ & $-1.292^{+0.500}_{-0.641}$ &
$-3.912^{+1.767}_{-0.813}$& --- \\
SS6 (210 km s$^{-1}$ $\leq \sigma < 243$ km s$^{-1}$: 53) & 116.000 & 2.320 & $0.309^{+0.006}_{-0.006}$ & $-2.205^{+0.668}_{-0.789}$ &
$-3.159^{+2.728}_{-1.369}$ & ---\\
SS7 (243 km s$^{-1}$$ \leq \sigma \leq 276$ km s$^{-1}$: 49)& 107.363  & 2.334 & $0.311^{+0.006}_{-0.006}$ & $-1.856^{+0.889}_{-1.062}$ &
$-1.208^{+3.130}_{-2.580}$ & ---\\
SS8 ($\sigma > 276$ km s$^{-1}: 38)^{U}$ & 108.338  & 3.095 &  $0.311^{+0.006}_{-0.006}$ & $-0.423^{+0.494}_{-0.552}$ &
$1.692^{+1.558}_{-1.438}$ & ---\\
SS9 ($D^{obs} \leq 1$ and $z_l < 0.2$: 52) & 110.539 & 2.256 & $0.311^{+0.006}_{-0.006}$ & $-3.107^{+0.857}_{-0.624}$ &
$-1.148^{+3.690}_{-2.743}$ & ---\\
SS10 ($D^{obs} \leq 1$ and $0.2 \leq z_l \leq 0.4$: 48) &  101.321 & 2.252 & $ 0.311^{+0.006}_{-0.006}$ & $-1.377^{+0.244}_{-0.284}$ &
$4.551^{+0.336}_{-0.781}$ & ---\\
SS11 ($D^{obs} \leq 1$ and $z_l>0.4$: 72) & 144.466& 2.094 & $0.311^{+0.006}_{-0.006}$ & $0.049^{+0.536}_{-0.711}$ &
$-3.672^{+1.567}_{-0.925}$ & ---\\
\multicolumn{7}{|c|}{}\\
\multicolumn{7}{|c|}{JBP model}\\
\multicolumn{7}{|c|}{}\\
FS (all systems: 204) & 569.624 & 2.834 & $0.309^{+0.006}_{-0.006}$ & $-2.312^{+0.592}_{-0.671}$ & $-2.220^{+3.368}_{-2.022}$  & -- \\
SS1 ($D^{obs}\leq 1$: 172) & 409.750 & 2.410 & $0.311^{+0.006}_{-0.006}$ & $-2.671^{+0.801}_{-0.811}$ & $0.201^{+3.209}_{-3.408}$  & --\\
FS (all systems with $f$: 204) & 559.147 & 2.796 & $0.311^{+0.006}_{-0.006}$ & $-2.124^{+0.960}_{-1.097}$ & $1.367^{+2.544}_{-3.659}$& $1.021^{+0.011}_{-0.006}$\\
SS2 ($D^{obs}<0.5: 29)^{U}$ & 50.465 & 1.941 & $0.311^{+0.006}_{-0.006}$ & $-0.055^{+0.421}_{-0.396}$ &
$3.105^{+1.400}_{-3.034}$ & --- \\
SS3 ($0.5 \leq D^{obs} \leq 1$: 143) & 258.744 & 1.848 & $0.308^{+0.006}_{-0.006}$ & $-1.047^{+0.324}_{-0.413}$ &
$-3.921^{+1.705}_{-0.805}$ & ---\\
SS4 ($D^{obs} > 1: 32)^{U}$ & 383.931 & 2.272 & $0.311^{+0.006}_{-0.006}$ & $-2.433^{+0.754}_{-0.851}$ &
$0.125^{+3.203}_{-3.335}$ & ---\\
SS5 ($\sigma < 210$ km s$^{-1}: 64)^{U}$ & 198.932 & 3.261 & $0.309^{+0.006}_{-0.006}$ & $-1.617^{+0.502}_{-0.645}$ &
$-3.419^{+2.520}_{-1.181}$ & ---\\
SS6 (210 km s$^{-1}$ $\leq \sigma < 243$ km s$^{-1}$: 53) & 116.777 & 2.336 & $0.309^{+0.006}_{-0.006}$ & $-2.370^{+0.648}_{-0.742}$ &
$-2.682^{+3.262}_{-1.718}$ & ---\\
SS7 (243 km s$^{-1}$$ \leq \sigma \leq 276$ km s$^{-1}$: 49)& 107.317  & 2.333 & $0.311^{+0.006}_{-0.006}$ & $-1.869^{+0.821}_{-0.985}$ &
$-0.752^{+3.553}_{-2.941}$ & ---\\
SS8 ($\sigma > 276$ km s$^{-1}: 38)^{U}$ & 108.660  & 3.105 &  $0.311^{+0.006}_{-0.006}$ & $-0.271^{+0.558}_{-0.404}$ &
$2.122^{+1.997}_{-2.838}$ & ---\\
SS9 ($D^{obs} \leq 1$ and $z_l < 0.2$: 52) & 110.569 & 2.257& $0.311^{+0.006}_{-0.006}$ & $-3.112^{+0.846}_{-0.619}$ &
$ -1.012^{+3.686}_{-2.823}$ & ---\\
SS10 ($D^{obs} \leq 1$ and $0.2 \leq z_l \leq 0.4$: 48) & 112.833 & 2.507 & $0.311^{+0.006}_{-0.006}$ &  $-3.529^{+0.803}_{-0.349}$ &
$-0.663^{+3.849}_{-3.098}$ & ---\\
SS11 ($D^{obs} \leq 1$ and $z_l>0.4$: 72) & 147.010& 2.131 & $0.311^{+0.006}_{-0.006}$ & 
$-0.589^{+0.510}_{-0.696}$ &
$-3.182^{+2.633}_{-1.341}$ & ---\\
\hline
\end{tabular}
\label{tab:wCDMCPLyJBPin}
\end{table*}

\begin{table*}
\caption{AIC, BIC and FOM values for the $\omega$CDM, CPL and JBP models derived from different test employing the SLS data.}
\centering
\begin{tabular}{|cccccccccc|}
\hline
Data set &  & $\omega$CDM &  &  & CPL &  &  & JBP & \\
\hline
 & AIC & BIC & FOM & AIC & BIC & FOM & AIC & BIC & FOM\\
\hline
\multicolumn{10}{|c|}{}\\
FS 204 &  574.359 & 580.995 & 343.671 & 575.317 & 585.271  & 143.0289 & 575.624 & 585.578  & 130.268 \\
FS with $f$ 204 & 565.619 & 575.573 & 18716.678 & 567.26 & 580.532  & 10122.341 & 567.147 & 580.419  &  7941.123\\
SS3 (143) & 267.795 & 273.720 & 584.736 & 261.552 & 270.441 & 586.979 & 264.744 &273.632  & 338.777\\
\hline
\end{tabular}
\label{tab:AICBICwCDMCPLyJBPin}
\end{table*}

\section{Conclusions and Outlooks} \label{Con}

In this paper we study three dark energy models: the $\omega$CDM model with a DE constant equation of state, and the CPL and the JBP parametrizations where DE evolves with time. To constrain the cosmological parameters of the three models we used a new compilation of strong gravitational lens systems (SLS) with a total of 204 objects, the largest sample to date (details of all systems can be found in Appendix \ref{tab:SL}). We test the models using different cases. First considering all the systems using the $D^{obs}$ estimated from the observed Einstein ring radius and velocity dispersion, secondly excluding those systems with a $D^{obs}>1$ (unphysical) value, and finally using the entire sample with a new free parameter $f$ that take into account systematics that might affect the observables. 
In addition, to assess the impact of some observables, we estimated the cosmological parameters using sub-samples of the SLS according to three different scenarios, considering distinct regions on the observational value of the lens equation ($D^{obs}$), the velocity dispersion ($\sigma$) and  the redshift interval probed by the
lens galaxy ($z_l$). We found weak constraints for some regions.

The $f$ parameter is consistent among the three models (within 5\% error), having similar values to those reported by \cite{Treu:2006ApJ,Cao:2012}. Assuming $f$ as an independent free parameter in each SLS, the cosmological constraints are consistent with those estimated assuming $f$ equal for all the SLS. However, the scatter (in comparisson with the Ofek estimations) of the value of $f$ seems to be larger for some measured systems. To study this deviation, we analyze a mock catalog of 788 SLS, that mimics the distribution of the observational data compiled in this work. When the Einstein radius of the simulated sample is compared with the one obtained from a SIE fitting, we found that the error is less than $10 \%$ for most of the objects (see Appendix \ref{APII}). Considering that the majority of the data in our SLS compilation comes from models assuming SIEs, this supports the range of the parameter used in our paper. However, when a power-law mass distribution  for the mock catalogue is assumed, the scatter of the Einstein radius increases (35$\%$ or less). 

We found that some of the sub-samples considered in this work provide values for the cosmological parameters that are inconsistent with other observations (SNe Ia, CMB). Nevertheless, improvements on the constraints for all the models are reflected in the $\chi^{2}_{red}$ value when we exclude the systems in the region of $D^{obs}>1$. This unphysical region 
\citep[also found by][]{Leaf:2018lfu}
seems to be related to those systems with different kind of uncertainties (e.g. not fully confirmed lenses, multiple arcs, uncertain redshifts, complex lens substructure, see \ref{tab:SLDgtone}). Thus, as a byproduct of our analysis, results with $D^{obs}>1$ point towards those systems with untrustworthy observed parameters or those which depart from our isothermal spherical mass distribution hypothesis (e.g. external shear and/or lens substructure). Most of the SLS  considered in the present work have been modeled assuming a SIE lens model. Thus, the value of $f$ obtained from the mock data deviates from the range proposed by \cite{Ofek:2003sp}  for some systems and a wider interval should be used. 

Regarding the velocity dispersion, some of the selected regions provide weaker constraints (larger values  in the $\chi^{2}_{red}$ function): $\sigma<210$~km s$^{-1}$ and $\sigma>276$~km s$^{-1}$.  \citet{Chen:2018jcf} also found weak constraints for the lens mass model parameters assuming different regions on the velocity dispersion. This could be due to an observational bias measuring the velocity dispersion $\sigma$, that can be related to a small $\theta_E$, or not measuring the entire lens mass, or  the lens  galaxy  has  close  companions.

We found that eight systems in the $D^{obs}<0.5$ region can not be modeled properly by the theoretical lens equation, 
obtaining double confidence contours for the cosmological parameters. Finally, the lowest  $\chi^{2}_{red}$ value for each model is achieved in the $0.5 \leq D^{obs} \leq 1$ region, with values for the cosmological parameters ($-1.653 \leq \omega_0 \leq -0.735$ and $-4.202 \leq \omega_1 \leq -3.921$) in agreement with those expected from other astrophysical observations (see for instance \citep{Magana:2014,Betoule_2014,Aghanim:2018eyx}).
Therefore, we favour the SS3 sample as the fiduciary sample to constrain DE cosmological parameters. The model selection criteria show that the CPL model is preferred from this sample constraints, i.e. these data point out towards a dynamical dark energy behavior consistent with the three different criteria presented in table \ref{tab:AICBICwCDMCPLyJBPin}, obtaining $\omega_0= -0.735^{+0.319}_{-0.395}$ and $\omega_1= -4.202^{+1.208}_{-0.592}$ for this region. \\
The estimation of the cosmological parameters presented in this paper, employing the strong lensing features of lens galaxies, provides constraints  which are consistent with other cosmological probes  \citep{Magana:2014,Betoule_2014, Scolnic:2017caz, Magana:2017usz,Aghanim:2018eyx}.  Nevertheless, a further  analysis should be done, in particular to consider systematic biases, that help us to more tightly estimate cosmological parameters and improve our method.


\section*{Acknowledgments}
We thank the anonymous referee for thoughtful remarks and suggestions. Authors acknowledges the enlightening conversations and valuable feedback with Karina Rojas and Mario Rodriguez. M.H.A. acknowledges support from CONACYT PhD fellow, Consejo Zacatecano de Ciencia, Tecnolog\'{\i}a e Innovaci\'on (COZCYT) and Centro de Astrof\'{\i}sica de Valpara\'{\i}so (CAV). M.H.A. thanks the staff of the Instituto de F\'{\i}sica y Astronom\'{\i}a of the Universidad de Valpara\'{\i}so where part of this work was done. J.M. acknowledges support from CONICYT project Basal AFB-170002 and CONICYT/FONDECYT 3160674. T.V. acknowledges support from PROGRAMA UNAM-DGAPA-PAPIIT IA102517.
M.A.G.-A. acknowledges support from CONACYT research fellow, Sistema Nacional de Investigadores (SNI), COZCYT and Instituto Avanzado de Cosmolog\'ia (IAC) collaborations. V.M. acknowledges support from Centro de Astrof\'{\i}sica de Valpara\'{\i}so (CAV) and CONICYT REDES (190147).

\section*{Data Availability}
The data underlying this article were accessed from the references presented in Table  \ref{tab:SL}.

\bibliographystyle{mnras}
\bibliography{librero0}

\begin{thebibliography}{}
\makeatletter
\relax
\def\mn@urlcharsother{\let\do\@makeother \do\$\do\&\do\#\do\^\do\_\do\%\do\~}
\def\mn@doi{\begingroup\mn@urlcharsother \@ifnextchar [ {\mn@doi@}
  {\mn@doi@[]}}
\def\mn@doi@[#1]#2{\def\@tempa{#1}\ifx\@tempa\@empty \href
  {http://dx.doi.org/#2} {doi:#2}\else \href {http://dx.doi.org/#2} {#1}\fi
  \endgroup}
\def\mn@eprint#1#2{\mn@eprint@#1:#2::\@nil}
\def\mn@eprint@arXiv#1{\href {http://arxiv.org/abs/#1} {{\tt arXiv:#1}}}
\def\mn@eprint@dblp#1{\href {http://dblp.uni-trier.de/rec/bibtex/#1.xml}
  {dblp:#1}}
\def\mn@eprint@#1:#2:#3:#4\@nil{\def\@tempa {#1}\def\@tempb {#2}\def\@tempc
  {#3}\ifx \@tempc \@empty \let \@tempc \@tempb \let \@tempb \@tempa \fi \ifx
  \@tempb \@empty \def\@tempb {arXiv}\fi \@ifundefined
  {mn@eprint@\@tempb}{\@tempb:\@tempc}{\expandafter \expandafter \csname
  mn@eprint@\@tempb\endcsname \expandafter{\@tempc}}}

\bibitem[\protect\citeauthoryear{Abbott et~al.}{Abbott
  et~al.}{2019}]{Abbott:2018wog}
Abbott T. M.~C.,  et~al., 2019, \mn@doi [Astrophys. J.]
  {10.3847/2041-8213/ab04fa}, 872, L30

\bibitem[\protect\citeauthoryear{Aghanim et~al.}{Aghanim
  et~al.}{2018a}]{Planck_CP:2018}
Aghanim N.,  et~al., 2018a

\bibitem[\protect\citeauthoryear{Aghanim et~al.}{Aghanim
  et~al.}{2018b}]{Aghanim:2018eyx}
Aghanim N.,  et~al., 2018b

\bibitem[\protect\citeauthoryear{{Agnello} et~al.,}{{Agnello}
  et~al.}{2015}]{Agnello2015}
{Agnello} A.,  et~al., 2015, \mn@doi [\mnras] {10.1093/mnras/stv2171}, \href
  {http://adsabs.harvard.edu/abs/2015MNRAS.454.1260A} {454, 1260}

\bibitem[\protect\citeauthoryear{{Akaike}}{{Akaike}}{1974}]{Akaike:1974}
{Akaike} H.,  1974, IEEE Transactions on Automatic Control, \href
  {http://adsabs.harvard.edu/abs/1974ITAC...19..716A} {19, 716}

\bibitem[\protect\citeauthoryear{{Alam} et~al.,}{{Alam}
  et~al.}{2017}]{Alam:2017}
{Alam} S.,  et~al., 2017, \mn@doi [\mnras] {10.1093/mnras/stx721}, \href
  {http://adsabs.harvard.edu/abs/2017MNRAS.470.2617A} {470, 2617}

\bibitem[\protect\citeauthoryear{{Albrecht} et~al.,}{{Albrecht}
  et~al.}{2006}]{Albrecht:2006}
{Albrecht} A.,  et~al., 2006, ArXiv Astrophysics e-prints, \href
  {http://adsabs.harvard.edu/abs/2006astro.ph..9591A} {}

\bibitem[\protect\citeauthoryear{{Auger}, {Treu}, {Bolton}, {Gavazzi},
  {Koopmans}, {Marshall}, {Bundy}  \& {Moustakas}}{{Auger}
  et~al.}{2009}]{Auger2009}
{Auger} M.~W.,  {Treu} T.,  {Bolton} A.~S.,  {Gavazzi} R.,  {Koopmans}
  L.~V.~E.,  {Marshall} P.~J.,  {Bundy} K.,   {Moustakas} L.~A.,  2009, \mn@doi
  [\apj] {10.1088/0004-637X/705/2/1099}, \href
  {http://adsabs.harvard.edu/abs/2009ApJ...705.1099A} {705, 1099}

\bibitem[\protect\citeauthoryear{Barnabe, Czoske, Koopmans, Treu  \&
  Bolton}{Barnabe et~al.}{2011}]{Barnabe:2011gb}
Barnabe M.,  Czoske O.,  Koopmans L. V.~E.,  Treu T.,   Bolton A.~S.,  2011,
  \mn@doi [Mon. Not. Roy. Astron. Soc.] {10.1111/j.1365-2966.2011.18842.x},
  415, 2215

\bibitem[\protect\citeauthoryear{{Bautista} et~al.,}{{Bautista}
  et~al.}{2017}]{Bautista:2017}
{Bautista} J.~E.,  et~al., 2017, \mn@doi [\aap] {10.1051/0004-6361/201730533},
  \href {http://adsabs.harvard.edu/abs/2017A%26A...603A..12B} {603, A12}

\bibitem[\protect\citeauthoryear{Betoule et~al.,}{Betoule
  et~al.}{2014}]{Betoule_2014}
Betoule M.,  et~al., 2014, \mn@doi [Astronomy & Astrophysics]
  {10.1051/0004-6361/201423413}, 568, A22

\bibitem[\protect\citeauthoryear{Biesiada}{Biesiada}{2006}]{Biesiada:2006zf}
Biesiada M.,  2006, \mn@doi [Phys. Rev.] {10.1103/PhysRevD.73.023006}, D73,
  023006

\bibitem[\protect\citeauthoryear{{Biesiada}, {Pi{\'o}rkowska}  \&
  {Malec}}{{Biesiada} et~al.}{2010}]{Biesiada:2010}
{Biesiada} M.,  {Pi{\'o}rkowska} A.,   {Malec} B.,  2010, \mn@doi [\mnras]
  {10.1111/j.1365-2966.2010.16725.x}, \href
  {http://adsabs.harvard.edu/abs/2010MNRAS.406.1055B} {406, 1055}

\bibitem[\protect\citeauthoryear{Bilic, Tupper  \& Viollier}{Bilic
  et~al.}{2002}]{Bilic:2001}
Bilic N.,  Tupper G.~B.,   Viollier R.~D.,  2002, \mn@doi [Phys. Lett.]
  {10.1016/S0370-2693(02)01716-1}, B535, 17

\bibitem[\protect\citeauthoryear{{Blake} et~al.,}{{Blake}
  et~al.}{2012}]{Blake:2012}
{Blake} C.,  et~al., 2012, \mn@doi [\mnras] {10.1111/j.1365-2966.2012.21473.x},
  \href {http://adsabs.harvard.edu/abs/2012MNRAS.425..405B} {425, 405}

\bibitem[\protect\citeauthoryear{{Bolton}, {Burles}, {Koopmans}, {Treu}  \&
  {Moustakas}}{{Bolton} et~al.}{2006}]{bolton2006}
{Bolton} A.~S.,  {Burles} S.,  {Koopmans} L.~V.~E.,  {Treu} T.,   {Moustakas}
  L.~A.,  2006, \mn@doi [\apj] {10.1086/498884}, \href
  {http://adsabs.harvard.edu/abs/2006ApJ...638..703B} {638, 703}

\bibitem[\protect\citeauthoryear{Bolton, Burles, Koopmans, Treu, Gavazzi,
  Moustakas, Wayth  \& Schlegel}{Bolton et~al.}{2008}]{Bolton:2008xf}
Bolton A.~S.,  Burles S.,  Koopmans L. V.~E.,  Treu T.,  Gavazzi R.,  Moustakas
  L.~A.,  Wayth R.,   Schlegel D.~J.,  2008, \mn@doi [Astrophys. J.]
  {10.1086/589327}, 682, 964

\bibitem[\protect\citeauthoryear{Brans \& Dicke}{Brans \&
  Dicke}{1961}]{PhysRev.124.925}
Brans C.,  Dicke R.~H.,  1961, \mn@doi [Phys. Rev.] {10.1103/PhysRev.124.925},
  124, 925

\bibitem[\protect\citeauthoryear{{Brownstein} et~al.,}{{Brownstein}
  et~al.}{2012a}]{brownstein2012}
{Brownstein} J.~R.,  et~al., 2012a, \mn@doi [\apj]
  {10.1088/0004-637X/744/1/41}, \href
  {http://adsabs.harvard.edu/abs/2012ApJ...744...41B} {744, 41}

\bibitem[\protect\citeauthoryear{Brownstein et~al.,}{Brownstein
  et~al.}{2012b}]{0004-637X-744-1-41}
Brownstein J.~R.,  et~al., 2012b, The Astrophysical Journal, 744, 41

\bibitem[\protect\citeauthoryear{Buchdahl}{Buchdahl}{1970}]{Buchdahl}
Buchdahl H.~A.,  1970, \mn@doi [Monthly Notices of the Royal Astronomical
  Society] {10.1093/mnras/150.1.1}, 150, 1

\bibitem[\protect\citeauthoryear{{Cabanac} et~al.,}{{Cabanac}
  et~al.}{2007}]{cabanac2007}
{Cabanac} R.~A.,  et~al., 2007, \mn@doi [\aap] {10.1051/0004-6361:20065810},
  \href {http://adsabs.harvard.edu/abs/2007A%26A...461..813C} {461, 813}

\bibitem[\protect\citeauthoryear{{Caldera-Cabral}, {Maartens}  \&
  {Ure{\~n}a-L{\'o}pez}}{{Caldera-Cabral} et~al.}{2009}]{Calrdera:2009}
{Caldera-Cabral} G.,  {Maartens} R.,   {Ure{\~n}a-L{\'o}pez} L.~A.,  2009,
  \mn@doi [\prd] {10.1103/PhysRevD.79.063518}, \href
  {http://adsabs.harvard.edu/abs/2009PhRvD..79f3518C} {79, 063518}

\bibitem[\protect\citeauthoryear{Caldwell}{Caldwell}{2002}]{Caldwell:2002}
Caldwell R.~R.,  2002, \mn@doi [Phys. Lett.] {10.1016/S0370-2693(02)02589-3},
  B545, 23

\bibitem[\protect\citeauthoryear{{Cao}, {Pan}, {Biesiada}, {Godlowski}  \&
  {Zhu}}{{Cao} et~al.}{2012}]{Cao:2012}
{Cao} S.,  {Pan} Y.,  {Biesiada} M.,  {Godlowski} W.,   {Zhu} Z.-H.,  2012,
  \mn@doi [\jcap] {10.1088/1475-7516/2012/03/016}, \href
  {http://adsabs.harvard.edu/abs/2012JCAP...03..016C} {3, 016}

\bibitem[\protect\citeauthoryear{Cao, Biesiada, Gavazzi, Piórkowska  \&
  Zhu}{Cao et~al.}{2015}]{Cao:2015qja}
Cao S.,  Biesiada M.,  Gavazzi R.,  Piórkowska A.,   Zhu Z.-H.,  2015, \mn@doi
  [Astrophys. J.] {10.1088/0004-637X/806/2/185}, 806, 185

\bibitem[\protect\citeauthoryear{{C{\'a}rdenas} \& {Rivera}}{{C{\'a}rdenas} \&
  {Rivera}}{2012}]{Cardenas:2012}
{C{\'a}rdenas} V.~H.,  {Rivera} M.,  2012, \mn@doi [Physics Letters B]
  {10.1016/j.physletb.2012.03.004}, \href
  {http://adsabs.harvard.edu/abs/2012PhLB..710..251C} {710, 251}

\bibitem[\protect\citeauthoryear{{C{\'a}rdenas}, {Bernal}  \&
  {Bonilla}}{{C{\'a}rdenas} et~al.}{2013}]{Cardenas:2013}
{C{\'a}rdenas} V.~H.,  {Bernal} C.,   {Bonilla} A.,  2013, \mn@doi [\mnras]
  {10.1093/mnras/stt983}, \href
  {http://adsabs.harvard.edu/abs/2013MNRAS.433.3534C} {433, 3534}

\bibitem[\protect\citeauthoryear{{Chaplygin}}{{Chaplygin}}{1904}]{Chaplygin}
{Chaplygin} S.,  1904, Sci. Mem. Mosc. Univ. Math. Phys, 21

\bibitem[\protect\citeauthoryear{Chen, Li  \& Shu}{Chen
  et~al.}{2018}]{Chen:2018jcf}
Chen Y.,  Li R.,   Shu Y.,  2018

\bibitem[\protect\citeauthoryear{Chevallier \& Polarski}{Chevallier \&
  Polarski}{2001}]{Chevallier:2000qy}
Chevallier M.,  Polarski D.,  2001, \mn@doi [Int. J. Mod. Phys.]
  {10.1142/S0218271801000822}, D10, 213

\bibitem[\protect\citeauthoryear{Chiba, Okabe  \& Yamaguchi}{Chiba
  et~al.}{2000}]{Chiba:2000}
Chiba T.,  Okabe T.,   Yamaguchi M.,  2000, \mn@doi [Phys. Rev.]
  {10.1103/PhysRevD.62.023511}, D62, 023511

\bibitem[\protect\citeauthoryear{Copeland, Sami  \& Tsujikawa}{Copeland
  et~al.}{2006}]{Copeland:2006wr}
Copeland E.~J.,  Sami M.,   Tsujikawa S.,  2006, \mn@doi [Int. J. Mod. Phys.]
  {10.1142/S021827180600942X}, D15, 1753

\bibitem[\protect\citeauthoryear{{Eisenstein} et~al.,}{{Eisenstein}
  et~al.}{2005}]{Eisenstein:2005}
{Eisenstein} D.~J.,  et~al., 2005, \mn@doi [\apj] {10.1086/466512}, \href
  {http://adsabs.harvard.edu/abs/2005ApJ...633..560E} {633, 560}

\bibitem[\protect\citeauthoryear{Foreman-Mackey, Hogg, Lang  \&
  Goodman}{Foreman-Mackey et~al.}{2013}]{ForemanMackey:2012ig}
Foreman-Mackey D.,  Hogg D.~W.,  Lang D.,   Goodman J.,  2013, \mn@doi [Publ.
  Astron. Soc. Pac.] {10.1086/670067}, 125, 306

\bibitem[\protect\citeauthoryear{Futamase \& Yoshida}{Futamase \&
  Yoshida}{2001}]{Futamase:2000qx}
Futamase T.,  Yoshida S.,  2001, \mn@doi [Prog. Theor. Phys.]
  {10.1143/PTP.105.887}, 105, 887

\bibitem[\protect\citeauthoryear{Galiautdinov \& Kopeikin}{Galiautdinov \&
  Kopeikin}{2016}]{Galiautdinov:2016qqy}
Galiautdinov A.,  Kopeikin S.~M.,  2016, \mn@doi [Phys. Rev.]
  {10.1103/PhysRevD.94.044015}, D94, 044015

\bibitem[\protect\citeauthoryear{Garc\'ia-Aspeitia \& Matos}{Garc\'ia-Aspeitia
  \& Matos}{2011}]{Aspeitia:2009bj}
Garc\'ia-Aspeitia M.~A.,  Matos T.,  2011, \mn@doi [Gen. Rel. Grav.]
  {10.1007/s10714-010-1093-2}, 43, 315

\bibitem[\protect\citeauthoryear{{Garc{\'{\i}}a-Aspeitia}, {Maga{\~n}a},
  {Hern{\'a}ndez-Almada}  \& {Motta}}{{Garc{\'{\i}}a-Aspeitia}
  et~al.}{2018a}]{Garcia:2018}
{Garc{\'{\i}}a-Aspeitia} M.~A.,  {Maga{\~n}a} J.,  {Hern{\'a}ndez-Almada} A.,
  {Motta} V.,  2018a, \mn@doi [International Journal of Modern Physics D]
  {10.1142/S0218271818500062}, \href
  {http://adsabs.harvard.edu/abs/2018IJMPD..2750006G} {27, 1850006}

\bibitem[\protect\citeauthoryear{Garc\'ia-Aspeitia, Hern\'andez-Almada,
  Maga\~na, Amante, Motta  \& Mart\'inez-Robles}{Garc\'ia-Aspeitia
  et~al.}{2018b}]{Garcia-Aspeitia:2018fvw}
Garc\'ia-Aspeitia M.~A.,  Hern\'andez-Almada A.,  Maga\~na J.,  Amante M.~H.,
  Motta V.,   Mart\'inez-Robles C.,  2018b, \mn@doi [Phys. Rev.]
  {10.1103/PhysRevD.97.101301}, D97, 101301

\bibitem[\protect\citeauthoryear{Garc\'ia-Aspeitia, Mart\'inez-Robles,
  Hern\'andez-Almada, Maga\~na  \& Motta}{Garc\'ia-Aspeitia
  et~al.}{2019}]{Garcia-Aspeitia:2019yni}
Garc\'ia-Aspeitia M.~A.,  Mart\'inez-Robles C.,  Hern\'andez-Almada A.,
  Maga\~na J.,   Motta V.,  2019, \mn@doi [Phys. Rev.]
  {10.1103/PhysRevD.99.123525}, D99, 123525

\bibitem[\protect\citeauthoryear{Gelman \& Rubin}{Gelman \&
  Rubin}{1992}]{gelman1992}
Gelman A.,  Rubin D.~B.,  1992, \mn@doi [Statist. Sci.]
  {10.1214/ss/1177011136}, 7, 457

\bibitem[\protect\citeauthoryear{{Grillo}, {Lombardi}  \& {Bertin}}{{Grillo}
  et~al.}{2008}]{2008A&A...477..397G}
{Grillo} C.,  {Lombardi} M.,   {Bertin} G.,  2008, \mn@doi [\aap]
  {10.1051/0004-6361:20077534}, \href
  {http://adsabs.harvard.edu/abs/2008A%26A...477..397G} {477, 397}

\bibitem[\protect\citeauthoryear{Hernandez-Almada, Magana, Garcia-Aspeitia  \&
  Motta}{Hernandez-Almada et~al.}{2018}]{Hernandez-Almada:2018osh}
Hernandez-Almada A.,  Magana J.,  Garcia-Aspeitia M.~A.,   Motta V.,  2018

\bibitem[\protect\citeauthoryear{{Hewett}, {Irwin}, {Foltz}, {Harding},
  {Corrigan}, {Webster}  \& {Dinshaw}}{{Hewett}
  et~al.}{1994}]{1994AJ....108.1534H}
{Hewett} P.~C.,  {Irwin} M.~J.,  {Foltz} C.~B.,  {Harding} M.~E.,  {Corrigan}
  R.~T.,  {Webster} R.~L.,   {Dinshaw} N.,  1994, \mn@doi [\aj]
  {10.1086/117174}, \href {http://adsabs.harvard.edu/abs/1994AJ....108.1534H}
  {108, 1534}

\bibitem[\protect\citeauthoryear{{Inada} et~al.,}{{Inada}
  et~al.}{2003}]{Inada2003}
{Inada} N.,  et~al., 2003, \mn@doi [\aj] {10.1086/375906}, \href
  {http://adsabs.harvard.edu/abs/2003AJ....126..666I} {126, 666}

\bibitem[\protect\citeauthoryear{{Inada} et~al.,}{{Inada}
  et~al.}{2005}]{Inada2005}
{Inada} N.,  et~al., 2005, \mn@doi [\aj] {10.1086/432930}, \href
  {http://adsabs.harvard.edu/abs/2005AJ....130.1967I} {130, 1967}

\bibitem[\protect\citeauthoryear{Jassal, Bagla  \& Padmanabhan}{Jassal
  et~al.}{2005}]{Jassal}
Jassal H.~K.,  Bagla J.~S.,   Padmanabhan T.,  2005, \mn@doi [Mon. Not. Roy.
  Astron. Soc.] {10.1111/j.1745-3933.2005.08577.x}, 356, L11

\bibitem[\protect\citeauthoryear{Jimenez \& Loeb}{Jimenez \&
  Loeb}{2002}]{Jimenez:2001gg}
Jimenez R.,  Loeb A.,  2002, \mn@doi [Astrophys. J.] {10.1086/340549}, 573, 37

\bibitem[\protect\citeauthoryear{{Jorgensen}, {Franx}  \&
  {Kjaergaard}}{{Jorgensen} et~al.}{1995a}]{1995MNRAS.273.1097J}
{Jorgensen} I.,  {Franx} M.,   {Kjaergaard} P.,  1995a, \mn@doi [\mnras]
  {10.1093/mnras/273.4.1097}, \href
  {http://adsabs.harvard.edu/abs/1995MNRAS.273.1097J} {273, 1097}

\bibitem[\protect\citeauthoryear{{Jorgensen}, {Franx}  \&
  {Kjaergaard}}{{Jorgensen} et~al.}{1995b}]{1995MNRAS.276.1341J}
{Jorgensen} I.,  {Franx} M.,   {Kjaergaard} P.,  1995b, \mn@doi [\mnras]
  {10.1093/mnras/276.4.1341}, \href
  {http://adsabs.harvard.edu/abs/1995MNRAS.276.1341J} {276, 1341}

\bibitem[\protect\citeauthoryear{{Jullo}, {Natarajan}, {Kneib}, {D'Aloisio},
  {Limousin}, {Richard}  \& {Schimd}}{{Jullo} et~al.}{2010}]{Jullo:2010Sci}
{Jullo} E.,  {Natarajan} P.,  {Kneib} J.~P.,  {D'Aloisio} A.,  {Limousin} M.,
  {Richard} J.,   {Schimd} C.,  2010, \mn@doi [Science]
  {10.1126/science.1185759}, \href
  {https://ui.adsabs.harvard.edu/abs/2010Sci...329..924J} {329, 924}

\bibitem[\protect\citeauthoryear{Kamenshchik, Moschella  \&
  Pasquier}{Kamenshchik et~al.}{2001}]{Kamenshchik:2001}
Kamenshchik A.~{\relax Yu}.,  Moschella U.,   Pasquier V.,  2001, \mn@doi
  [Phys. Lett.] {10.1016/S0370-2693(01)00571-8}, B511, 265

\bibitem[\protect\citeauthoryear{{Keeton}}{{Keeton}}{2011}]{2011ascl.soft02003K}
{Keeton} C.~R.,  2011, {GRAVLENS: Computational Methods for Gravitational
  Lensing} (\mn@eprint {ascl} {1102.003})

\bibitem[\protect\citeauthoryear{{Kochanek}}{{Kochanek}}{1995}]{kochanek1995}
{Kochanek} C.~S.,  1995, \mn@doi [\apj] {10.1086/175721}, \href
  {http://adsabs.harvard.edu/abs/1995ApJ...445..559K} {445, 559}

\bibitem[\protect\citeauthoryear{{Koopmans} \& {Treu}}{{Koopmans} \&
  {Treu}}{2003a}]{koopmans2003}
{Koopmans} L.~V.~E.,  {Treu} T.,  2003a, \mn@doi [\apj] {10.1086/345423}, \href
  {http://adsabs.harvard.edu/abs/2003ApJ...583..606K} {583, 606}

\bibitem[\protect\citeauthoryear{Koopmans \& Treu}{Koopmans \&
  Treu}{2003b}]{Koopmans:2002qh}
Koopmans L. V.~E.,  Treu T.,  2003b, \mn@doi [Astrophys. J.] {10.1086/345423},
  583, 606

\bibitem[\protect\citeauthoryear{{Lacy}, {Gregg}, {Becker}, {White}, {Glikman},
  {Helfand}  \& {Winn}}{{Lacy} et~al.}{2002}]{Lacy2002}
{Lacy} M.,  {Gregg} M.,  {Becker} R.~H.,  {White} R.~L.,  {Glikman} E.,
  {Helfand} D.,   {Winn} J.~N.,  2002, \mn@doi [\aj] {10.1086/340568}, \href
  {http://adsabs.harvard.edu/abs/2002AJ....123.2925L} {123, 2925}

\bibitem[\protect\citeauthoryear{Langlois, Saito, Yamauchi  \& Noui}{Langlois
  et~al.}{2018}]{Langlois:2017dyl}
Langlois D.,  Saito R.,  Yamauchi D.,   Noui K.,  2018, \mn@doi [Phys. Rev.]
  {10.1103/PhysRevD.97.061501}, D97, 061501

\bibitem[\protect\citeauthoryear{Leaf \& Melia}{Leaf \&
  Melia}{2018}]{Leaf:2018lfu}
Leaf K.,  Melia F.,  2018, \mn@doi [Mon. Not. Roy. Astron. Soc.]
  {10.1093/mnras/sty1365}, 478, 5104

\bibitem[\protect\citeauthoryear{{Lehar}, {Cooke}, {Lawrence}, {Silber}  \&
  {Langston}}{{Lehar} et~al.}{1996}]{Lehar1996}
{Lehar} J.,  {Cooke} A.~J.,  {Lawrence} C.~R.,  {Silber} A.~D.,   {Langston}
  G.~I.,  1996, \mn@doi [\aj] {10.1086/117919}, \href
  {http://adsabs.harvard.edu/abs/1996AJ....111.1812L} {111, 1812}

\bibitem[\protect\citeauthoryear{{Leier}, {Ferreras}, {Saha}  \&
  {Falco}}{{Leier} et~al.}{2011}]{Leier2011}
{Leier} D.,  {Ferreras} I.,  {Saha} P.,   {Falco} E.~E.,  2011, \mn@doi [\apj]
  {10.1088/0004-637X/740/2/97}, \href
  {http://adsabs.harvard.edu/abs/2011ApJ...740...97L} {740, 97}

\bibitem[\protect\citeauthoryear{{Li}, {Li}, {Wang}  \& {Wang}}{{Li}
  et~al.}{2011}]{Li:2011}
{Li} M.,  {Li} X.-D.,  {Wang} S.,   {Wang} Y.,  2011, \mn@doi [Communications
  in Theoretical Physics] {10.1088/0253-6102/56/3/24}, \href
  {http://adsabs.harvard.edu/abs/2011CoTPh..56..525L} {56, 525}

\bibitem[\protect\citeauthoryear{Linder}{Linder}{2003}]{Linder:2002dt}
Linder E.~V.,  2003, \mn@doi [Phys. Rev.] {10.1103/PhysRevD.68.083503}, D68,
  083503

\bibitem[\protect\citeauthoryear{{Maga{\~n}a}, {C{\'a}rdenas}  \&
  {Motta}}{{Maga{\~n}a} et~al.}{2014}]{Magana:2014}
{Maga{\~n}a} J.,  {C{\'a}rdenas} V.~H.,   {Motta} V.,  2014, \mn@doi [\jcap]
  {10.1088/1475-7516/2014/10/017}, \href
  {http://adsabs.harvard.edu/abs/2014JCAP...10..017M} {10, 017}

\bibitem[\protect\citeauthoryear{{Maga{\~n}a}, {Motta}, {C{\'a}rdenas},
  {Verdugo}  \& {Jullo}}{{Maga{\~n}a} et~al.}{2015}]{Magana:2015ApJ}
{Maga{\~n}a} J.,  {Motta} V.,  {C{\'a}rdenas} V.~H.,  {Verdugo} T.,   {Jullo}
  E.,  2015, \mn@doi [\apj] {10.1088/0004-637X/813/1/69}, \href
  {https://ui.adsabs.harvard.edu/abs/2015ApJ...813...69M} {813, 69}

\bibitem[\protect\citeauthoryear{Maga\~na, Motta, Cardenas  \& Foex}{Maga\~na
  et~al.}{2017}]{Magana:2017usz}
Maga\~na J.,  Motta V.,  Cardenas V.~H.,   Foex G.,  2017, \mn@doi [Mon. Not.
  Roy. Astron. Soc.] {10.1093/mnras/stx750}, 469, 47

\bibitem[\protect\citeauthoryear{{Maga{\~n}a}, {Acebr{\'o}n}, {Motta},
  {Verdugo}, {Jullo}  \& {Limousin}}{{Maga{\~n}a}
  et~al.}{2018}]{Magana:2018ApJ}
{Maga{\~n}a} J.,  {Acebr{\'o}n} A.,  {Motta} V.,  {Verdugo} T.,  {Jullo} E.,
  {Limousin} M.,  2018, \mn@doi [\apj] {10.3847/1538-4357/aada7d}, \href
  {https://ui.adsabs.harvard.edu/abs/2018ApJ...865..122M} {865, 122}

\bibitem[\protect\citeauthoryear{Maga{\~n}a, Amante, Garcia-Aspeitia  \&
  Motta}{Maga{\~n}a et~al.}{2018}]{Magana:2017nfs}
Maga{\~n}a J.,  Amante M.~H.,  Garcia-Aspeitia M.~A.,   Motta V.,  2018,
  \mn@doi [Mon. Not. Roy. Astron. Soc.] {10.1093/mnras/sty260}, 476, 1036

\bibitem[\protect\citeauthoryear{Melia, Wei  \& Wu}{Melia
  et~al.}{2015}]{Melia:2014oqa}
Melia F.,  Wei J.-J.,   Wu X.-F.,  2015, \mn@doi [Astron. J.]
  {10.1088/0004-6256/149/1/2}, 149, 2

\bibitem[\protect\citeauthoryear{{Moresco} et~al.,}{{Moresco}
  et~al.}{2016}]{Moresco:2016JCAP}
{Moresco} M.,  et~al., 2016, \mn@doi [Journal of Cosmology and Astro-Particle
  Physics] {10.1088/1475-7516/2016/05/014}, \href
  {https://ui.adsabs.harvard.edu/abs/2016JCAP...05..014M} {2016, 014}

\bibitem[\protect\citeauthoryear{Morgan, Becker, Gregg, Schechter  \&
  White}{Morgan et~al.}{2001}]{Morgan:2000ap}
Morgan N.~D.,  Becker R.~H.,  Gregg M.~D.,  Schechter P.~L.,   White R.~L.,
  2001, \mn@doi [Astron. J.] {10.1086/318744}, 121, 611

\bibitem[\protect\citeauthoryear{{Morgan}, {Kochanek}, {Pevunova}  \&
  {Schechter}}{{Morgan} et~al.}{2005}]{Morgan2005}
{Morgan} N.~D.,  {Kochanek} C.~S.,  {Pevunova} O.,   {Schechter} P.~L.,  2005,
  \mn@doi [\aj] {10.1086/430145}, \href
  {http://adsabs.harvard.edu/abs/2005AJ....129.2531M} {129, 2531}

\bibitem[\protect\citeauthoryear{{Mu{\~n}oz}, {Falco}, {Kochanek}, {Leh{\'a}r},
  {McLeod}, {Impey}, {Rix}  \& {Peng}}{{Mu{\~n}oz} et~al.}{1998}]{munoz1998}
{Mu{\~n}oz} J.~A.,  {Falco} E.~E.,  {Kochanek} C.~S.,  {Leh{\'a}r} J.,
  {McLeod} B.~A.,  {Impey} C.~D.,  {Rix} H.-W.,   {Peng} C.~Y.,  1998, \mn@doi
  [\apss] {10.1023/A:1002120921330}, \href
  {http://adsabs.harvard.edu/abs/1998Ap%26SS.263...51M} {263, 51}

\bibitem[\protect\citeauthoryear{{Mu{\~n}oz}, {Kochanek}  \&
  {Keeton}}{{Mu{\~n}oz} et~al.}{2001}]{munoz2001}
{Mu{\~n}oz} J.~A.,  {Kochanek} C.~S.,   {Keeton} C.~R.,  2001, \mn@doi [\apj]
  {10.1086/322314}, \href {http://adsabs.harvard.edu/abs/2001ApJ...558..657M}
  {558, 657}

\bibitem[\protect\citeauthoryear{Ofek, Rix  \& Maoz}{Ofek
  et~al.}{2003}]{Ofek:2003sp}
Ofek E.~O.,  Rix H.-W.,   Maoz D.,  2003, \mn@doi [Mon. Not. Roy. Astron. Soc.]
  {10.1046/j.1365-8711.2003.06707.x}, 343, 639

\bibitem[\protect\citeauthoryear{{Ofek}, {Maoz}, {Rix}, {Kochanek}  \&
  {Falco}}{{Ofek} et~al.}{2006}]{Ofek2006}
{Ofek} E.~O.,  {Maoz} D.,  {Rix} H.-W.,  {Kochanek} C.~S.,   {Falco} E.~E.,
  2006, \mn@doi [\apj] {10.1086/500403}, \href
  {http://adsabs.harvard.edu/abs/2006ApJ...641...70O} {641, 70}

\bibitem[\protect\citeauthoryear{Perez \& Sudarsky}{Perez \&
  Sudarsky}{2017}]{Perez:2017krv}
Perez A.,  Sudarsky D.,  2017

\bibitem[\protect\citeauthoryear{Perlmutter, Aldering, Goldhaber, Knop, Nugent,
  others  \& Project}{Perlmutter et~al.}{1999}]{Perlmutter}
Perlmutter S.,  Aldering G.,  Goldhaber G.,  Knop R.~A.,  Nugent P.,  others
  Project T. S.~C.,  1999, The Astrophysical Journal, 517, 565

\bibitem[\protect\citeauthoryear{{Pindor} et~al.,}{{Pindor}
  et~al.}{2004}]{Pindor2004}
{Pindor} B.,  et~al., 2004, \mn@doi [\aj] {10.1086/381904}, \href
  {http://adsabs.harvard.edu/abs/2004AJ....127.1318P} {127, 1318}

\bibitem[\protect\citeauthoryear{{Planck Collaboration} et~al.,}{{Planck
  Collaboration} et~al.}{2016}]{Planck:2016}
{Planck Collaboration} et~al., 2016, \mn@doi [\aap]
  {10.1051/0004-6361/201525830}, \href
  {http://adsabs.harvard.edu/abs/2016A%26A...594A..13P} {594, A13}

\bibitem[\protect\citeauthoryear{{Qi}, {Cao}, {Zhang}, {Biesiada}, {Wu}  \&
  {Zhu}}{{Qi} et~al.}{2018}]{Qi:2018arXiv}
{Qi} J.-Z.,  {Cao} S.,  {Zhang} S.,  {Biesiada} M.,  {Wu} Y.,   {Zhu} Z.-H.,
  2018, preprint, \href {http://adsabs.harvard.edu/abs/2018arXiv180301990Q} {}
  (\mn@eprint {arXiv} {1803.01990})

\bibitem[\protect\citeauthoryear{{Ratra} \& {Peebles}}{{Ratra} \&
  {Peebles}}{1988}]{Ratra:1988}
{Ratra} B.,  {Peebles} P.~J.~E.,  1988, \mn@doi [\prd]
  {10.1103/PhysRevD.37.3406}, \href
  {http://adsabs.harvard.edu/abs/1988PhRvD..37.3406R} {37, 3406}

\bibitem[\protect\citeauthoryear{Riess, Filippenko, Challis, Clocchiatti,
  Diercks  et~al.}{Riess et~al.}{1998}]{Riess}
Riess A.~G.,  Filippenko A.~V.,  Challis P.,  Clocchiatti A.,  Diercks A.,
  et~al., 1998, The Astronomical Journal, 116, 1009

\bibitem[\protect\citeauthoryear{Riess et~al.}{Riess
  et~al.}{2016}]{Riess:2016jrr}
Riess A.~G.,  et~al., 2016, \mn@doi [Astrophys. J.]
  {10.3847/0004-637X/826/1/56}, 826, 56

\bibitem[\protect\citeauthoryear{{Riess}, {Casertano}, {Yuan}, {Macri}  \&
  {Scolnic}}{{Riess} et~al.}{2019}]{Riess:2019}
{Riess} A.~G.,  {Casertano} S.,  {Yuan} W.,  {Macri} L.~M.,   {Scolnic} D.,
  2019, arXiv e-prints, \href
  {http://adsabs.harvard.edu/abs/2019arXiv190307603R} {}

\bibitem[\protect\citeauthoryear{Ritondale, Auger, Vegetti  \&
  McKean}{Ritondale et~al.}{2018}]{10.1093/mnras/sty2833}
Ritondale E.,  Auger M.~W.,  Vegetti S.,   McKean J.~P.,  2018, \mn@doi
  [Monthly Notices of the Royal Astronomical Society] {10.1093/mnras/sty2833},
  482, 4744

\bibitem[\protect\citeauthoryear{{Rusin}, {Norbury}, {Biggs}, {Marlow},
  {Jackson}, {Browne}, {Wilkinson}  \& {Myers}}{{Rusin}
  et~al.}{2002}]{rusin2002}
{Rusin} D.,  {Norbury} M.,  {Biggs} A.~D.,  {Marlow} D.~R.,  {Jackson} N.~J.,
  {Browne} I.~W.~A.,  {Wilkinson} P.~N.,   {Myers} S.~T.,  2002, \mn@doi
  [\mnras] {10.1046/j.1365-8711.2002.05043.x}, \href
  {http://adsabs.harvard.edu/abs/2002MNRAS.330..205R} {330, 205}

\bibitem[\protect\citeauthoryear{{Rusin} et~al.,}{{Rusin}
  et~al.}{2003a}]{rusin2003}
{Rusin} D.,  et~al., 2003a, \mn@doi [\apj] {10.1086/346206}, \href
  {http://adsabs.harvard.edu/abs/2003ApJ...587..143R} {587, 143}

\bibitem[\protect\citeauthoryear{{Rusin}, {Kochanek}  \& {Keeton}}{{Rusin}
  et~al.}{2003b}]{rusin2003b}
{Rusin} D.,  {Kochanek} C.~S.,   {Keeton} C.~R.,  2003b, \mn@doi [\apj]
  {10.1086/377356}, \href {http://adsabs.harvard.edu/abs/2003ApJ...595...29R}
  {595, 29}

\bibitem[\protect\citeauthoryear{Schmidt, Suntzeff, Phillips, Schommer,
  Clocchiatti  et~al.}{Schmidt et~al.}{1998}]{Schmidt}
Schmidt B.~P.,  Suntzeff N.~B.,  Phillips M.~M.,  Schommer R.~A.,  Clocchiatti
  A.,   et~al., 1998, The Astrophysical Journal, 507, 46

\bibitem[\protect\citeauthoryear{{Schneider}, {Ehlers}  \& {Falco}}{{Schneider}
  et~al.}{1992}]{Schneider_book:1992}
{Schneider} P.,  {Ehlers} J.,   {Falco} E.~E.,  1992, {Gravitational Lenses},
  \mn@doi{10.1007/978-3-662-03758-4.
}

\bibitem[\protect\citeauthoryear{{Schwarz}}{{Schwarz}}{1978}]{Schwarz:1978}
{Schwarz} G.,  1978, Annals of Statistics, \href
  {http://adsabs.harvard.edu/abs/1978AnSta...6..461S} {6, 461}

\bibitem[\protect\citeauthoryear{Scolnic et~al.}{Scolnic
  et~al.}{2018}]{Scolnic:2017caz}
Scolnic D.~M.,  et~al., 2018, \mn@doi [Astrophys. J.]
  {10.3847/1538-4357/aab9bb}, 859, 101

\bibitem[\protect\citeauthoryear{Shi, Huang  \& Lu}{Shi
  et~al.}{2012}]{Shi:2012ma}
Shi K.,  Huang Y.,   Lu T.,  2012, \mn@doi [Mon. Not. Roy. Astron. Soc.]
  {10.1111/j.1365-2966.2012.21784.x}, 426, 2452

\bibitem[\protect\citeauthoryear{Shu et~al.,}{Shu
  et~al.}{2016}]{0004-637X-833-2-264}
Shu Y.,  et~al., 2016, The Astrophysical Journal, 833, 264

\bibitem[\protect\citeauthoryear{{Shu} et~al.,}{{Shu}
  et~al.}{2017}]{2017ApJ...851...48S}
{Shu} Y.,  et~al., 2017, \mn@doi [\apj] {10.3847/1538-4357/aa9794}, \href
  {http://adsabs.harvard.edu/abs/2017ApJ...851...48S} {851, 48}

\bibitem[\protect\citeauthoryear{Sonnenfeld, Gavazzi, Suyu, Treu  \&
  Marshall}{Sonnenfeld et~al.}{2013a}]{Sonnenfeld:2013cha}
Sonnenfeld A.,  Gavazzi R.,  Suyu S.~H.,  Treu T.,   Marshall P.~J.,  2013a,
  \mn@doi [Astrophys. J.] {10.1088/0004-637X/777/2/97}, 777, 97

\bibitem[\protect\citeauthoryear{Sonnenfeld, Treu, Gavazzi, Suyu, Marshall,
  Auger  \& Nipoti}{Sonnenfeld et~al.}{2013b}]{Sonnenfeld:2013xga}
Sonnenfeld A.,  Treu T.,  Gavazzi R.,  Suyu S.~H.,  Marshall P.~J.,  Auger
  M.~W.,   Nipoti C.,  2013b, \mn@doi [Astrophys. J.]
  {10.1088/0004-637X/777/2/98}, 777, 98

\bibitem[\protect\citeauthoryear{Sonnenfeld, Treu, Marshall, Suyu, Gavazzi,
  Auger  \& Nipoti}{Sonnenfeld et~al.}{2015}]{Sonnenfeld:2014gpa}
Sonnenfeld A.,  Treu T.,  Marshall P.~J.,  Suyu S.~H.,  Gavazzi R.,  Auger M.,
   Nipoti C.,  2015, \mn@doi [Astrophys. J.] {10.1088/0004-637X/800/2/94}, 800,
  94

\bibitem[\protect\citeauthoryear{Sotiriou \& Faraoni}{Sotiriou \&
  Faraoni}{2010}]{Sotiriou:2008rp}
Sotiriou T.~P.,  Faraoni V.,  2010, \mn@doi [Rev. Mod. Phys.]
  {10.1103/RevModPhys.82.451}, 82, 451

\bibitem[\protect\citeauthoryear{{Spingola}, {McKean}, {Auger}, {Fassnacht},
  {Koopmans}, {Lagattuta}  \& {Vegetti}}{{Spingola}
  et~al.}{2018}]{2018MNRAS.478.4816S}
{Spingola} C.,  {McKean} J.~P.,  {Auger} M.~W.,  {Fassnacht} C.~D.,  {Koopmans}
  L.~V.~E.,  {Lagattuta} D.~J.,   {Vegetti} S.,  2018, \mn@doi [\mnras]
  {10.1093/mnras/sty1326}, \href
  {https://ui.adsabs.harvard.edu/abs/2018MNRAS.478.4816S} {478, 4816}

\bibitem[\protect\citeauthoryear{Starobinsky}{Starobinsky}{1980}]{STAROBINSKY198099}
Starobinsky A.,  1980, \mn@doi [Physics Letters B]
  {https://doi.org/10.1016/0370-2693(80)90670-X}, 91, 99

\bibitem[\protect\citeauthoryear{{Tonry}}{{Tonry}}{1998}]{tonry1998}
{Tonry} J.~L.,  1998, \mn@doi [\aj] {10.1086/300170}, \href
  {http://adsabs.harvard.edu/abs/1998AJ....115....1T} {115, 1}

\bibitem[\protect\citeauthoryear{Treu \& Koopmans}{Treu \&
  Koopmans}{2002}]{Treu:2002ee}
Treu T.,  Koopmans L.,  2002, \mn@doi [Astrophys. J.] {10.1086/341216}, 575, 87

\bibitem[\protect\citeauthoryear{{Treu} \& {Koopmans}}{{Treu} \&
  {Koopmans}}{2004}]{Treu2004}
{Treu} T.,  {Koopmans} L.~V.~E.,  2004, \mn@doi [\apj] {10.1086/422245}, \href
  {http://adsabs.harvard.edu/abs/2004ApJ...611..739T} {611, 739}

\bibitem[\protect\citeauthoryear{{Treu}, {Koopmans}, {Bolton}, {Burles}  \&
  {Moustakas}}{{Treu} et~al.}{2006}]{Treu:2006ApJ}
{Treu} T.,  {Koopmans} L.~V.,  {Bolton} A.~S.,  {Burles} S.,   {Moustakas}
  L.~A.,  2006, \mn@doi [\apj] {10.1086/500124}, \href
  {http://adsabs.harvard.edu/abs/2006ApJ...640..662T} {640, 662}

\bibitem[\protect\citeauthoryear{{Treu} et~al.,}{{Treu}
  et~al.}{2018}]{treu2018}
{Treu} T.,  et~al., 2018, \mn@doi [\mnras] {10.1093/mnras/sty2329}, \href
  {http://adsabs.harvard.edu/abs/2018MNRAS.481.1041T} {481, 1041}

\bibitem[\protect\citeauthoryear{{Turner}, {Ostriker}  \& {Gott}}{{Turner}
  et~al.}{1984}]{1984ApJ...284....1T}
{Turner} E.~L.,  {Ostriker} J.~P.,   {Gott} J.~R. I.,  1984, \mn@doi [\apj]
  {10.1086/162379}, \href
  {https://ui.adsabs.harvard.edu/abs/1984ApJ...284....1T} {284, 1}

\bibitem[\protect\citeauthoryear{Vegetti, Koopmans, Auger, Treu  \&
  Bolton}{Vegetti et~al.}{2014}]{Vegetti:2014lqa}
Vegetti S.,  Koopmans L. V.~E.,  Auger M.~W.,  Treu T.,   Bolton A.~S.,  2014,
  \mn@doi [Mon. Not. Roy. Astron. Soc.] {10.1093/mnras/stu943}, 442, 2017

\bibitem[\protect\citeauthoryear{{Wang}}{{Wang}}{2008}]{Wang:2008}
{Wang} Y.,  2008, \mn@doi [\prd] {10.1103/PhysRevD.77.123525}, \href
  {http://adsabs.harvard.edu/abs/2008PhRvD..77l3525W} {77, 123525}

\bibitem[\protect\citeauthoryear{{Wang}, {Hu}, {Li}  \& {Li}}{{Wang}
  et~al.}{2016}]{Wang:2016ApJ}
{Wang} S.,  {Hu} Y.,  {Li} M.,   {Li} N.,  2016, \mn@doi [\apj]
  {10.3847/0004-637X/821/1/60}, \href
  {http://adsabs.harvard.edu/abs/2016ApJ...821...60W} {821, 60}

\bibitem[\protect\citeauthoryear{Weinberg}{Weinberg}{1989}]{Weinberg}
Weinberg S.,  1989, Reviews of Modern Physics, 61

\bibitem[\protect\citeauthoryear{Wetterich}{Wetterich}{1988}]{WETTERICH:1988}
Wetterich C.,  1988, \mn@doi [Nuclear Physics B]
  {https://doi.org/10.1016/0550-3213(88)90193-9}, 302, 668

\bibitem[\protect\citeauthoryear{Yennapureddy \& Melia}{Yennapureddy \&
  Melia}{2018}]{Yennapureddy:2018rdz}
Yennapureddy M.~K.,  Melia F.,  2018, \mn@doi [Eur. Phys. J.]
  {10.1140/epjc/s10052-018-5746-8}, C78, 258

\bibitem[\protect\citeauthoryear{Zeldovich}{Zeldovich}{1968}]{Zeldovich}
Zeldovich Y.~B.,  1968, Soviet Physics Uspekhi, 11

\bibitem[\protect\citeauthoryear{{Zhang} \& {Xia}}{{Zhang} \&
  {Xia}}{2018}]{Zhang:2018}
{Zhang} M.-J.,  {Xia} J.-Q.,  2018, \mn@doi [Nuclear Physics B]
  {10.1016/j.nuclphysb.2018.02.020}, \href
  {http://adsabs.harvard.edu/abs/2018NuPhB.929..438Z} {929, 438}

\makeatother
\end{thebibliography}

\appendix
\section{Strong-lensing systems compilation} \label{AP1}

\begin{itemize}
\item In Table \ref{tab:SL}  presents the compilation of SLS with 204 points.

\item Table \ref{tab:SLDgtone} shows the 32 systems with $D^{obs}>1$. Many of these systems appear flagged, which means that such objects are: not confirmed lenses, or have complex source structures with multiple arcs and counter-arcs, or the foreground lens is clearly composed of two distinct components, have uncertain redshift measurements, or the arcs (rings) are embebed in the light of the foreground lens. We refer the interested reader to the references presented in Table \ref{tab:SL}. We suggest that these systems with $D^{obs}>1$, shoul not be used in cosmological parameter estimation \citep[see also][]{Leaf:2018lfu}.

\end{itemize}

\begin{table*}
\caption{Compilation  of 204 strong-lensing measurements. Here, the $*$ indicates that the uncertainties were estimated.}
\label{tab:SL}
\begin{tabular}{|lllllll|}
\hline
System Name	&	Survey 	&	$z_l$  &	$z_s$  &	$\theta_{E} (\prime\prime)$  &	$\sigma_0$ (Km s$^{-1}$) &		Reference       \\
\hline   
SDSSJ0819+4534 &	SLACS 	&	0.194 &	0.446 &	0.85 &	225 $\pm$15 	 &	\cite{Auger2009}     						  \\
SDSSJ0959+4416 &	SLACS 	&	0.237 &	0.531 &	0.96 &	244 $\pm$19  &	\cite{Bolton:2008xf}  \\
SDSSJ1029+0420 &	SLACS 	&	0.104 &	0.615 &	1.01 &	210 $\pm$11 &	\cite{Bolton:2008xf} \\
SDSSJ1103+5322 &	SLACS 	&	0.158 &	0.735 &	1.02 &	196 $\pm$12 & \cite{Bolton:2008xf}  \\
SDSSJ1306+0600 &	SLACS 	&	0.173 &	0.472 &	1.32 &	237 $\pm$17  &		\cite{Auger2009} \\
SDSSJ1313+4615 &	SLACS 	&	0.185 &	0.514 &	1.37 &	221 $\pm$17 	&	\cite{Auger2009}\\
SDSSJ1318-0313 &	SLACS 	&	0.240 &	1.300 &	1.58 &	213 $\pm$18  & 	\cite{Auger2009}   \\
SDSSJ1420+6019 &	SLACS 	&	0.063 &	0.535 &	1.04 &	205 $\pm$10  &	\cite{Bolton:2008xf} \\
SDSSJ1443+0304 &	SLACS 	&	0.134 &	0.419 &	0.81 &	209 $\pm$11 & \cite{Bolton:2008xf} \\
SDSSJ1614+4522 &	SLACS 	&	0.178 &	0.811 &	0.84 &	182 $\pm$13 &	\cite{Bolton:2008xf} \\
SDSSJ1644+2625 &	SLACS 	&	0.137 &	0.610 &	1.27 &	229 $\pm$12 	&	\cite{Auger2009}\\
SDSSJ1719+2939 &	SLACS 	&	0.181 &	0.578 &	1.28 &	286 $\pm$15 	&	\cite{Auger2009} \\
HE0047-1756 &	CASTLES &	0.408 &	1.670 &	0.80 &	190 $\pm$27$^*$   &	\cite{Ofek2006} \\ 
HE0230-2130 &	CASTLES &	0.522 &	2.162 &	0.87 &	240 $\pm$34$^*$  &		\cite{Ofek2006}      \\
J0246-0825 &	CASTLES &	0.723 &	1.686 &	0.53 &	265 $\pm$37$^*$   &		\cite{Inada2005}\\
HE0435-1223 &	CASTLES &	0.454 &	1.689 &	1.22 &	257 $\pm$36$^*$  &		\cite{Morgan2005}     \\
SDSSJ092455.87+021924.9 &	CASTLES &	0.393 &	1.523 &	0.88 &	230 $\pm$32$^*$ &		\cite{Inada2003} \\
LBQS1009-0252 & CASTLES &	0.871 &	2.739 &	0.77 &	245 $\pm$34$^*$  &	\cite{1994AJ....108.1534H} \\
J1004+1229 &	CASTLES &	0.950 &	2.640 &	0.83 &	240 $\pm$34$^*$  &		\cite{Lacy2002}  \\
SDSSJ115517.35+634622.0  &	CASTLES &	0.176 &	2.888 &	0.76 &	190 $\pm$27$^*$ & \cite{Pindor2004} \\
FBQ1633+3134 &	CASTLES &	0.684 &	1.518 &	0.35 &	160 $\pm$22$^*$ 	&	\cite{Morgan:2000ap}      \\
MG1654+1346 &	CASTLES &	0.254 &	1.740 &	1.05 &	200$^{+120}_{-80}$  & \cite{rusin2003}   \\
DES J2146-0047 & DES	&	0.799 &	2.38 &	 0.68 &	215 $\pm$21$^*$ 	&	\cite{Agnello2015}  \\
SDSSJ0151+0049 & BELLS & 0.517 & 1.364 & 0.68 & 219 $\pm$39  & \cite{0004-637X-744-1-41}       \\
SDSSJ0747+5055 & BELLS & 0.438 & 0.898 & 0.75 & 328 $\pm$60  & \cite{0004-637X-744-1-41}       \\
SDSSJ0747+4448 & BELLS & 0.437 & 0.897 & 0.61 & 281 $\pm$52 & \cite{0004-637X-744-1-41}        \\
SDSSJ0801+4727 & BELLS & 0.483 & 1.518 & 0.49 & 98 $\pm$24 & \cite{0004-637X-744-1-41}        \\
SDSSJ0830+5116 & BELLS & 0.53 & 1.332  & 1.14 & 268 $\pm$36  & \cite{0004-637X-744-1-41}    \\
SDSSJ0944$-$0147 & BELLS & 0.539 & 1.179 & 0.72 & 204 $\pm$34 & \cite{0004-637X-744-1-41}   \\
SDSSJ1159$-$0007 & BELLS & 0.579 & 1.346 & 0.68 & 165 $\pm$41 &  \cite{0004-637X-744-1-41}  \\
SDSSJ1215+0047 & BELLS & 0.642 & 1.297 & 1.37 & 262 $\pm$45 &  \cite{0004-637X-744-1-41}     \\
SDSSJ1221+3806 & BELLS & 0.535 & 1.284 & 0.7 & 187 $\pm$48 & \cite{0004-637X-744-1-41}  \\
SDSSJ1234$-$0241 & BELLS & 0.49  & 1.016 & 0.53 & 122 $\pm$31 & \cite{0004-637X-744-1-41}   \\
SDSSJ1318$-$0104 & BELLS & 0.659 & 1.396 & 0.68 & 177 $\pm$27 & \cite{0004-637X-744-1-41}  \\
SDSSJ1337+3620 & BELLS & 0.564 & 1.182 & 1.39 & 225 $\pm$35 & \cite{0004-637X-744-1-41}     \\
SDSSJ1349+3612 & BELLS & 0.44  & 0.893 & 0.75 & 178 $\pm$18 & \cite{0004-637X-744-1-41}      \\
SDSSJ1352+3216 & BELLS & 0.463 & 1.034 & 1.82 & 161 $\pm$21 & \cite{0004-637X-744-1-41}     \\
SDSSJ1522+2910 & BELLS & 0.555 & 1.311 & 0.74 & 166 $\pm$27 & \cite{0004-637X-744-1-41}     \\
SDSSJ1541+1812 & BELLS & 0.56 & 1.113  & 0.64 & 174 $\pm$24 & \cite{0004-637X-744-1-41} \\
SDSSJ1542+1629 & BELLS & 0.352 & 1.023 & 1.04 & 210 $\pm$16 & \cite{0004-637X-744-1-41}      \\
SDSSJ1545+2748 & BELLS & 0.522 & 1.289 & 1.21 & 250 $\pm$37 & \cite{0004-637X-744-1-41}      \\
SDSSJ1601+2138 & BELLS & 0.544 & 1.446 & 0.86 & 207 $\pm$36 & \cite{0004-637X-744-1-41}      \\
SDSSJ1631+1854 & BELLS & 0.408 & 1.086 & 1.63 & 272 $\pm$14 & \cite{0004-637X-744-1-41}      \\
SDSSJ2122+0409 & BELLS & 0.626 & 1.452 & 1.58 & 324 $\pm$56 & \cite{0004-637X-744-1-41}      \\
SDSSJ2125+0411 & BELLS & 0.363 & 0.978 & 1.2  & 247 $\pm$17 & \cite{0004-637X-744-1-41}      \\
SDSSJ2303+0037 & BELLS & 0.458 & 0.936 & 1.02 & 274 $\pm$31 & \cite{0004-637X-744-1-41}      \\
SDSSJ0008-0004 & SLACS & 0.44  & 1.192 & 1.16  & 193 $\pm$36  & \cite{Auger2009}  \\
SDSSJ0029-0055 & SLACS & 0.227 & 0.931 & 0.96 & 229 $\pm$18  & \cite{Auger2009}  \\
SDSSJ0037-0942  & SLACS & 0.196 & 0.632 & 1.53 & 279 $\pm$10  & \cite{Auger2009} \\
SDSSJ0044+0113  & SLACS & 0.12  & 0.196 & 0.79  & 266 $\pm$13 & \cite{Auger2009}   \\
SDSSJ0109+1500 & SLACS & 0.294 & 0.525 & 0.69 & 251 $\pm$19 & \cite{Auger2009}  \\
SDSSJ0157-0056 & SLACS & 0.513 & 0.924 & 0.79 & 295 $\pm$47 & \cite{Auger2009}  \\
SDSSJ0216-0813  & SLACS & 0.332 & 0.524 & 1.16 & 333 $\pm$23 & \cite{Auger2009}   \\
SDSSJ0252+0039 & SLACS & 0.28  & 0.982 & 1.04  & 164 $\pm$12 & \cite{Auger2009}    \\
SDSSJ0330-0020 & SLACS & 0.351 & 1.071 & 1.1  & 212 $\pm$21 & \cite{Auger2009}   \\
SDSSJ0405-0455 & SLACS & 0.075 & 0.81  & 0.8   & 160 $\pm$7  & \cite{Auger2009}   \\
SDSSJ0728+3835 & SLACS & 0.206 & 0.688 & 1.25 & 214 $\pm$11  & \cite{Auger2009}  \\
SDSSJ0737+3216 & SLACS & 0.322 & 0.581 & 1    & 338 $\pm$16 & \cite{Auger2009}  \\
SDSSJ0808+4706 & SLACS & 0.219 & 1.025 & 1.23 & 236 $\pm$11 & \cite{Auger2009}  \\
SDSSJ0822+2652 & SLACS & 0.241 & 0.594 & 1.17 & 259 $\pm$15 & \cite{Auger2009}  \\
SDSSJ0841+3824 & SLACS & 0.116 & 0.657 & 1.41 & 225 $\pm$8 & \cite{Auger2009}   \\
SDSSJ0903+4116 & SLACS & 0.43  & 1.065 & 1.29  & 223 $\pm$27 & \cite{Auger2009}  \\
SDSSJ0912+0029  & SLACS & 0.164 & 0.324 & 1.63 & 326 $\pm$12 & \cite{Auger2009}  \\
\hline
\end{tabular}
\end{table*}

\begin{table*}
\contcaption{}
 \label{tab:continued}
\begin{tabular}{|lllllll|}
\hline
System Name	&	Survey 	&	$z_l$  &	$z_s$  &	$\theta_{E} (\prime\prime)$  &	$\sigma_{0}$ (Km s$^{-1}$) &		Reference \\
\hline   
SDSSJ0936+0913  & SLACS & 0.19  & 0.588 & 1.09  & 243 $\pm$11 & \cite{Auger2009}  \\
SDSSJ0946+1006  & SLACS & 0.222 & 0.608 & 1.38 & 263 $\pm$21 & \cite{Auger2009} \\
SDSSJ0956+5100 & SLACS & 0.24  & 0.47  & 1.33   & 334 $\pm$15 & \cite{Auger2009}   \\
SDSSJ0959+0410 & SLACS & 0.126 & 0.535 & 0.99 & 197 $\pm$13 & \cite{Auger2009} \\
SDSSJ1016+3859 & SLACS & 0.168 & 0.439 & 1.09 & 247 $\pm$13 & \cite{Auger2009}  \\
SDSSJ1020+1122 & SLACS & 0.282 & 0.553 & 1.2  & 282 $\pm$18 & \cite{Auger2009} \\
SDSSJ1023+4230 & SLACS & 0.191 & 0.696 & 1.41 & 242 $\pm$15 & \cite{Auger2009}  \\
SDSSJ1100+5329 & SLACS & 0.317 & 0.858 & 1.52 & 187 $\pm$23 & \cite{Auger2009} \\
SDSSJ1106+5228 & SLACS & 0.096 & 0.407 & 1.23 & 262 $\pm$9 & \cite{Auger2009}  \\
SDSSJ1112+0826 & SLACS & 0.273 & 0.63  & 1.49  & 320 $\pm$20 &\cite{Auger2009} \\
SDSSJ1134+6027 & SLACS & 0.153 & 0.474 & 1.1  & 239 $\pm$11 & \cite{Auger2009}  \\
SDSSJ1142+1001 & SLACS & 0.222 & 0.504 & 0.98 & 221 $\pm$22 & \cite{Auger2009}   \\
SDSSJ1143-0144 & SLACS & 0.106 & 0.402 & 1.68 & 269 $\pm$5 & \cite{Auger2009} \\
SDSSJ1153+4612 & SLACS & 0.18  & 0.875 & 1.05  & 269 $\pm$15 & \cite{Auger2009}   \\
SDSSJ1204+0358 & SLACS & 0.164 & 0.631 & 1.31 & 267 $\pm$17 & \cite{Auger2009}  \\
SDSSJ1205+4910 & SLACS & 0.215 & 0.481 & 1.22 & 281 $\pm$13 & \cite{Auger2009} \\
SDSSJ1213+6708 & SLACS & 0.123 & 0.64  & 1.42  & 292 $\pm$11 & \cite{Auger2009} \\
SDSSJ1218+0830 & SLACS & 0.135 & 0.717 & 1.45 & 219 $\pm$10 & \cite{Auger2009} \\
SDSSJ1250+0523 & SLACS & 0.232 & 0.795 & 1.13 & 252 $\pm$14 & \cite{Auger2009}  \\
SDSSJ1330-0148 & SLACS & 0.081 & 0.712 & 0.87 & 185 $\pm$9  & \cite{Auger2009}  \\
SDSSJ1402+6321 & SLACS & 0.205 & 0.481 & 1.35 & 267 $\pm$17 & \cite{Auger2009}  \\
SDSSJ1403+0006 & SLACS & 0.189 & 0.473 & 0.83 & 213 $\pm$17 & \cite{Auger2009} \\
SDSSJ1416+5136 & SLACS & 0.299 & 0.811 & 1.37 & 240 $\pm$25 & \cite{Auger2009}  \\
SDSSJ1430+4105 & SLACS & 0.285 & 0.575 & 1.52 & 322 $\pm$32 & \cite{Auger2009} \\
SDSSJ1436-0000 & SLACS & 0.285 & 0.805 & 1.12 & 281 $\pm$19 & \cite{Auger2009} \\
SDSSJ1451-0239 & SLACS & 0.125 & 0.52  & 1.04  & 223 $\pm$14 & \cite{Auger2009} \\
SDSSJ1525+3327 & SLACS & 0.358 & 0.717 & 1.31 & 264 $\pm$26 & \cite{Auger2009}  \\
SDSSJ1531-0105 & SLACS & 0.16  & 0.744 & 1.71  & 279 $\pm$12 & \cite{Auger2009}  \\
SDSSJ1538+5817 & SLACS & 0.143 & 0.531 & 1    & 189 $\pm$12 & \cite{Auger2009}   \\
SDSSJ1621+3931 & SLACS & 0.245 & 0.602 & 1.29 & 236 $\pm$20 & \cite{Auger2009}   \\
SDSSJ1627-0053  & SLACS & 0.208 & 0.524 & 1.23 & 290 $\pm$14 & \cite{Auger2009}   \\
SDSSJ1630+4520 & SLACS & 0.248 & 0.793 & 1.78 & 276 $\pm$16 & \cite{Auger2009}    \\
SDSSJ1636+4707 & SLACS & 0.228 & 0.674 & 1.09 & 231 $\pm$15 & \cite{Auger2009}   \\
SDSSJ2238-0754 & SLACS & 0.137 & 0.713 & 1.27 & 198 $\pm$11 & \cite{Auger2009}   \\
SDSSJ2300+0022 & SLACS & 0.228 & 0.464 & 1.24 & 279 $\pm$17 & \cite{Auger2009}   \\
SDSSJ2303+1422 & SLACS & 0.155 & 0.517 & 1.62 & 255 $\pm$16 & \cite{Auger2009}  \\
SDSSJ2321-0939 & SLACS & 0.082 & 0.532 & 1.6  & 249 $\pm$8  & \cite{Auger2009} \\
SDSSJ2341+0000 & SLACS & 0.186 & 0.807 & 1.44 & 207 $\pm$13 & \cite{Auger2009} \\
Q0047-2808 & LSD & 0.485  & 3.595  & 1.34 $\pm$0.01   & 229 $\pm$15  & \cite{Koopmans:2002qh}   \\
CFRS03-1077 & LSD & 0.938 & 2.941 & 1.24 $\pm$0.06  & 251 $\pm$19  & \cite{Treu2004}      \\
HST14176+5226  & LSD & 0.81 & 3.399 & 1.41 $\pm$0.08      & 224 $\pm$15  &  \cite{Treu2004}  \\
HST15433 & LSD & 0.497 & 2.092 & 0.36 $\pm$0.04    & 116 $\pm$10  &\cite{Treu2004}  \\
SL2SJ021247$-$055552 & SL2S & 0.75 & 2.74 & 1.27 $\pm$0.04    & 273 $\pm$22  & \cite{Sonnenfeld:2013xga}  \\
SL2SJ021325$-$074355 & SL2S & 0.717 & 3.48  & 2.39 $\pm$0.07   & 293 $\pm$34 & \cite{Sonnenfeld:2013xga} \\
SL2SJ021411$-$040502 & SL2S & 0.609 & 1.88  & 1.41 $\pm$0.04   & 287 $\pm$47 & \cite{Sonnenfeld:2013xga}  \\
SL2SJ021737$-$051329 & SL2S & 0.646 & 1.847 & 1.27 $\pm$0.04  & 239 $\pm$27 & \cite{Sonnenfeld:2013xga}  \\
SL2SJ021902$-$082934 & SL2S & 0.389 & 2.15  & 1.30 $\pm$0.04    & 289 $\pm$23 & \cite{Sonnenfeld:2013xga}  \\
SL2SJ022346$-$053418 & SL2S & 0.499 & 1.44  & 1.22 $\pm$0.11   & 288 $\pm$28 & \cite{Sonnenfeld:2013xga}  \\
SL2SJ022511$-$045433 & SL2S & 0.238 & 1.199 & 1.76 $\pm$0.05  & 234 $\pm$21 & \cite{Sonnenfeld:2013xga}   \\
SL2SJ022610$-$042011 & SL2S & 0.494 & 1.232 & 1.19 $\pm$0.04  &263 $\pm$24  & \cite{Sonnenfeld:2013xga} \\
SL2SJ023251$-$040823 & SL2S & 0.352 & 2.34  & 1.04 $\pm$0.03  & 281 $\pm$26 & \cite{Sonnenfeld:2013xga} \\
SL2SJ084847$-$035103 & SL2S & 0.682 & 1.55  & 0.85 $\pm$0.07  & 197 $\pm$21 & \cite{Sonnenfeld:2013xga} \\
SL2SJ084909$-$041226 & SL2S & 0.722 & 1.54  & 1.10 $\pm$0.03    & 320 $\pm$24 & \cite{Sonnenfeld:2013xga}   \\
SL2SJ084959$-$025142 & SL2S & 0.274 & 2.09  & 1.16 $\pm$0.04   & 276 $\pm$35 & \cite{Sonnenfeld:2013xga}  \\
SL2SJ085019$-$034710 & SL2S & 0.337 & 3.25  & 0.93 $\pm$0.03  & 290 $\pm$24 & \cite{Sonnenfeld:2013xga}  \\
SL2SJ085540$-$014730 & SL2S & 0.365 & 3.39  & 1.03 $\pm$0.04   & 222 $\pm$25 & \cite{Sonnenfeld:2013xga}  \\
SL2SJ085559$-$040917 & SL2S & 0.419 & 2.95  & 1.36 $\pm$0.10   & 281 $\pm$22 & \cite{Sonnenfeld:2013xga}  \\
SL2SJ090407$-$005952 & SL2S & 0.611 & 2.36  & 1.40 $\pm$0.04    & 183 $\pm$21 & \cite{Sonnenfeld:2013xga}  \\
SL2SJ095921+020638 & SL2S & 0.552 & 3.35  & 0.74 $\pm$0.02  & 188 $\pm$22 & \cite{Sonnenfeld:2013xga} \\
SL2SJ135949+553550 & SL2S & 0.783 & 2.77  & 1.14 $\pm$0.03   & 228 $\pm$29 & \cite{Sonnenfeld:2013xga}  \\
SL2SJ140454+520024 & SL2S & 0.456 & 1.59  & 2.55 $\pm$0.08  & 342 $\pm$20 & \cite{Sonnenfeld:2013xga} \\
SL2SJ140546+524311 & SL2S & 0.526 & 3.01  & 1.51 $\pm$0.05  & 284 $\pm$21 & \cite{Sonnenfeld:2013xga} \\
SL2SJ140650+522619 & SL2S & 0.716 & 1.47  & 0.94 $\pm$0.03  & 253 $\pm$19 & \cite{Sonnenfeld:2013xga} \\
SL2SJ141137+565119 & SL2S & 0.322 & 1.42  & 0.93 $\pm$0.03  & 214 $\pm$23 & \cite{Sonnenfeld:2013xga}  \\
SL2SJ142031+525822 & SL2S & 0.38  & 0.99  & 0.96 $\pm$0.14   & 246 $\pm$23 & \cite{Sonnenfeld:2013xga} \\
SL2SJ142059+563007 & SL2S & 0.483 & 3.12  & 1.40 $\pm$0.04    & 228 $\pm$19 & \cite{Sonnenfeld:2013xga} \\
\hline
\end{tabular}
\end{table*}

\begin{table*}
\contcaption{}
 \label{tab:continued}
\begin{tabular}{|lllllll|}
\hline
System Name	&	Survey 	&	$z_l$  &	$z_s$  &	$\theta_{E} (\prime\prime)$  &	$\sigma_{0}$ (Km s$^{-1}$) &		Reference \\
\hline   
SL2SJ220329+020518 & SL2S & 0.4   & 2.15  & 1.95 $\pm$0.06    & 213 $\pm$21 & \cite{Sonnenfeld:2013xga} \\
SL2SJ220506+014703 & SL2S & 0.476 & 2.53  & 1.66 $\pm$0.06   & 317 $\pm$30 & \cite{Sonnenfeld:2013xga}  \\
SL2SJ221326$-$000946 & SL2S & 0.338 & 3.45  & 1.07 $\pm$0.03  & 165 $\pm$20 & \cite{Sonnenfeld:2013xga} \\
SL2SJ221929$-$001743 & SL2S & 0.289 & 1.02  & 0.52 $\pm$0.13  & 189 $\pm$20 & \cite{Sonnenfeld:2013xga}   \\
SL2SJ222012+010606 & SL2S & 0.232 & 1.07  & 2.16 $\pm$0.07  & 127 $\pm$15 & \cite{Sonnenfeld:2013xga}   \\
SL2SJ222148+011542 & SL2S & 0.325 & 2.35  & 1.40 $\pm$0.05    & 222 $\pm$23 & \cite{Sonnenfeld:2013xga}  \\
SL2SJ222217+001202 & SL2S & 0.436 & 1.36  & 1.44 $\pm$0.15   & 221 $\pm$22 & \cite{Sonnenfeld:2013xga}   \\
SDSSJ002927.38+254401.7 &  BELLS  &        0.5869  &   2.4504   &    1.34     &            241 $\pm$45  &  \cite{0004-637X-833-2-264}\\
SDSSJ020121.39+322829.6 &  BELLS &        0.3957  &   2.8209   &    1.70     &            256 $\pm$20  &  \cite{0004-637X-833-2-264}      \\
SDSSJ023740.63$-$064112.9 &  BELLS  &        0.4859  &   2.2491   &    0.65     &            290 $\pm$89  &  \cite{0004-637X-833-2-264}      \\
SDSSJ074249.68+334148.9 &  BELLS  &        0.4936  &   2.3633   &    1.22     &            218 $\pm$28  &  \cite{0004-637X-833-2-264}   \\
SDSSJ075523.52+344539.5 &  BELLS  &        0.7224  &   2.6347   &    2.05     &            272 $\pm$52  &  \cite{0004-637X-833-2-264}       \\
SDSSJ085621.59+201040.5 &  BELLS  &        0.5074  &   2.2335   &    0.98     &            334 $\pm$54  & \cite{0004-637X-833-2-264}    \\
SDSSJ091807.86+451856.7 &  BELLS  &        0.5238  &   2.3440   &    0.77     &            119 $\pm$61  &  \cite{0004-637X-833-2-264}   \\
SDSSJ091859.21+510452.5 &  BELLS  &        0.5811  &   2.4030   &    1.60     &            298 $\pm$49  &  \cite{0004-637X-833-2-264}   \\
SDSSJ111027.11+280838.4 &  BELLS  &        0.6073  &   2.3999   &    0.98     &            191 $\pm$39  &  \cite{0004-637X-833-2-264}     \\
SDSSJ111634.55+091503.0 &  BELLS  &        0.5501  &   2.4536   &    1.03     &            274 $\pm$55  &  \cite{0004-637X-833-2-264}    \\
SDSSJ114154.71+221628.8 &  BELLS  &        0.5858  &   2.7624   &    1.27     &            285 $\pm$44  &  \cite{0004-637X-833-2-264}   \\
SDSSJ120159.02+474323.2 &  BELLS  &        0.5628  &   2.1258   &    1.18     &            239 $\pm$43  & \cite{0004-637X-833-2-264}    \\
SDSSJ122656.45+545739.0 &  BELLS  &        0.4980  &   2.7322   &    1.37     &            248 $\pm$26  &  \cite{0004-637X-833-2-264}   \\
SDSSJ222825.76+120503.9 &  BELLS  &        0.5305  &   2.8324   &    1.28     &            255 $\pm$50  &  \cite{0004-637X-833-2-264}   \\
SDSSJ234248.68$-$012032.5 &  BELLS  &        0.5270  &   2.2649   &    1.11     &            274 $\pm$43  &  \cite{0004-637X-833-2-264}     \\
SL2SJ020457-110309   &     SL2S  &     0.609    &  1.89      &  0.54 $\pm$0.07           &       250 $\pm$30   &   \cite{Sonnenfeld:2014gpa}  \\
SL2SJ020524-093023   &     SL2S  &     0.557    &  1.33      &  0.76 $\pm$0.09           &       276 $\pm$37   &       \cite{Sonnenfeld:2014gpa}  \\
SL2SJ021801-080247   &     SL2S  &     0.884    &  2.06      &  1.00 $\pm$0.03              &       246 $\pm$48   &       \cite{Sonnenfeld:2014gpa,Sonnenfeld:2013cha}  \\
SL2SJ023307-043838   &     SL2S  &     0.671    &  1.87      &  1.77 $\pm$0.06           &       204 $\pm$21   &       \cite{Sonnenfeld:2014gpa} \\
SDSSJ0143$-$1006 & SLACS & 0.2210 & 1.1046 &  1.23 &  203 $\pm$17  &  \cite{2017ApJ...851...48S} \\
SDSSJ0159$-$0006 & SLACS & 0.1584 & 0.7477 & 0.92 & 216 $\pm$18  &  \cite{2017ApJ...851...48S} \\
SDSSJ0324+0045 & SLACS & 0.3210 & 0.9199 & 0.55 & 183 $\pm$19 & \cite{2017ApJ...851...48S} \\
SDSSJ0324$-$0110 & SLACS & 0.4456 & 0.6239 & 0.63 & 310 $\pm$38 & \cite{2017ApJ...851...48S} \\
SDSSJ0753+3416 & SLACS & 0.1371 & 0.9628 & 1.23 & 208 $\pm$12 & \cite{2017ApJ...851...48S} \\
SDSSJ0754+1927 & SLACS & 0.1534 & 0.7401 & 1.04 & 193 $\pm$16 & \cite{2017ApJ...851...48S} \\
SDSSJ0757+1956 & SLACS & 0.1206 & 0.8326 & 1.62 & 206 $\pm$11 & \cite{2017ApJ...851...48S} \\
SDSSJ0826+5630 & SLACS & 0.1318 & 1.2907 & 1.01 & 163 $\pm$8  & \cite{2017ApJ...851...48S}\\
SDSSJ0847+2348 & SLACS & 0.1551 & 0.5327 & 0.96 & 199 $\pm$16 & \cite{2017ApJ...851...48S} \\
SDSSJ0851+0505 & SLACS & 0.1276 & 0.6371 & 0.91 & 175 $\pm$11 & \cite{2017ApJ...851...48S} \\
SDSSJ0920+3028 & SLACS & 0.2881 & 0.3918 & 0.70 & 297 $\pm$17 & \cite{2017ApJ...851...48S} \\
SDSSJ0955+3014 & SLACS & 0.3214 & 0.4671 & 0.54 & 271 $\pm$33 & \cite{2017ApJ...851...48S} \\
SDSSJ0956+5539 & SLACS & 0.1959 & 0.8483 & 1.17 & 188 $\pm$11 & \cite{2017ApJ...851...48S} \\
SDSSJ1010+3124 & SLACS & 0.1668 & 0.4245 & 1.14 & 221 $\pm$11 & \cite{2017ApJ...851...48S} \\
SDSSJ1031+3026 & SLACS & 0.1671 & 0.7469 & 0.88 & 197 $\pm$13 & \cite{2017ApJ...851...48S} \\
SDSSJ1040+3626 & SLACS & 0.1225 & 0.2846 & 0.59 & 186 $\pm$10 & \cite{2017ApJ...851...48S} \\
SDSSJ1041+0112 & SLACS & 0.1006 & 0.2172 & 0.60 & 200 $\pm$7  & \cite{2017ApJ...851...48S} \\
SDSSJ1048+1313 & SLACS & 0.1330 & 0.6679 & 1.18 & 195 $\pm$10 & \cite{2017ApJ...851...48S} \\
SDSSJ1051+4439 & SLACS & 0.1634 & 0.5380 & 0.99 & 216 $\pm$16 & \cite{2017ApJ...851...48S} \\
SDSSJ1056+4141 & SLACS & 0.1343 & 0.8318 & 0.72 & 157 $\pm$10 & \cite{2017ApJ...851...48S} \\
SDSSJ1101+1523 & SLACS & 0.1780 & 0.5169 & 1.18 & 270 $\pm$15 & \cite{2017ApJ...851...48S}\\
SDSSJ1116+0729 & SLACS & 0.1697 & 0.6860 & 0.82 & 190 $\pm$11 & \cite{2017ApJ...851...48S} \\
SDSSJ1127+2312 & SLACS & 0.1303 & 0.3610 & 1.25 & 230 $\pm$9  & \cite{2017ApJ...851...48S} \\
SDSSJ1137+1818 & SLACS & 0.1241 & 0.4627 & 1.29 & 222 $\pm$8  & \cite{2017ApJ...851...48S}\\
SDSSJ1142+2509 & SLACS & 0.1640 & 0.6595 & 0.79 & 159 $\pm$10 & \cite{2017ApJ...851...48S} \\
SDSSJ1144+0436 & SLACS & 0.1036 & 0.2551 & 0.76 & 207 $\pm$14 & \cite{2017ApJ...851...48S} \\
SDSSJ1213+2930 & SLACS & 0.0906 & 0.5954 & 1.35 & 232 $\pm$7  & \cite{2017ApJ...851...48S} \\
SDSSJ1301+0834 & SLACS & 0.0902 & 0.5331 & 1.00 & 178 $\pm$8  & \cite{2017ApJ...851...48S} \\
SDSSJ1330+1750 & SLACS & 0.2074 & 0.3717 & 1.01 & 250 $\pm$12 & \cite{2017ApJ...851...48S} \\
SDSSJ1403+3309 & SLACS & 0.0625 & 0.7720 & 1.02 & 190 $\pm$6  & \cite{2017ApJ...851...48S} \\
SDSSJ1430+6104 & SLACS & 0.1688 & 0.6537 & 1.00 & 180 $\pm$15 & \cite{2017ApJ...851...48S} \\
SDSSJ1433+2835 & SLACS & 0.0912 & 0.4115 & 1.53 & 230 $\pm$6  & \cite{2017ApJ...851...48S} \\
SDSSJ1541+3642 & SLACS & 0.1406 & 0.7389 & 1.17 & 194 $\pm$11 & \cite{2017ApJ...851...48S} \\
SDSSJ1543+2202 & SLACS & 0.2681 & 0.3966 & 0.78 & 285 $\pm$16 & \cite{2017ApJ...851...48S} \\
SDSSJ1550+2020 & SLACS & 0.1351 & 0.3501 & 1.01 & 243 $\pm$9  & \cite{2017ApJ...851...48S} \\
SDSSJ1553+3004 & SLACS & 0.1604 & 0.5663 & 0.84 & 194 $\pm$15 & \cite{2017ApJ...851...48S} \\
SDSSJ1607+2147 & SLACS & 0.2089 & 0.4865 & 0.57 & 197 $\pm$16 & \cite{2017ApJ...851...48S} \\
SDSSJ1633+1441 & SLACS & 0.1281 & 0.5804 & 1.39 & 231 $\pm$9  & \cite{2017ApJ...851...48S} \\
SDSSJ2309$-$0039 & SLACS & 0.2905 & 1.0048 & 1.14 & 184 $\pm$13 & \cite{2017ApJ...851...48S} \\
\hline
\end{tabular}
\end{table*}

\begin{table*}
\contcaption{}
 \label{tab:continued}
\begin{tabular}{|lllllll|}
\hline
system name	&	survey 	&	zl  &	zs  &	$\theta_{E} (\prime\prime)$  &	$\sigma_{0}$ (Km s$^{-1}$)&		reference   \\
\hline   
SDSSJ2324+0105 & SLACS & 0.1899 & 0.2775 & 0.59 & 245 $\pm$15 & \cite{2017ApJ...851...48S}  \\
Q0142-100  &	CASTLES &	0.491 &	2.719 &	1.12 &	246$^{+17}_{-47}$ &	\cite{rusin2003} \citet{Leier2011}   \\
MG0414+0534  &	CASTLES &	0.958 &	2.639 &	1.19 &	247$^{+10}_{-13}$ &	\cite{rusin2003} \citet{Leier2011} \\
B0712+472  &	CASTLES &	0.406 &	1.339 &	0.71 &	164$\pm$7 & \cite{rusin2003}	\citet{Leier2011} \\
HS0818+1227  &	CASTLES &	0.390 &	3.115 &	1.42 &	246$^{+17}_{-42}$ &	\cite{rusin2003} \citet{Leier2011} \\
B1030+074 &	CASTLES &	0.599 &	1.535 &	0.78 &	257$^{+10}_{-18}$ & \citet{rusin2003}	\citet{Leier2011}  \\
HE1104-1805 &	CASTLES &	0.729 &	2.303 &	1.595 &	303$^{+11}_{-10}$ & \cite{rusin2003}	\citet{Leier2011} \\
B1608+656  &	CASTLES &	0.630 &	1.394 &	1.135 &	267$^{+6}_{-15}$ & \citet{rusin2003}	\citet{Leier2011} \\
HE2149-2745  &	CASTLES &	0.603 &	2.033 &	0.85 &	191$^{+7}_{-9}$	& \cite{rusin2003}	\citet{Leier2011}   \\
MG1549+3047 &	CASTLES &	0.111 &	1.170 &	1.15 &	227$\pm$18 &  \cite{rusin2003}	\cite{Lehar1996}   \\
\hline
\end{tabular}
\end{table*}

\begin{table*}
\begin{tabular}{|llll|}
\hline
System Name	&	D$^{obs}$ & $\theta_E (\prime\prime)$ & Flag	\\
\hline 
SDSSJ1318-0313 &	1.207548476671243 & 1.58 &\\
SDSSJ0801+4727&	 1.7690928325399644 & 0.49 & $\dagger$\\
SDSSJ1234$-$0241&	 1.234704134938535& 0.53 & $\dagger$\\
SDSSJ1352+3216 &	2.4345927827771154 &  1.82 & $\dagger$\\
SDSSJ0008-0004 & 
1.079816763956848
 &  1.16 & $\dagger$\\
SDSSJ0252+0039 &	1.3407639908720494 & 1.04 & $\dagger$\\
SDSSJ0405-0455 &	1.0835693599307283 & 0.8 & $\dagger$\\
SDSSJ1100+5329 &	1.5071867558989571 & 1.52 &\\
SDSSJ1218+0830 &	1.0483021267443504 & 1.45 &\\
SDSSJ2238-0754 &	1.123259330363861 & 1.27 & $\dagger$\\
SDSSJ2341+0000 &	1.1652751780813544 &  1.44 & $\dagger$\\
SL2SJ022511$-$045433 &	1.1145194380761674 & 1.76 &\\
SL2SJ090407$-$005952 &	1.4495478313743804 & 1.40 & \\
SL2SJ220329+020518&	 1.4903288161448882 & 1.95 & $\dagger$\\
SL2SJ221326$-$000946 &	1.362770060019399 & 1.07 & $\dagger$\\
SL2SJ222012+010606 &	4.643580764983071 & 2.16 & $\dagger$\\
SL2SJ222217+001202 &	1.022314778682008 & 1.44 & $\dagger$\\
SDSSJ091807.86+451856.7 &	1.8853999737796163 & 0.77 & $\dagger$\\
SL2SJ023307-043838 &	1.4747541461571618 & 1.77 &\\
SDSSJ0143$-$1006 &	1.0349508604157698 & 1.23 &\\
SDSSJ0757+1956 &	1.323692987529667 & 1.62 & $\dagger$\\
SDSSJ0826+5630 &	1.318113655499309 & 1.01 & $\dagger$\\
SDSSJ0851+0505 &	1.0303196656712754 & 0.91 & $\dagger$\\
SDSSJ0956+5539 &	1.147828113281079 & 1.17  & $\dagger$\\
SDSSJ1048+1313 &	1.0760178574880814 & 1.18 &\\
SDSSJ1056+4141 &	1.0128377643232576 & 0.72 & $\dagger$\\
SDSSJ1142+2509 &	1.0835264989141573 & 0.79 & $\dagger$\\
SDSSJ1301+0834 &	1.094376326151474 & 1.0 & $\dagger$\\
SDSSJ1430+6104 &	1.0701919604254106 & 1.0 & $\dagger$\\
SDSSJ1433+2835 &	1.0028649501362656 & 1.53 & $\dagger$\\
SDSSJ1541+3642 &	1.0779263693220975 & 1.17 & $\dagger$\\
SDSSJ2309$-$0039 &	1.1675511061635446 & 1.14 &\\
\hline
\end{tabular}
\caption{32 gravitational lens systems with $D^{\rm{obs}}>1$.}
\label{tab:SLDgtone}
\end{table*}

\section{The effect of $f$ as an individual and free parameter for each SLS measurement} \label{Fseparada}

In the third test (FS with $f$) presented in Table \ref{tab:wCDMCPLyJBPin}, we estimated cosmological parameters under the hypothesis that all the measurements has the same value for the $f$ parameter in Eq. \eqref{Df}. This corrective parameter, has been introduced to explain unknown systematics that can be produced by some of the assumptions mentioned in section \ref{data}. However, each SLS has been study under different circumstances and employing different telescopes and instruments. Therefore, the unknown systematics should not be the same, i.e the value for the corrective parameter $f$ must be different in each case. To quantify this, we
carried out an optimization for the parameters considering $f$ as an individual and free parameter for each SLS. We performed the optimization for the $\omega$CDM model through the differential evolution method from scipy python package. We assume the same priors for the $\omega$ ($-4 < \omega < 1$) and $\Omega_{m0}$ (gaussian prior of $\Omega_{0m}=0.3111 \pm 0.0056$) parameters, nevertheless, to assess the impact of the possible systematic errors, we consider a bigger interval for the $f$ parameter within the following range $0.6<f<1.5$.

\begin{figure}
\centering
\includegraphics[width=8cm,scale=0.45]{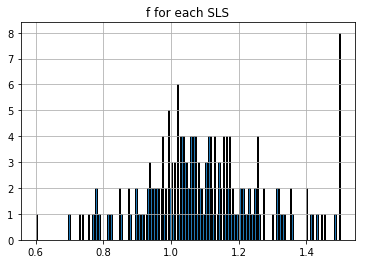} 
\includegraphics[width=8cm,scale=0.45]{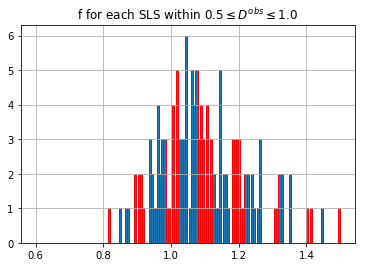} 
\caption{Probability density function (PDF) histogram of the best fit values of $f$ for the complete sample (top panel) and the SS3 sample $0.5<D^{obs}<1$ (bottom panel) for the $\omega$CDM model, assuming $f$ as an individual and free parameter for each SLS.}
\label{fig:fseparada}
\end{figure}

\begin{figure}
\centering
\includegraphics[width=8cm,scale=0.45]{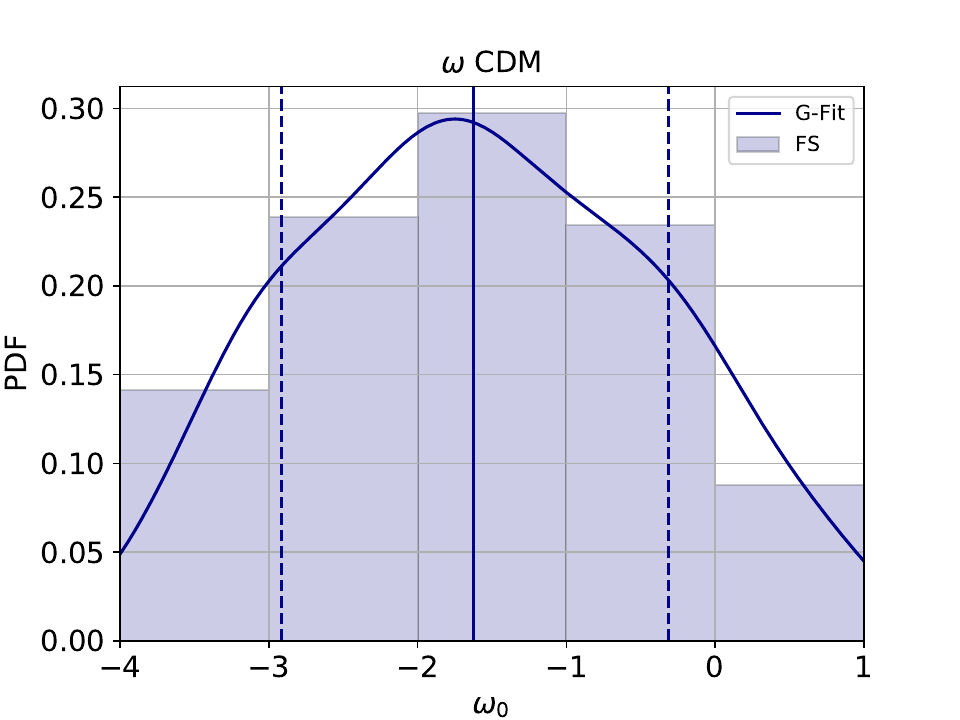} 
\caption{Histogram of the best fit values of $\omega_{0}$ for the FS sample  ($\omega$CDM model). The continuous line represents a Gaussian kernel fit. The mean value is indicated with a vertical solid line, and with dashed vertical lines the $1\sigma$ values.}
\label{fig:w_fsep}
\end{figure}

\begin{figure}
\centering
\includegraphics[width=8cm,scale=0.45]{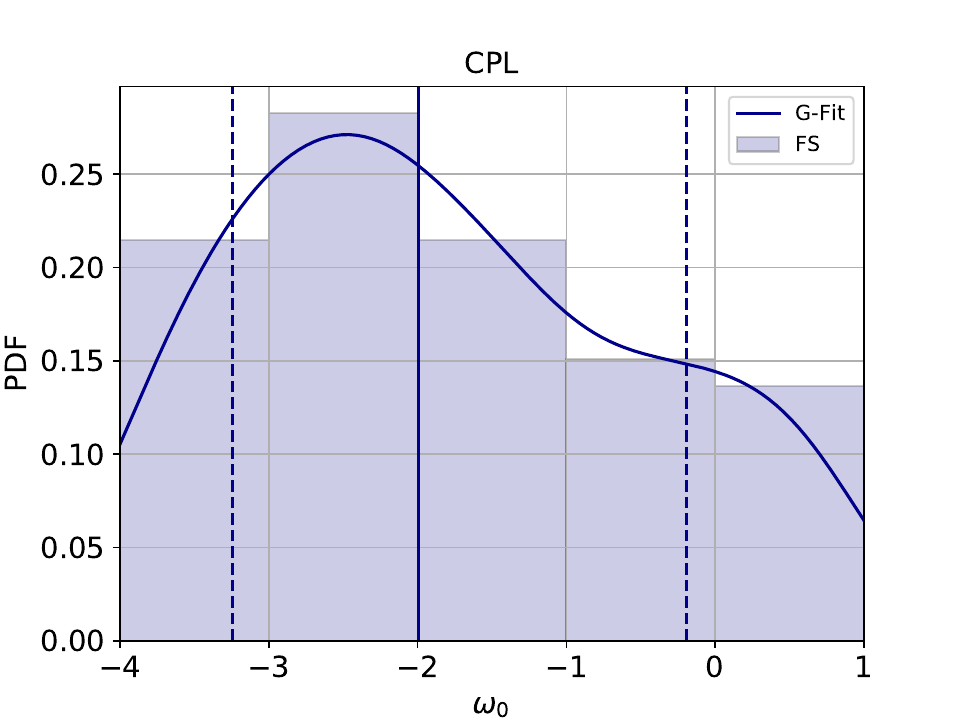} \\
\includegraphics[width=8cm,scale=0.45]{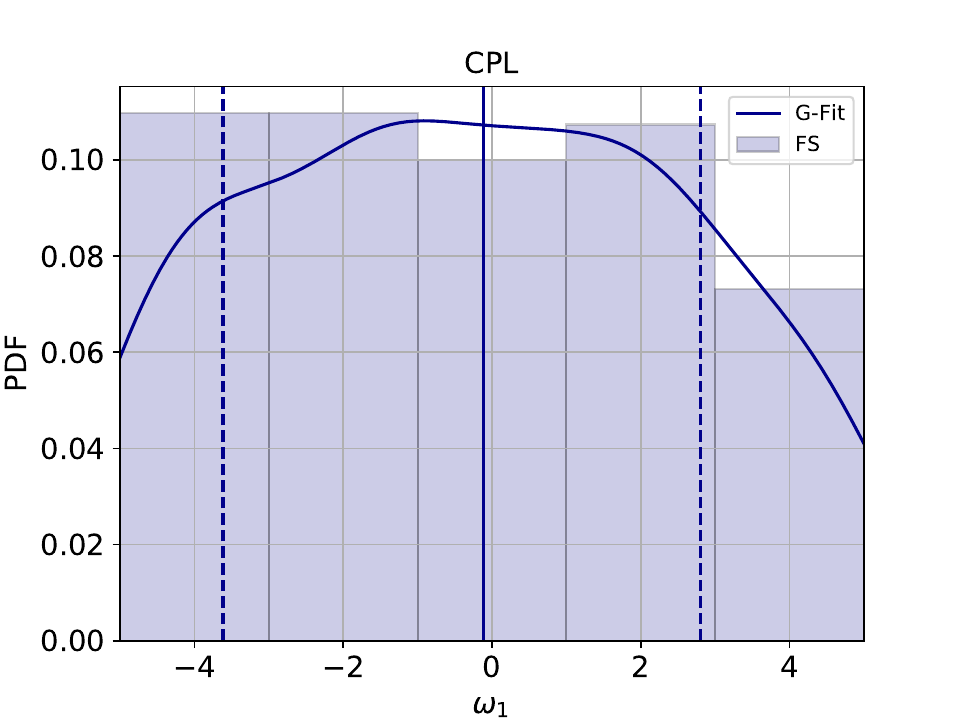} 
\caption{Histogram of the best fit values of $\omega_{0}$ (top panel) and $\omega_{1}$ (bottom panel) of the FS in the CPL model. The continuous line represents a Gaussian kernel fit. The mean value is indicated with a vertical solid line, and with dashed vertical lines the $1\sigma$ values.}
\label{fig:cpl_fsep}
\end{figure}

\begin{figure}
\centering
\includegraphics[width=8cm,scale=0.45]{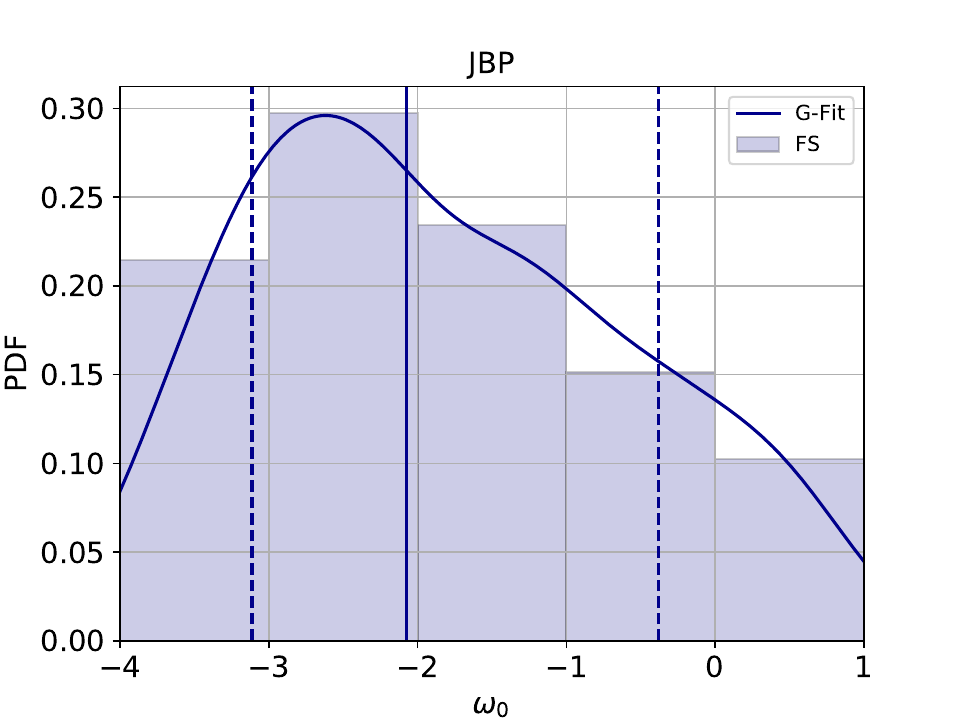} \\
\includegraphics[width=8cm,scale=0.45]{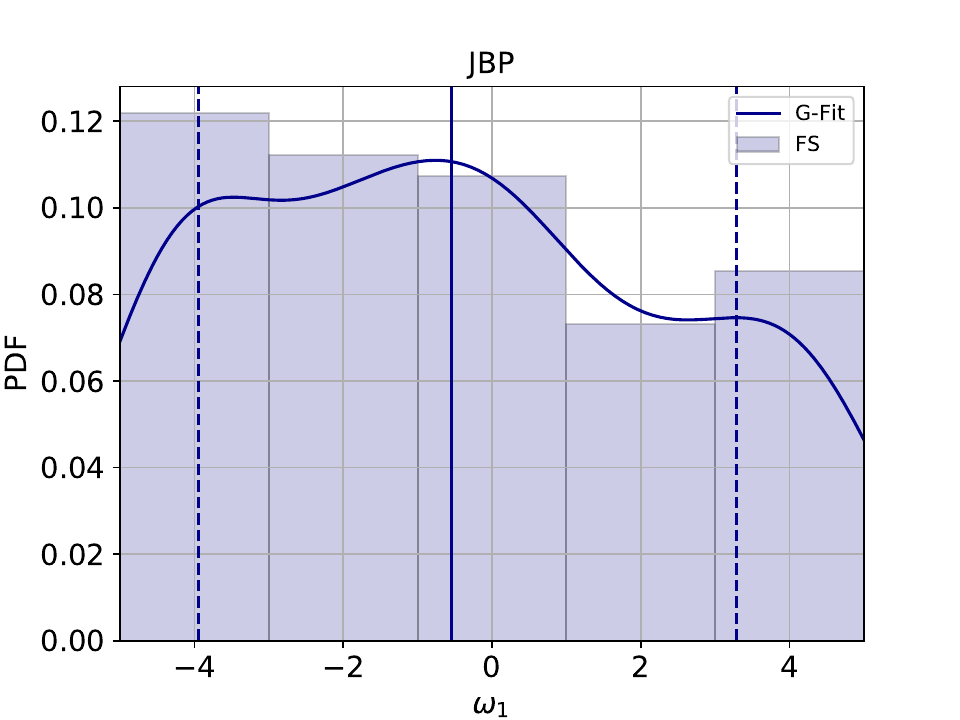} 
\caption{Histogram of the best fit values of $\omega_{0}$ (top panel) and $\omega_{1}$ (bottom panel) of the FS for the JBP model. The continuous line represents a Gaussian kernel fit. The mean value is indicated with a vertical solid line, and with dashed vertical lines the $1\sigma$ values.}
\label{fig:jbp_fsep}
\end{figure}

Figure \ref{fig:fseparada} shows the histogram of the values obtained for the $f$ parameter when the complete and SS3 sample ($0.5 \leq D^{obs} \leq 1$) are considered. 
For the complete sample, this result show that many  systems are outside of the region proposed by \citet{Treu:2006ApJ}, and \citet{Ofek:2003sp}. In particular nine systems have extreme $f$ values, eight with  $f=1.5$ and one at $f=0.6$. Which suggests that some measured systems have larger systematic errors. However, when we use the SS3 sample, only one system take a value of $f=1.5$ and the remain systems that are beyond the mentioned region are considerably less. This result indicates that the corrective  parameter $f$ decreases when the SS3 sample is considered, and it is consistent with the results obtained for the three cosmological models, obtaining the lowest values for $\chi^{2}_{red}$ (see table \ref{tab:wCDMCPLyJBPin}). Nevertheless, the two samples have similar $f$ mean values, obtaining $f=1.10$ and $f=1.09$ for the complete and SS3 sample respectively. Following the same procedure, we optimize the cosmological parameters for the CPL, and JBP models along with $f$ (as individual free parameter in each SLS). Obtaining the same behavior as in the $\omega$CDM scenario for both samples. Therefore,  under  the  assumption that  each  SLS has its own associated $f$,  we  do  not  expect significant differences in the cosmological constraints. Indeed, in Figures \ref{fig:w_fsep}-\ref{fig:jbp_fsep} we show the histograms of the cosmological parameter best fits when $f$ takes different values in each data point on the FS sample. To compare with the MCMC analysis of section \ref{Res}, we fit a Gaussian kernel to these histograms and calculate the mean values and their uncertainties. We obtain for the $w$CDM model $w_{0}=-1.62^{+1.31}_{-1.28}$. This result is consistent with the value reported in Table \ref{tab:wCDMCPLyJBPin} (FS+$f$) within $\sim1.02\sigma$. For the optimization in the CPL model, we obtain the constraints $w_{0}=-1.99^{+1.80}_{-1.25}$ and $w_{1}=-0.10^{+2.91}_{-3.50}$, which are also consistent with the MCMC  results within $\sim 0.83\sigma$, and $\sim 0.32\sigma$ respectively. Similarly, for the JBP model, the individual optimization gives $w_{0}=-2.07^{+1.69}_{-1.03}$, and $w_{1}=-0.53^{+3.82}_{-3.40}$. These values are in agrement within $\sim 0.58\sigma$, and $\sim 0.43\sigma$, with the values reported in Table \ref{tab:wCDMCPLyJBPin}.

\section{Mock data with Strong-lensing systems} \label{APII}

In  this  section,  we  use  a  mock  data  set  of  SLS  to estimate the effect on the $f$ parameter when we use the $\theta_E$ values reported in the literature.

\begin{figure}
\centering
\includegraphics[width=8.5cm,scale=0.65]{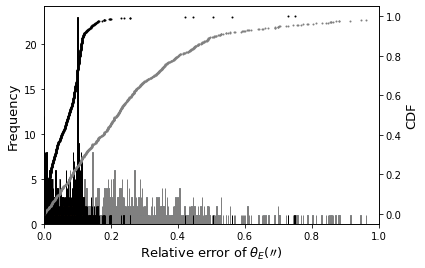} 
\includegraphics[width=8.5cm,scale=0.65]{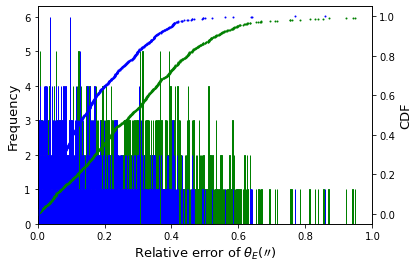} 
\caption{Relative error's histograms. \textit{Top panel.-} Calculated using the simulated $\theta_E$ asuming a SIE and SIE$+ \alpha$, (black and gray colors respectively), and the $\theta_E$ from a SIE model fitting.  \textit{Bottom panel.-} Calculated using the simulated $\theta_E$ (asuming a SIS and SIS$+ \alpha$, blue and green colors respectively) and the $\theta_E$ from a SIE model fitting. The dotted line in both panels shows the cumulative distribution function (CDF).}
\label{fig:Mock}
\end{figure}

To simulate our data we procede as follow: mimicking the same distribution of $z_{s}$ of our SLS sample (within the range $0.196< z_{s} < 3.595$), we generate an aleatory sample of 1000 sources.  Then, for each source we calculate the most probably redshift of the lens, $z_{l}$, following the work of \citet{1984ApJ...284....1T}. Since not all the obtained $z_{l}$ were physically possible, the final sample was reduced to 788 lenses.  Later, we randomly associate to each one of the 788 lenses, their parameters, i. e., velocity dispersion ($98 < \sigma < 342$), ellipticity (0.5 $< \epsilon <$ 0.9), position angle ($0^{\circ} < \theta < 180^{\circ}$), and external shear (0. $<  \gamma <$ 0.2). In addition, following the same procedure, we have simulated SLS assuming an isothermal and ellipitical power-law mass distribution, with a slope $\alpha$ between $1.6<\alpha<2.6$ obtaining $\sim 788$ systems. We considered the definition for the Einstein radius following the work of \citet{Vegetti:2014lqa} for both approaches. Finally, these mock  SLS were fitted using the GRAVLENS code \citep{2011ascl.soft02003K} using a SIE model, and assuming a  $\Lambda$CDM Cosmology ($\Lambda$= 0.73, $\Omega_{m}$ = 0.27, $h$ = 0.7). Figure \ref{fig:Mock} shows the histograms of the relative error between the simulated Einstein radius (from the SIE and SIE$+\alpha$ lens models, and assuming a SIS SIS$+\alpha$) and the fitted SIE Einstein radius. In addition, we overplot the cumulative distribution function (CDF) for all the cases. The comparison (through the relative error) between the simulated SIE $\theta_{E}$, and the fitted one, gives as a result that approximately the 80$\%$ of the sample have an error less than the 10$\%$. However, comparing the simulated SIE$+\alpha$ $\theta_{E}$ and the fitted SIE $\theta_{E}$, gives an error of $\sim 35\%$ or less for the majority of the data (80$\%$) On the other hand, when we compare the simulated $\theta_E$ (assuming a SIS) with the $\theta_E$ obtained from the fit, we obtain that  approximately the 80$\%$ of the objects have an error less than the 25$\%$. Moreover, when the SIS$+\alpha$ $\theta_{E}$ is compared with the fitted SIE, the error increases to 45$\%$ or less for the 80$\%$ of the objects. We can interprate these results as follows. If we adopt a simple SIS $\theta_{E}$ to characterize a lens, we could have a worse estimate of the lens mass, i.e. we will  have a bigger scatter on the value of the velocity dispersion of the lens. This scatter could be inherited to the corrective parameter $f$ used in Eq. \eqref{Df}. Moreover, assuming a power-law mass distribution with a SIS or SIE lens model, the lens mass can not be modeled accurately enough obtaining a worst estimate on the value of the velocity dispersion. These results can be related to the values obtained for the $f$ parameter considered as an independent free parameter for each system (see appendix \ref{Fseparada}), in wich some systems are outside the range proposed by \cite{Ofek:2003sp}. 
This problem could be addressed by both including an external shear parameter in complex systems (i.e. quadruply lensed quasars, systems with lensed arcs) and by estimating cosmological parameters and lens modelling simultaneously. Even though some simulated systems present a bigger scatter with respect to the fitted SIE $\theta_E$, neither of these systems shows a nonphysical value for the observational lens equation.

\bsp	
\label{lastpage}
\end{document}